\documentclass[zpreprint,zbstdefault]{./LaTeX/zeus/zeus_paper}
\usepackage[english]{babel}

\newcommand{\ZcoosysB}{%
The ZEUS coordinate system is a right-handed Cartesian system, with the $Z$
axis pointing in the proton beam direction, referred to as the ``forward
direction'', and the $X$ axis pointing left towards the centre of HERA.
The coordinate origin is at the nominal interaction point.\xspace}
\newcommand{\Zpsrap}{%
The pseudorapidity is defined as $\eta=-\ln\left(\tan\frac{\theta}{2}\right)$,
where the polar angle, $\theta$, is measured with respect to the proton beam
direction.\xspace}

\newcommand{\ZcoosysfnBeta}{\footnote{\ZcoosysB\Zpsrap}}

\chardef\usc=95
\chardef\til=126
\catcode`\@=11 
\DeclareRobustCommand\xdotspace{\futurelet\@let@token\@xdotspace}
\def\@xdotspace{%
  \ifx\@let@token.\else
  \ifx\@let@token\bgroup.\else
  \ifx\@let@token\egroup.\else
  \ifx\@let@token\/.\else
  \ifx\@let@token\ .\else
  \ifx\@let@token~.\else
  \ifx\@let@token!.\else
  \ifx\@let@token,.\else
  \ifx\@let@token:.\else
  \ifx\@let@token;.\else
  \ifx\@let@token?.\else
  \ifx\@let@token/.\else
  \ifx\@let@token'.\else
  \ifx\@let@token).\else
  \ifx\@let@token-.\else
  \ifx\@let@token\@xobeysp.\else
  \ifx\@let@token\space.\else
  \ifx\@let@token\@sptoken.\else
   .\space
   \fi\fi\fi\fi\fi\fi\fi\fi\fi\fi\fi\fi\fi\fi\fi\fi\fi\fi}
\catcode`\@=12 

\newcommand{\stru}[2]{%
   \relax\ifmmode\hbox{\vrule height#1 depth#2 width0pt}%
   \else\vrule height#1 depth#2 width0pt\fi}

\newcommand{\Ronum}[1]{\uppercase\expandafter{\romannumeral#1}}
\newcommand{\ronum}[1]{\expandafter{\romannumeral#1}}
\DeclareRobustCommand{\LaTeXZ}{%
  \LaTeX\kern-.05em4\kern-.1em
  {\raisebox{-0.2ex}{$\scriptstyle\text{ZEUS}$}}\xspace}

\newcommand{\eq}[1]{Eq.~(\ref{eq-#1})}
\newcommand{\eqsand}[2]{Eqs.~(\ref{eq-#1}) and~(\ref{eq-#2})}

\newcommand{\fig}[1]{Fig.~\ref{fig-#1}}
\newcommand{\Fig}[1]{Figure~\ref{fig-#1}}

\newcommand{\tab}[1]{Table~\ref{tab-#1}}

\newcommand{\taband}[2]{Tables~\ref{tab-#1} and~\ref{tab-#2}}

\newcommand{\sect}[1]{Section~\ref{sec-#1}}

\DeclareMathAlphabet{\mathbf}{OT1}{cmr}{bx}{sl}
\newcommand{\eVdist}{\kern-0.06667em}

\newcommand{\Gev}{{\text{Ge}\eVdist\text{V\/}}}

\newcommand{\mev}{{\,\text{Me}\eVdist\text{V\/}}}
\newcommand{\gev}{{\,\text{Ge}\eVdist\text{V\/}}}

\newcommand{\pb}{\,\text{pb}}

\newcommand{\mm}{\,\text{mm}}
\newcommand{\cm}{\,\text{cm}}

\newcommand{\ns}{\,\text{ns}}

\newcommand{\rad}{\,\text{rad}}

\newcommand{\mrad}{\,\text{mrad}}

\newcommand{\Tesla}{\,\text{T}}

\newcommand{\slashfrac}[2]{%
  \raisebox{0.5ex}{\ensuremath #1}\kern-0.12em/\kern-0.08em
  \raisebox{-.8ex}{\ensuremath #2}}

\newcommand{\sqr}[3]{%
    {\vcenter{\hrule height.#3ex\hbox{\vrule width.#2ex height#1ex
     \kern#1ex\vrule width.#3ex}\hrule height.#2ex}}}

\catcode`\@=11 
\newcommand{\parenbar}{\mathpalette\p@renb@r}
\def\p@renb@r#1#2{\vbox{%
  \ifx#1\scriptscriptstyle \dimen@.7em\dimen@ii.2em\else
  \ifx#1\scriptstyle \dimen@.8em\dimen@ii.25em\else
  \dimen@1em\dimen@ii.4em\fi\fi \offinterlineskip
  \ialign{\hfill##\hfill\cr
    \vbox{\hrule width\dimen@ii}\cr
    \noalign{\vskip-.3ex}%
    \hbox to\dimen@{$\mathchar300\hfil\mathchar301$}\cr
    \noalign{\vskip-.3ex}%
    $#1#2$\cr}}}
\catcode`\@=12 

\newcommand{\gh}{{\gamma_h}}

\newcommand{\IP}{{\rm I$\kern-0.01667em$P}\xspace}
\newcommand{\JB}{{\rm JB}}

\newcommand{\had}{{\rm had}}

\mathchardef\qsm=63
\mathchardef\pls=43
\mathchardef\mns=512
\mathchardef\plm=518
\mathchardef\eql=61
\mathchardef\smallleft=300
\mathchardef\smallright=301
\mathchardef\les=316
\mathchardef\gre=318
\mathchardef\leq=532
\mathchardef\grq=533

\catcode`\@=11 
\newcounter{pict@width}
\newcounter{pict@height}
\newlength{\pict@scale}
\newcommand{\psfigadd}[4]{%
\setcounter{pict@width}{1*\ratio{#2+\pict@scale/2}{\pict@scale}}
\setcounter{pict@height}{1*\ratio{#3+\pict@scale/2}{\pict@scale}}
\setlength{\unitlength}{\pict@scale}
\hbox to #2{\hspace{-\fill}\begin{picture}(\thepict@width,\thepict@height)
\put(0,0){\psfig{figure=#1,width=#2,height=#3,clip=}}
\SetScale{0.283466457}
\SetWidth{1.763889}
{#4}
\end{picture}}
}
\newcounter{pict@widthfst}
\newcounter{pict@widthscd}
\newcounter{pict@widthtot}
\newcommand{\psfigaddtwo}[7]{%
\setcounter{pict@widthfst}{1*\ratio{#2+\pict@scale/2}{\pict@scale}}
\setcounter{pict@widthscd}{1*\ratio{#2+#4+\pict@scale/2}{\pict@scale}}
\setcounter{pict@widthtot}{1*\ratio{#2+#4+#6+\pict@scale/2}{\pict@scale}}
\setcounter{pict@height}{1*\ratio{#3+\pict@scale/2}{\pict@scale}}
\setlength{\unitlength}{\pict@scale}
\hbox{\hspace{-\fill}\begin{picture}(\thepict@widthtot,\thepict@height)
\put(0,0){\psfig{figure=#1,width=#2,height=#3,clip=}}
\put(\thepict@widthscd,0){\psfig{figure=#5,width=#6,height=#3,clip=}}
\SetScale{0.283466457}
\SetWidth{1.763889}
{#7}
\end{picture}}
}
\newcommand{\psfigror}[4]{%
\setcounter{pict@width}{1*\ratio{#2+\pict@scale/2}{\pict@scale}}
\setcounter{pict@height}{1*\ratio{#3+\pict@scale/2}{\pict@scale}}
\setlength{\unitlength}{\pict@scale}
\hbox{\begin{picture}(\thepict@width,\thepict@height)
\put(0,\thepict@height){\psfig{figure=#1,width=#3,height=#2,clip=,angle=270}}
\SetScale{0.283466457}
\SetWidth{1.763889}
{#4}
\end{picture}}
}
\newcommand{\psfigrol}[4]{%
\setcounter{pict@width}{1*\ratio{#2+\pict@scale/2}{\pict@scale}}
\setcounter{pict@height}{1*\ratio{#3+\pict@scale/2}{\pict@scale}}
\setlength{\unitlength}{\pict@scale}
\hbox{\begin{picture}(\thepict@width,\thepict@height)
\put(0,0){\psfig{figure=#1,width=#3,height=#2,clip=,angle=90}}
\SetScale{0.283466457}
\SetWidth{1.763889}
{#4}
\end{picture}}
}
\catcode`\@=12 
\newlength\listtextwidth

\catcode`\@=11 
\newlength{\@tabfninsert}
\newlength{\@tabfnwidth}
\newcommand{\tabfootnote}[2]{%
  \setlength{\@tabfninsert}{0.8em}
  \setlength{\@tabfnwidth}{\textwidth}
  \addtolength{\@tabfnwidth}{-\@tabfninsert}
  \addtolength{\@tabfnwidth}{-0.4em}
  \noindent\makebox[\@tabfninsert][r]{\footnotesize$^{#1}$\hfil}\hfill%
  \parbox[t]{\@tabfnwidth}{\footnotesize #2\hfill}}
\catcode`\@=12 

\def\@evenhead{\underline{\protect\parbox{\textwidth}{\bf\boldmath\protect\rule[-0.2cm]{0pt}{0.2cm}{\Large\thepage}\hfill\leftmark\hfill~}}}%
\def\@oddhead{\underline{\protect\parbox{\textwidth}{\bf\boldmath\protect\rule[-0.2cm]{0pt}{0.2cm}~\hfill\rightmark\hfill{\Large\thepage}}}}%

\newcommand{\etamax}{{\eta_{\rm{max}}}}

\newcommand{\trk}{{\rm trk}}

\newcommand{\Pom}{{\text{I\kern-0.12em P}}}

\newcommand{\qq}       {\mbox{$Q^{2}$}}
\newcommand{\onehalf}  {\mbox{${\textstyle{\frac{1}{2}}}$}}
\newcommand{\chiz}     {\mbox{$\chi_{_{Z}}$}}
\newcommand{\sthw}     {\mbox{$\sin^2\!\thw$}}
\newcommand{\thw}      {\mbox{$\theta_{\scriptscriptstyle W}$}}
\newcommand{\cthw}     {\mbox{$\cos^2\!\thw$}}
\newcommand{\Mzq}      {\mbox{$M_Z^{2}$}}

\newcommand{\sigqq}    {\mbox{$d\sigma/d\qq$}}
\newcommand{\gevv}     {\mbox{${\rm Ge\kern -0.1em V}^{2}$}}
\newcommand{\Et}       {\mbox{$E_{T}$}}

\newcommand{\Pt}       {\mbox{$P_{T}$}}
\newcommand{\qqda}     {\mbox{$Q^{2}\dab$}}
\newcommand{\xda}      {\mbox{$x\dab$}}
\newcommand{\yda}      {\mbox{$y\dab$}}
\newcommand{\dab}      {_{\rm \scriptscriptstyle DA}}
\newcommand{\jbb}      {_{\rm \scriptscriptstyle JB}}

\newcommand{\sigx}     {\mbox{$d\sigma/dx$}}
\newcommand{\sigy}     {\mbox{$d\sigma/dy$}}

\newcommand{\qqc}      {\mbox{$Q^{2}_{c}$}}
\newcommand{\lw}[1]{\smash{\lower1.8ex\hbox{#1}}}

\def\RefDJango{{\cite{%
proc:hera:1991:1419,*spi:www:djangoh11%
}}\xspace}
\def\RefJetSet{{\cite{%
cpc:39:347,*cpc:43:367,*cpc:82:74%
}}\xspace}
\def\RefRecMethod{{\cite{%
proc:hera:1991:23,*proc:hera:1991:43%
}}\xspace}
\def\citeCTD{{\cite{%
nim:a279:290,*npps:b32:181,*nim:a338:254%
}}\xspace}
\def\citeCAL{{\cite{%
nim:a309:77,*nim:a309:101,*nim:a321:356,*nim:a336:23%
}}\xspace}

\begin{document}
\title{ \boldmath Measurement of high-$Q^2$ \\ 
$e^-p$ neutral current cross sections at HERA \\
and the extraction of $xF_3$}
                    
\author{ZEUS Collaboration}
\draftversion{5.1}
\date{\today}

\date{}
\abstract{Cross sections for $e^-p$ neutral current deep inelastic
scattering have been measured at a centre-of-mass energy of $318 \gev$
using an integrated luminosity of $15.9 \pb^{-1}$ collected with the
ZEUS detector at HERA. Results on the double-differential
cross-section $d^2\sigma / dx\,dQ^2$ in the range $185 < Q^2 < 50\,000
\gev^2$ and $0.0037 < x < 0.75$, as well as the single-differential
cross-sections $d\sigma / dQ^2$, $d\sigma / dx$ and $d\sigma / dy$ for
$Q^2 > 200 \gev^2$, are presented. To study the effect of $Z$-boson
exchange, $d\sigma / dx$ has also been measured for $Q^2 > 10\,000
\gev^2$. The structure function $xF_3$ has been extracted by combining
the $e^-p$ results presented here with the recent ZEUS measurements of
$e^+p$ neutral current deep inelastic scattering. All results agree
well with the predictions of the Standard Model. }

\prepnum{DESY-02-113}
%
\makezeustitle
\pagenumbering{arabic} 
\pagestyle{plain}
%
%
%
%
%
\pagenumbering{Roman}                                                                              
\begin{center}                                                                                     
{                      \Large  The ZEUS Collaboration              }                               
\end{center}                                                                                       
  S.~Chekanov,                                                                                     
  D.~Krakauer,                                                                                     
  S.~Magill,                                                                                       
  B.~Musgrave,                                                                                     
  J.~Repond,                                                                                       
  R.~Yoshida\\                                                                                     
 {\it Argonne National Laboratory, Argonne, Illinois 60439-4815}~$^{n}$                            
\par \filbreak                                                                                     
  M.C.K.~Mattingly \\                                                                              
 {\it Andrews University, Berrien Springs, Michigan 49104-0380}                                    
\par \filbreak                                                                                     
  P.~Antonioli,                                                                                    
  G.~Bari,                                                                                         
  M.~Basile,                                                                                       
  L.~Bellagamba,                                                                                   
  D.~Boscherini,                                                                                   
  A.~Bruni,                                                                                        
  G.~Bruni,                                                                                        
  G.~Cara~Romeo,                                                                                   
  L.~Cifarelli,                                                                                    
  F.~Cindolo,                                                                                      
  A.~Contin,                                                                                       
  M.~Corradi,                                                                                      
  S.~De~Pasquale,                                                                                  
  P.~Giusti,                                                                                       
  G.~Iacobucci,                                                                                    
  A.~Margotti,                                                                                     
  R.~Nania,                                                                                        
  F.~Palmonari,                                                                                    
  A.~Pesci,                                                                                        
  G.~Sartorelli,                                                                                   
  A.~Zichichi  \\                                                                                  
  {\it University and INFN Bologna, Bologna, Italy}~$^{e}$
% ...
\newpage                                                                                           
% ...
                                                           %
                                                           %
\newpage   
                                                           %
                                                           %
\begin{tabular}[h]{rp{14cm}}                                                                       
$^{a}$ &  supported by the Natural Sciences and Engineering Research                               
          Council of Canada (NSERC) \\                                                             
$^{b}$ &  supported by the German Federal Ministry for Education and                               
          Research (BMBF), under contract numbers HZ1GUA 2, HZ1GUB 0, HZ1PDA 5, HZ1VFA 5\\         
$^{c}$ &  supported by the MINERVA Gesellschaft f\"ur Forschung GmbH, the                          
          Israel Science Foundation, the U.S.-Israel Binational Science                            
          Foundation, the Israel Ministry of Science and the Benozyio Center                       
          for High Energy Physics\\                                                                
$^{d}$ &  supported by the German-Israeli Foundation, the Israel Science                           
          Foundation, and by the Israel Ministry of Science\\                                      
$^{e}$ &  supported by the Italian National Institute for Nuclear Physics (INFN) \\                
$^{f}$ &  supported by the Japanese Ministry of Education, Science and                             
          Culture (the Monbusho) and its grants for Scientific Research\\                          
$^{g}$ &  supported by the Korean Ministry of Education and Korea Science                          
          and Engineering Foundation\\                                                             
$^{h}$ &  supported by the Netherlands Foundation for Research on Matter (FOM)\\                   
$^{i}$ &  supported by the Polish State Committee for Scientific Research,                         
          grant no. 620/E-77/SPUB-M/DESY/P-03/DZ 247/2000-2002\\                                   
$^{j}$ &  partially supported by the German Federal Ministry for Education                         
          and Research (BMBF)\\                                                                    
$^{k}$ &  supported by the Fund for Fundamental Research of Russian Ministry                       
          for Science and Edu\-cation and by the German Federal Ministry for                       
          Education and Research (BMBF)\\                                                          
$^{l}$ &  supported by the Spanish Ministry of Education and Science                               
          through funds provided by CICYT\\                                                        
$^{m}$ &  supported by the Particle Physics and Astronomy Research Council, UK\\                   
$^{n}$ &  supported by the US Department of Energy\\                                               
$^{o}$ &  supported by the US National Science Foundation\\                                        
$^{p}$ &  supported by the Polish State Committee for Scientific Research,                         
          grant no. 112/E-356/SPUB-M/DESY/P-03/DZ 301/2000-2002, 2 P03B 13922\\                    
$^{q}$ &  supported by the Polish State Committee for Scientific Research,                         
          grant no. 115/E-343/SPUB-M/DESY/P-03/DZ 121/2001-2002, 2 P03B 07022\\                    
\end{tabular}                                                                                      
                                                           %
                                                           %

\clearpage
\pagenumbering{arabic} 
%
\section{Introduction}
\label{sec-Introduction}

Studies of deep inelastic scattering (DIS) have played a key role in
the development of the Standard Model (SM) and in understanding the
structure of nucleons. The HERA $e^\pm p$ collider allows the
measurement of DIS over a kinematic region that extends to large
values of the negative of the four-momentum-transfer squared, $Q^2$,
as well as to low Bjorken $x$. The SM describes $e^\pm p$ neutral
current (NC) DIS in terms of the space-like exchange of a virtual
photon and a virtual $Z$ boson. When $Q^2$ is much smaller than the
square of the $Z$-boson mass, $M_Z^2$, the $Z$-exchange contribution
is negligible. For $Q^2 \sim M_Z^2$, the $Z$-exchange contribution is
comparable to that of photon exchange.

Using data collected during $e^+p$ running from 1994 to 1997, when
HERA ran at a centre-of-mass energy, $\sqrt{s}$, of $300 \gev$, the
ZEUS and H1 collaborations have measured the NC DIS cross section up
to $Q^2$ values as high as $40\,000 \gev^2$
\cite{epj:c11:427,epj:c13:609}. The measured $e^+p$ NC DIS cross
sections for $Q^2$ values larger than $10\,000 \gev^2$ are well
described at next-to-leading order (NLO) in quantum chromodynamics
(QCD) by the SM prediction including both photon- and $Z$-exchange
contributions. The effect of the parity-violating part of the $Z$
exchange to $e^+p$ NC scattering is to decrease the cross section
below that expected for photon exchange alone. In $e^-p$ NC DIS, the
sign of this contribution is reversed, so that the SM cross section is
larger than that expected from pure photon exchange. The comparison of
the $e^-p$ with the $e^+p$ NC cross section, therefore, provides a
direct test of the electroweak sector of the SM. Cross sections for
$e^-p$ NC DIS were reported recently by the H1 collaboration
\cite{epj:c19:269}.

This paper presents the NC $e^-p$ DIS cross-sections $d^2\sigma / dx\,
dQ^2$ for $185 < Q^2 < 50\,000 \gev^2$ and $0.0037 < x < 0.75$,
together with measurements of $d\sigma / dQ^2$, $d\sigma / dx$ and
$d\sigma / dy$ for $Q^2 > 200 \gev^2$, where $y = Q^2/sx$, neglecting
the proton mass. To exhibit the effect of $Z$-boson exchange, $d\sigma
/ dx$ is also evaluated for $Q^2 > 10\,000 \gev^2$.  The cross
sections were obtained using the $e^-p$ data collected in 1998/99 at
$\sqrt{s} = 318 \gev$, corresponding to an integrated luminosity of
$(15.9 \pm 0.3) \pb^{-1}$.  The results are compared to recent ZEUS
measurements of the $e^+p$ NC cross sections \cite{epj:c21:443} and
the parity-violating structure function $xF_3$ is extracted.


%

\section{Standard Model cross sections}
\label{sec-KineXSect}

The electroweak Born-level NC DIS unpolarised cross sections for the
reactions $e^\pm p \rightarrow e^\pm X$ can be expressed as
\cite{pl:b201:369,ijmp:a13:3385}
\begin{equation}
  \frac{d^2\sigma_{\rm Born}(e^\pm p)}{dx \,dQ^2} =  
  \frac{2 \pi \alpha^2}{x Q^4}
  \left[
    Y_+ F_2 \left(x, Q^2 \right) \mp Y_- xF_3 \left(x, Q^2 \right) - 
    y^2 F_L \left(x, Q^2 \right)
  \right] \ ,
  \label{eq-Born}
\end{equation}
where $\alpha$ is the fine-structure constant and $Y_{\pm} \equiv 1
\pm (1-y)^2$. At leading order (LO) in QCD, the structure functions
$F_2$ and $xF_3$ can be written as products of electroweak couplings
and parton density functions (PDFs) as follows:
\begin{alignat*}{2}
  F_2  &=& \; \onehalf  &  x \sum\limits_{f}\,
  \left[ 
    (V_f^L)^2+(V_f^R)^2+(A_f^L)^2+(A_f^R)^2
  \right]
  (q_f + \bar{q}_f) \ ,                                    \nonumber \\
  &&&                                                      \nonumber \\[-2.2em]
  &&&                                                                \\[-0.8em]
  xF_3 &=& & x \sum\limits_{f}\,
  \left[ 
    V_f^LA_f^L-V_f^RA_f^R 
  \right]
  (q_f - \bar{q}_f) \ ,                                    \nonumber
\end{alignat*}
where $x q_f(x,\qq)$ are the quark and $x {\bar q_f}(x,\qq)$ the
anti-quark PDFs, and $f$ runs over the five quark flavours $u,...,b$.
The functions $V_f$ and $A_f$ can be written in terms of the fermion
vector and axial-vector couplings as
\begin{alignat*}{3}
  V_{f}^L(Q^2) & = & \; e_{f} & -(v_{e} + a_{e})\,v_{f}\ \chiz(\qq) \ , \nonumber \\[2mm]
  V_{f}^R(Q^2) & = & \; e_{f} & -(v_{e} - a_{e})\,v_{f}\ \chiz(\qq) \ , \nonumber \\[2mm]
  A_{f}^L(Q^2) & = & & -(v_{e} + a_{e})\,a_{f}\ \chiz(\qq) \ , \nonumber \\[2mm]
  A_{f}^R(Q^2) & = & & -(v_{e} - a_{e})\,a_{f}\ \chiz(\qq) \ , \nonumber 
\end{alignat*}
where $L$ and $R$ refer to the left- and right-handed quark states,
respectively. The weak couplings $a_{e,f}$ and $v_{e,f}$ ($a_{e,f}=T^3_{e,f}$
and $v_{e,f}=T^3_{e,f} - 2e_{e,f}\, \sthw$) are functions of the weak isospin,
$T^3_{e,f}$ ($T^3_{e,f} = \frac{1}{2}$ ($-\frac{1}{2}$) for $\nu,u$ ($e,d$)),
and the weak mixing angle, \thw ($\sthw = 0.232$\cite{epj:c15:1}), $e_f$ is the
electric charge of the quark in units of the positron charge and \chiz\ is
given by
\begin{equation*}
  \chiz(Q^2) =\frac{1}{4\, \sthw \cthw} \ \frac{\qq}{\qq + \Mzq} \ .
  \label{eq-chi} 
\end{equation*}

The reduced cross section, $\tilde{\sigma}$, is defined as
\begin{equation}
  \tilde{\sigma} = 
                      \frac{x \, Q^4}{2 \pi \alpha^2 Y_+} \
                      \frac{d^2\sigma_{\rm Born}}{dx\, dQ^2} \ .
                      \label{eq-red}
\end{equation}

All cross-section calculations presented in this paper have been performed using
NLO QCD, in which $F_L$ is non-zero ~\cite{ijmp:a13:3385}. These calculations
predict that the contribution of $F_L$ to $d^2\sigma_{\rm Born}/dx\, dQ^2$ is
approximately $1.3\%$, averaged over the kinematic range considered in this
paper. However, in the region of small $x$, near $Q^2 = 200 \gev^2$, the $F_L$
contribution to the cross section can be as large as $10\%$.

%
\section{The ZEUS experiment at HERA}
\label{sec-zeus}

HERA accelerates electrons to an energy of $E_e = 27.5 \gev$ and
protons to an energy of $E_p = 920\gev$, yielding $\sqrt{s} = 318
\gev$. The inter-bunch spacing in the electron and proton beams is $96
\ns$. In normal running, some bunches in both the electron and the
proton rings are left empty (pilot bunches). The pilot bunches are used
to study the single-beam backgrounds.

A detailed description of the ZEUS detector can be found
elsewhere~\cite{zeus:1993:bluebook}. A brief outline of the components that are
most relevant for this analysis is given below.

The high-resolution uranium--scintillator calorimeter (CAL)~\citeCAL consists of
three parts: the forward (FCAL), the barrel (BCAL) and the rear (RCAL)
calorimeters. Each part is subdivided into towers and each tower is
longitudinally segmented into one electromagnetic section (EMC) and either one
(in RCAL) or two (in BCAL and FCAL) hadronic sections (HAC). The smallest
subdivision of the calorimeter is called a cell.  The CAL energy resolutions,
measured under test-beam conditions, are $\sigma(E)/E=0.18/\sqrt{E}$ for
electrons and $\sigma(E)/E=0.35/\sqrt{E}$ for hadrons ($E$ in $\Gev$). The
timing resolution of the CAL is $\sim 1 \ns$ for energy deposits greater than
$4.5 \gev$.

Presampler detectors are mounted in front of the CAL.  They consist of
scintillator tiles matching the calorimeter towers and measure signals
from particle showers created by interactions in the material lying
between the interaction point and the calorimeter.

Charged particles are tracked in the central tracking detector
(CTD)~\citeCTD, which operates in a magnetic field of $1.43\Tesla$
provided by a thin superconducting coil. The CTD consists of
72~cylindrical drift chamber layers, organised in 9~superlayers
covering the polar-angle\ZcoosysfnBeta\ region
\mbox{$15^\circ<\theta<164^\circ$}.  The transverse-momentum
resolution for full-length tracks is $\sigma(p_t)/p_t=0.0058 \,
p_t\oplus0.0065\oplus0.0014/p_t$, with $p_t$ in $\Gev$.

The RCAL is instrumented with a layer of $3 \times 3 \cm^2$
silicon-pad detectors at a depth of 3.3 radiation lengths. This
hadron-electron separator (HES) \cite{nim:a277:176} is used to improve
the electron angle measurements.

The luminosity is measured using the Bethe-Heitler reaction $ep
\rightarrow e\gamma p$ \cite{desy-92-066,*zfp:c63:391,*acpp:b32:2025}.
The resulting small-angle photons are measured by the luminosity
monitor, a lead-scintillator calorimeter placed in the HERA tunnel 107
m from the interaction point in the electron beam direction.

%
\section{Monte Carlo simulation}
\label{sec-MC}

Monte Carlo (MC) simulations were used to evaluate the efficiency for selecting
events, to determine the accuracy of the kinematic reconstruction, to estimate
the background rate, and to extrapolate the measured cross sections to the full
kinematic range. A sufficient number of events was generated to ensure that
statistical errors from the MC samples are negligible in comparison to those of
the data. The MC samples were normalised to the total integrated luminosity of
the data.

Events from NC DIS were simulated including radiative effects, using
the HERACLES 4.6.1~\cite{cpc:69:155} program with the DJANGOH 1.1
~\RefDJango interface to the hadronisation programs and using CTEQ5D
\cite{epj:c12:375} PDFs. In HERACLES, $\mathcal{O}(\alpha)$
electroweak corrections for initial- and final-state radiation, vertex
and propagator corrections and two-boson exchange are included. The
colour-dipole model of ARIADNE 4.10~\cite{cpc:71:15} was used to
simulate the $\mathcal{O}(\alpha_S)$ plus leading-logarithmic
corrections to the quark-parton model. As a systematic check, the MEPS
model of LEPTO 6.5 ~\cite{cpc:101:108} was used.  Both programs use
the Lund string model of JETSET 7.4~\RefJetSet for the hadronisation.
Diffractive events, characterised by having no particle production
between the current jet and the proton remnant, were generated using
RAPGAP~2.08/06~\cite{cpc:86:147} and appropriately mixed with the
non-diffractive NC DIS sample. The contribution of diffractive events,
originally generated with the same $x$-$Q^2$ distribution as
non-diffractive events, was obtained by fitting the $\etamax$
distribution\footnote{The quantity $\etamax$ is defined as the
  pseudorapidity of the CAL energy deposit with the lowest polar angle
  and an energy above $400 \mev$.} of the data with a linear
combination of non-diffractive and diffractive MC samples while
maintaining the overall normalisation \cite{thesis:kappes:2001}.  The
fit was carried out in each of the $x$-$Q^2$ bins used in the
measurement of the double-differential cross section (see
\sect{Binning}). The fitted fractions exhibited no dependence on $Q^2$
and an exponential function was used to parameterise the $x$
dependence. The diffractive fraction falls from 10\% at $x=0.005$ to
2\% at $x=0.05$.  Photoproduction (PHP) backgrounds, including both
direct and resolved processes, were simulated at LO using
HERWIG~6.1~\cite{cpc:67:465}.

The ZEUS detector response was simulated using a program based on
GEANT 3.13~\cite{tech:cern-dd-ee-84-1}. The generated events were
passed through the detector simulation, subjected to the same trigger
requirements as the data and processed by the same reconstruction and
analysis programs.

The vertex distribution in data is a crucial input to ensure the
accuracy of the evaluation of the event-selection efficiency of the
MC. The shape of the $Z$-vertex distribution was determined from a
sample of NC DIS events for which the event-selection efficiency was
independent of $Z$.

%
%
%
\section{Event characteristics and kinematic reconstruction}
\label{sec-Char}

Neutral current events with $\qq > 200 \gev^2$ are characterised by the presence
of a high-energy isolated electron. Many of these electrons have an energy close
to the beam energy and are scattered into the RCAL. As $Q^2$ increases, the
scattered electrons are produced with higher energies, up to several hundred
$\Gev$, and at smaller polar angles, so that they are measured in the BCAL or the
FCAL.

The variables $\delta$, \Pt\ and \Et\ are used in the event selection.  The
quantity $\delta$ is defined by
\begin{equation}
  \delta \equiv \sum\limits_{i} (E-p_Z)_{i} = \sum\limits_{i} ( E_i - E_i \cos
  \theta_{i} ) \ ,
  \label{eq-Delta}
\end{equation}
where the sum runs over all calorimeter energy deposits $E_i$ (uncorrected in
the trigger but corrected in the offline analysis, as discussed below) with
polar angles $\theta_i$.  Conservation of energy and longitudinal momentum,
$p_z$, requires $\delta = 2E_e= 55 \gev$ if all final-state particles are
detected and perfectly measured. Undetected particles that escape through the
forward beam hole have a negligible effect on $\delta$. However, particles lost
through the rear beam hole, as in the case of PHP, where the electron
emerges at very small scattering angles, or in events with an initial-state
bremsstrahlung photon, can lead to a substantial reduction in
$\delta$.

The net transverse momentum, \Pt, and the transverse energy, \Et, are defined by
\begin{alignat}{2}
  P_T^2 & = & P_X^2 + P_Y^2 = & \left( \sum\limits_{i} E_i \sin \theta_i \cos
    \phi_i \right)^2+ \left( \sum\limits_{i} E_i \sin \theta_i \sin \phi_i
  \right)^2,
  \label{eq-PT2}\\ 
  E_T & = & \sum\limits_{i} E_i \sin \theta_i \ , \nonumber
\end{alignat}
where $\phi_i$ is the azimuthal angle, and the sums run over all energy deposits
in the calorimeter.

The CAL energy deposits were separated into those associated with the
scattered electron and all other energy deposits. The sum of the
latter is referred to as the hadronic energy.  The spatial
distribution of the hadronic energy, together with the reconstructed
vertex position, were used to evaluate the hadronic polar angle,
$\gamma_h$, which, in the naive quark-parton model, is the polar angle
of the struck quark.

The reconstruction of $x$, $Q^2$ and $y$ was performed using the
double angle (DA) method~\RefRecMethod. The DA estimators are given by
\begin{alignat*}{2}
  \qqda &=& \, 4E_e^{2} & \frac{\sin\gamma_h(1+\cos\theta_e)}
  {\sin\gamma_h + \sin\theta_e - \sin(\gamma_h+\theta_e)} \ , \\[2mm]
  \xda &=& \, \frac{E_e}{E_p} & \frac{\sin\gamma_h + \sin\theta_e +
    \sin(\gamma_h+\theta_e)}
  {\sin\gamma_h + \sin\theta_e - \sin(\gamma_h+\theta_e)} \ , \\[2mm]
  \yda &=& & \frac{\sin\theta_e(1-\cos\gamma_h)} {\sin\gamma_h + \sin\theta_e -
    \sin(\gamma_h+\theta_e)} \ ,
\end{alignat*}
where $\theta_e$ is the polar angle of the scattered electron.

The DA method is insensitive to uncertainties in the overall energy
scale of the calorimeter. However, it is sensitive to initial-state
QED radiation and, in addition, an accurate simulation of the hadronic
final state is necessary.

The relative resolution in $\qq$ was $3\%$ over the entire kinematic
range covered. The relative resolution in $x$ varied from $15\%$ in
the lowest $\qq$ bins to $4\%$ in the highest $\qq$ bins. The
relative resolution in $y$ was $10 \%$ in the lowest $\qq$ bins,
decreasing to $1\%$ for high $y$ values in the highest $Q^2$ bins
(see \sect{Binning}).

In the event selection, $y$ calculated using the electron method ($y_e$) and the
Jacquet-Blondel method \cite{proc:epfacility:1979:391} ($y_\JB$) were also used.
These variables are defined by
\begin{alignat*}{1}
  y_e   &= 1 - \frac{E_e'}{2 E_e} \left( 1 - \cos \theta_e \right) \, , \\[2mm]
  y\jbb &= \frac{\delta_h}{2 E_e} \; ,
\end{alignat*}
where $E_e'$ is the energy of the scattered electron and $\delta_h$
was calculated from \eq{Delta} using only the hadronic energy.

%
\section{Electron reconstruction}
\label{sec-RecoEFinder}

\subsection{Electron identification}
In order to identify and reconstruct the scattered electron, an
algorithm that combines calorimeter and CTD information was used
\cite{epj:c11:427}. The algorithm starts by identifying CAL clusters
that are topologically consistent with an electromagnetic shower. Each
cluster had to have an energy of at least $10 \gev$ and, if the
electron candidate fell within the acceptance of the CTD, a track was
required that, when extrapolated, had to pass within $10 \cm$ of a
cluster centre at the shower maximum. Such a track is referred to as a
matched track.  An electron candidate was considered to lie within the
CTD acceptance if a matched track from the reconstructed event vertex
traversed at least four of the nine superlayers of the CTD. For the
nominal interaction point, i.e.  $Z = 0$, this requirement corresponds
to the angular range of $23^\circ < \theta_e < 156^\circ$.

Monte Carlo studies showed that the overall efficiency for finding the
scattered electron was about 95\% for $E'_e \geq 10 \gev$ and $Q^2 <
15\,000 \gev^2$, decreasing to about $85\%$ for $Q^2 > 30\,000 \gev^2$.
The electron identification efficiency was checked with a data sample
of NC DIS events selected using independent requirements such as high
$E_T$ in the trigger and an isolated high-$p_t$ track associated with
the scattered electron. The efficiency curves from data and MC
simulation agreed to better than $0.5\%$. An alternative
electron-finding algorithm \cite{epj:c21:443} was also used:
differences in the measured cross sections were negligible.

\subsection{Electron-energy determination}
\label{sec-EePrime}

The scattered-electron energy was determined from the calorimeter
deposit since, above $10 \gev$, the calorimeter energy resolution is
better than the momentum resolution of the CTD. The measured energy
was corrected for the energy lost in inactive material in front of the
CAL. The presampler was used in the RCAL, while in the B/FCAL a
detailed material map was used \cite{epj:c21:443}. To render the
energy response uniform across the face of the calorimeter,
corrections, obtained by smoothing the non-uniform response functions
in data and the MC simulation, were used \cite{epj:c11:427}. The
corrections were determined separately for the BCAL \cite{epj:c11:427}
and the RCAL \cite{epj:c21:443}. Too few electrons were scattered into
the FCAL for such a correction to be derived.

After applying the corrections described above, the electron-energy
resolution was 10\% at $E'_e = 10 \gev$ falling to 5\% for $E'_e > 20
\gev$.  The scale uncertainty on the energy of the scattered electron
detected in the BCAL was $\pm 1\%$.  For electrons detected in the
RCAL, the scale uncertainty was $\pm 2\%$ at $8 \gev$, falling
linearly to $\pm 1 \%$ for electrons with energies of $15 \gev$ and
above \cite{epj:c21:443}. A scale uncertainty of $\pm 3\%$ was
assigned to electrons reconstructed in the FCAL.

\subsection{Determination of the electron polar angle}

The polar angle of the scattered electron can be determined either
from the cluster position within the calorimeter using the
reconstructed event vertex, or from the polar angle of the track
matched to the cluster. Studies \cite{thesis:amaya:2001-tmp-3cb44d33}
showed that, inside the acceptance of the CTD, the angular resolution
for tracks is superior to that for calorimeter clusters. Hence, in the
CTD acceptance region, which contains $98.8\%$ of the events,
$\theta_e$ was determined from the track. For candidates outside this
region, the position of the cluster was used together with the event
vertex.

To determine the CAL alignment, the positions of the calorimeter cell
boundaries were obtained by using the energy-deposition pattern of
electron tracks extrapolated into the CAL. This allowed the BCAL to
be aligned in $Z$ ($\phi$) with respect to the CTD to $\pm 0.3 \mm$
($\pm0.6 \mrad$) \cite{thesis:kappes:2001}.  For the alignment of the
RCAL, the position of the extrapolated track was compared to that
determined by the HES\cite{thesis:kappes:2001}.  The precision of the
alignment was $\pm 0.3 \mm$ ($\pm 0.6 \mm$) in $X$ ($Z$) and $\pm 0.9
\mrad$ in $\phi$.  In all cases, the precision is sufficient to render
the resulting systematic uncertainties on the cross sections negligible.

The resolution in $\theta_e$ was obtained by comparing the
MC-generated angle to that obtained after applying the detector
simulation, reconstruction and correction algorithms. The resulting
resolution for electrons outside the CTD acceptance was $\pm 5 \mrad$
in the RCAL and $\pm 2 \mrad$ in the FCAL. For tracks inside the CTD
acceptance, the resolution was $\pm 3 \mrad$.

%
\section{Reconstruction of the hadronic system}
\label{sec-HadRecons}

\subsection{Hadronic-energy determination}
\label{sec-HadEne}

The hadronic-energy deposits were corrected for energy loss in the
material between the interaction point and the calorimeter using the
material maps implemented in the detector-simulation package.

After applying all corrections, the measured resolution for the
hadronic transverse momentum, $p^\had_T$, was about $13\%$ ($11\%$) at
$p^\had_T = 20 \gev$ in BCAL (FCAL), decreasing to 8\% (7.5\%) at
$p^\had_T = 60 \gev$.  The uncertainties in the hadronic energy scales
of the FCAL and the BCAL were $\pm 1\%$, while for the RCAL the
uncertainty was $\pm 2\%$ \cite{pl:b539:197}.

\subsection{Determination of the hadronic polar angle, $\gamma_h$}
\label{sec-GammaHad}
 The angle $\gamma_h$ is given by \cite{proc:hera:1991:23}
\begin{equation*}
  \cos \gamma_h = \frac{P^2_{T,h} - \delta_h^2}{P^2_{T,h} + \delta_h^2},
\end{equation*}
where $P^2_{T,h}$ was calculated from \eq{PT2} using only the hadronic
energy. Particles interacting in the material between the primary
vertex and the CAL generate energy deposits that bias the
reconstructed value of $\gamma_h$.  To minimise this bias, an
algorithm was developed in which CAL clusters with energies below $3
\gev$ and with polar angles larger than an angle $\gamma_{\rm max}$
were removed \cite{epj:c11:427}.  The value of $\gamma_{\rm max}$ was
derived from a NC MC sample by requiring that the bias in the
reconstructed hadronic variables was minimised.

The resolution of $\gamma_h$ was below $15 \mrad$ for $\gamma_h < 0.2
\rad$, increasing to $100 \mrad$ at $\gamma_h \sim 2 \rad$. 
These resolutions dominate the errors on the kinematic variables.

%
\section{Event selection}
\label{Sect:EvtSel}

\subsection{Trigger}
\label{sec-Trigger}

ZEUS operates a three-level trigger system~\cite{zeus:1993:bluebook}. For the
measurements presented in this paper, the first-level trigger required
an ``OR'' of the following:
\begin{itemize}
\item a total electromagnetic energy of at least $3.4 \gev$ in the EMC cells of
  the RCAL;
\item $4.8 \gev$ in the EMC cells of the BCAL and a ``good track'',
  defined as a charged track consistent with emanating from the IP;
\item an isolated energy deposit of at least $2 \gev$ in the EMC section of
  the RCAL;
\item $15 \gev$ summed over the entire EMC cells of the CAL;
\item $E_T^{\prime\prime} > 12 \gev$ and a good track, where
  $E_T^{\prime\prime}$ is the total transverse energy excluding the two rings of
  FCAL towers nearest to the forward beampipe.
\end{itemize}
The $E_T^{\prime\prime}$ requirement was designed to tag high-\qq\ 
events by their large \Et\ while rejecting beam-gas background. The
latter is characterised by large energy deposits at low polar angles.
The major requirement at the second trigger level was $\delta +
2E_\gamma > 29 \gev$, where $E_\gamma$ is the energy measured in the
luminosity monitor.  This requirement suppresses photoproduction
events.  Backgrounds were further reduced at the second level by
removing events with calorimeter timing inconsistent with an $ep$
interaction.  At the third level, events were fully reconstructed. The
requirements were similar to, but looser than, the offline cuts
described below; a simpler and generally more efficient (but less
pure) electron finder was used.

The main uncertainty in the trigger chain comes from the first level.
The data and MC simulation agree to within $\sim 0.5\%$ and the
overall efficiency is close to $100\%$. Therefore, uncertainties on
the measured cross sections coming from the trigger simulation are
small.

\clearpage

\subsection{Offline selection}
\label{sec-selec:offline}

The following criteria were applied offline: 
\begin{itemize}
\item electrons, identified as described in \sect{RecoEFinder}, were required to
  satisfy the following criteria:
  \begin{itemize}       
    \item to ensure high purity, the electron was required to have an
      energy of at least $10 \gev$; 
    \item to reduce background, isolated electrons were selected by
      requiring that less than $5 \gev$, not associated with the
      scattered electron, be deposited in calorimeter cells inside an
      $\eta$-$\phi$ cone of radius $R_{\rm cone} = 0.8$ centred on the
      electron. The quantity $R_{\rm cone}$ is defined as $R_{\rm
        cone} = \sqrt{(\Delta \phi)^2 + (\Delta \eta)^2}$, where
      $\Delta \phi$ (in radians) is the azimuthal angle between the
      CAL energy deposit and the scattered electron and $\Delta \eta$
      is the difference in pseudorapidity between the scattered
      electron and the energy deposit;
    \item each electron cluster within the CTD acceptance ($23^\circ
      \lesssim \theta_e \lesssim 156^\circ$) had to be matched to a
      track with a momentum, $p_\trk$, of at least $5 \gev$. The
      distance of closest approach (DCA) of the extrapolated track to
      the centre of the CAL cluster had to be less than $10 \cm$;
    \item for electrons outside of the forward tracking acceptance of the CTD
      ($\theta_e \lesssim 23^\circ$), the tracking requirement in the electron
      selection was replaced by a cut on the transverse momentum of the
      electron, $p_t^e > 30 \gev$;
    \item for electrons outside the backward tracking acceptance of the CTD
      ($\theta_e \gtrsim 156^\circ$), no track was required;
    \item a fiducial-volume cut was applied to the electron. It
      excluded the upper part of the central RCAL area ($20 \times 80
      \cm^2$), which is occluded by the cryogenic supply for the
      solenoid magnet, as well as the transition regions between the
      three parts of the CAL, corresponding to scattered-electron
      polar angles of $35.6^\circ < \theta_e < 37.3^\circ$ and
      $128.2^\circ < \theta_e < 140.2^\circ$;
  \end{itemize} 
\item to ensure that event quantities were accurately determined, a
  reconstructed vertex with $-50 < Z < 50 \cm$ was required, a range
  consistent with the $ep$ interaction region. A small fraction of the
  proton current was contained in satellite bunches, which were
  separated by $\pm 4.8 \ns$ with respect to the nominal
  bunch-crossing time, resulting in some of the $ep$ interactions
  occurring $\pm 72 \cm$ from the nominal interaction point. This cut
  rejects $ep$ events from these regions;
\item to suppress PHP events, in which the scattered electron escaped through
  the beam hole in the RCAL, $\delta$ was required to be greater than $38 \gev$.
  This cut also reduces the number of events with initial-state QED radiation.
  The requirement $\delta<65 \gev$ removed ``overlay'' events in which a normal
  DIS event coincided with additional energy deposits in the RCAL from some
  other source. For electrons outside the forward tracking acceptance of the
  CTD, the lower $\delta$ cut was raised to $44 \gev$;
\item to reduce further the background from PHP, $y_e$ was required to
  satisfy $y_e < 0.95$;
\item the net transverse momentum, \Pt, is expected to be close to
  zero for true NC events and was measured with an error approximately
  proportional to $\sqrt{\Et}$.  To remove cosmic rays and beam-related
  background, $P_T / \sqrt{E_T}$ was required to be less than $4\sqrt{\Gev}$;
\item to reduce the contribution from QED radiative corrections,
  elastic Compton scattering events ($e p \rightarrow e \gamma p$)
  were removed. This was done using an algorithm that searched for an
  additional photon candidate and discarded the event if the sum of
  the energies associated with the electron and photon candidates was
  within $2 \gev$ of the total energy in the calorimeter;
\item in events with low $\gamma_h$, a large amount of energy is
  deposited near the inner edges of the FCAL or escapes through the
  forward beampipe. As the MC simulation of this forward energy flow
  is somewhat uncertain, events for which $\gamma_h$, extrapolated to the
  FCAL surface, lay within a circle of $20 \cm$ around the forward
  beam line were removed. For an interaction at the nominal interaction
  point, this
  circle cut corresponds to a lower $\gamma_h$ cut of $90 \mrad$;
\item the kinematic range over which the MC simulation is valid does not extend
  to very low $y$ at high $x$. To avoid these regions of phase space,
  $y\jbb(1-x\dab)^2$ was required to be greater than $0.004$
  \cite{cpc:81:381,*spi:www:django6}.
\end{itemize}

\begin{figure}
  \vspace*{-2.0cm}
  \begin{center}
    \includegraphics[width=0.95\textwidth]{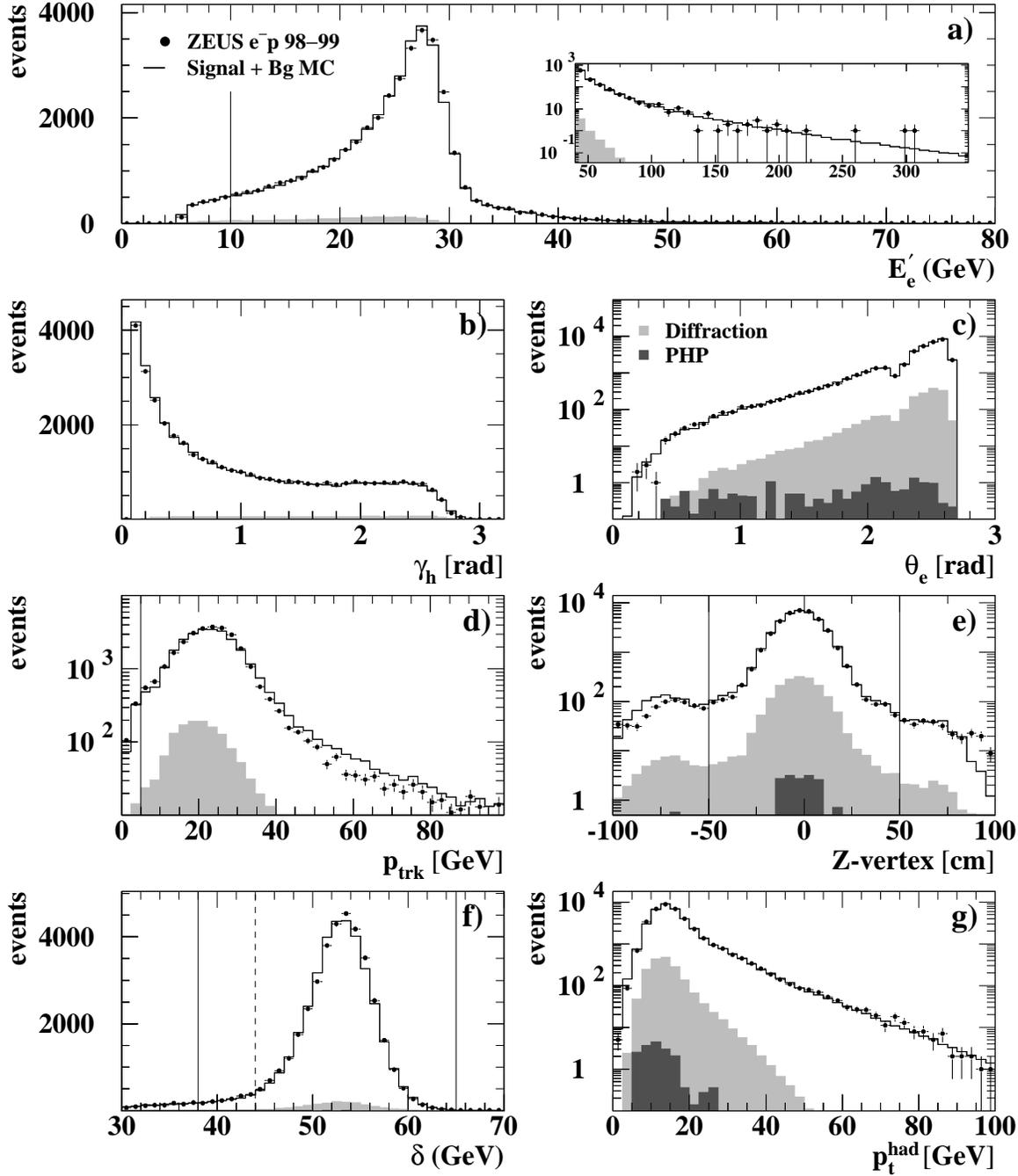}
  \end{center}
  \caption{
    Comparison of $e^-p$ data (points) and MC simulation (histograms)
    for a) $E^{\prime}_e$, the energy of the scattered electron (the
    inset shows the high-energy part of the distribution), b)
    $\gamma_h$, the angle of the hadronic system, c) $\theta_e$, the
    angle of the scattered electron, d) $p_\trk$, the momentum of the
    track matched to the electron, e) $Z$ coordinate of the event
    vertex, f) $\delta = \sum_i (E-p_{Z})_i$ and g) $p_t^\had$, the
    transverse momentum of the hadronic system. The darkest shaded
    area shows the PHP contribution. The histogram shows the sum of
    the background and the diffractive (lightest shading) and the
    non-diffractive NC signal MC samples.  The vertical lines indicate
    the cut boundaries described in the text (the dashed line
    represents the tightened cut for electrons outside the forward CTD
    acceptance).}
  \label{fig-CtrlPlts}
\end{figure}

A total of $38\,411$ events with $\qqda > 185 \gev^2$ satisfied the
above criteria. Distributions from data and the sum of the signal and
PHP-background MC samples are compared in \fig{CtrlPlts}.  Good
agreement between data and MC simulation is seen over the full range
of most variables.  Disagreements between data and MC simulation occur
in the region of the kinematic peak ($E_e' \approx E_e$) in the
electron energy distribution, at high and low values of the momentum
of the electron track, $p_\trk$, and in the peak region of the
distribution of $\delta$.  The differences suggest that there are
simulation errors in some aspects of either or both the fragmentation
and the detector response. The uncertainties caused by these
disagreements were included in the systematic uncertainties (see
\sect{SysErr}).

The PHP background averages $\sim 0.3\%$ over the kinematic range
covered, rising to $\sim 1.3\%$ at high $y$. Background from
prompt-photon events is negligible \cite{zfp:c74:207}. An upper limit
to the background associated with non-$ep$ collisions is given by the
absence of any events from pilot bunches. Taking into account
the relative currents in the pilot and colliding bunches yields a 90\%
C.L. upper limit of 70 events. Backgrounds from sources not
related to $ep$ collisions were therefore neglected.

%
\section{Results}
\label{Sect:Results}

\subsection{Binning, acceptance and cross-section determination}
\label{sec-Binning}

The bin sizes used for the determination of the single- and
double-differential cross sections were chosen commensurate with the
resolutions. \Fig{KinePlan} shows the kinematic region used in
extracting the $e^-p$ double-differential cross section. The number of
events per bin decreases from $\sim 1\,800$ in the lowest $Q^2$ bins
to four in the bin at the highest $Q^2$ and $x$. The efficiency,
defined as the number of events generated and reconstructed in a bin
divided by the number of events that were generated in that bin,
varied between 50\% and 80\%, apart from the region between the R/BCAL
at $\theta_e = 2.25 \rad$. The purity, defined as the number of events
reconstructed and generated in a bin divided by the total number of
events reconstructed in that bin, ranged from 50\% to 80\%. The
acceptance, ${\cal A}$, listed in the tables, is defined as the
efficiency divided by the purity.

The value of the cross section at a fixed point within a bin was
obtained from the ratio of the number of observed events, after
background subtraction, to the number of events estimated from the MC
simulation in that bin, multiplied by the appropriate cross section
obtained from \eq{red} using the CTEQ5D PDFs. In this way, the
$d\sigma/dQ^2$ and $d\sigma/dx$ measurements were extrapolated to the
full range of $y$ and corrected for initial- and final-state
radiation. Using the ZEUS NLO QCD (ZEUS-S) fit \cite{desy-02-105} in
the extraction of the cross sections instead of CTEQ5D typically
changed the single-differential cross sections by less than $\pm 1\%$;
only in the highest $Q^2$ bins was the effect as large as $3\%$.  The
change in the double-differential cross section was typically below
$\pm 1\%$ and increased to $\pm 2\%$ only in the upper $\qq$ bins at
high $x$.

The statistical uncertainties on the cross sections were calculated,
using Poisson statistics, from the numbers of events observed in the
bins, taking into account the statistical uncertainty from the MC
simulations.

\subsection{Systematic uncertainties}
\label{sec-SysErr}

Systematic uncertainties associated with deficiencies in the
simulation were estimated by re-calculating the cross sections after
tuning the MC simulation. Values of the event-selection cuts were
varied where this method was not applicable. The positive and negative
deviations from the nominal cross-section values were added in
quadrature separately to obtain the total positive and negative
systematic uncertainty. The uncertainty on the luminosity of the
combined 1998--99 $e^-p$ sample of 1.8\% was not included in the total
systematic uncertainty.  The other uncertainties are discussed in
detail below.

\subsubsection{Uncorrelated systematic uncertainties}

The following systematic uncertainties exhibit no bin-to-bin
correlations:
\begin{itemize}
\item variation of the electron-energy resolution in the MC simulation
  --- the effect on the cross sections of changing the CAL
  energy resolution for the scattered electron in the MC by $\pm 1\%$
  was negligible over nearly the full kinematic range. The effect
  increased to about $\pm 1\%$ only for bins at high $y$ and for
  double-differential bins in the upper $Q^2$ range;
\item electron angle --- differences between data and MC simulation in
  the electron scattering angle due to uncertainties in the simulation
  of the CTD were at most $\pm 1 \mrad$. Typically, the variations were
  below $\pm 1\%$; the effect increased to as much as $\pm 2\%$ in
  only a few double-differential bins.
  
  For electrons outside the forward acceptance of the CTD, the FCAL
  position was varied by $\pm 3 \mm$ in $X$, $Y$ and $Z$, covering
  the uncertainty on the FCAL alignment. Typically, the changes in the
  cross sections were below $\pm 1\%$ and reached $\pm 2\%$ in only 
  a few double-differential bins at high $x$;
\item hadronic angle --- the uncertainty associated with the
  reconstruction of $\gh$\ was investigated by varying the
  calorimeter-energy scale for the hadronic final state
  \cite{pl:b539:197} and by varying the $\gamma_{\rm max}$
  parameter in the correction to the hadronic energy given in
  \sect{GammaHad} in a range for which the reconstruction of
  $\gamma_h$ remains close to optimal.  This resulted in a systematic
  uncertainty in the single-differential cross sections of less than
  $\pm 1\%$ in most bins, increasing to about $\pm2\%$ in individual
  bins. For the double-differential bins, the effect was generally
  below $\pm 2\%$ at low and medium $Q^2$, occasionally reaching $\pm
  4\%$. In the highest $Q^2$ region, the effect was as large as $\pm
  7\%$;
\item FCAL circle cut --- the cut at $20 \cm$ was varied by $\pm 3
  \cm$.  The resulting changes in the cross sections were typically
  below $\pm 1\%$. Only for the highest $x$ bins of the
  double-differential cross section did the effect increase to $\pm
  6\%$ and in two bins at $Q^2 = 1\,200 \gev^2$ and $1\,500 \gev^2$
  ($x = 0.4$) to $-17 \%$ and $+40\%$, respectively;
\item background estimation --- systematic uncertainties arising from
  the normalisation of the PHP background were estimated by doubling
  and halving the background predicted by the MC simulation, resulting
  in negligible changes in the single-differential cross sections over
  the full kinematic range and small variations of at most $\pm 1\%$
  in the double-differential bins;
\item variation of selection thresholds --- the DCA cut was lowered
  from $10 \cm$ to $8 \cm$. The uncertainties in the cross sections
  associated with this changes were below $\pm 2\%$ over the full
  kinematic range, except for two double-differential bins at high $x$,
  where the effect reached $- 6\%$.
  
  The upper $\delta$ cut at $65 \gev$ was varied by $\pm 2 \gev$. The
  effect on the cross sections was generally below $\pm 1\%$ but
  became as large as $\pm 15\%$ in a few bins.
  
  The $P_T/\sqrt{E_T}$ cut was varied by $\pm 1 \sqrt{\Gev}$. The
  cross-section uncertainties were below about $\pm 1\%$ over the full
  kinematic range;
\item diffractive contribution --- the fraction of diffractive events
  was varied within the errors determined by the fit described in
  \sect{MC}. The resulting uncertainties were typically below $\pm
  1\%$, rising to about $\pm2\%$ at high $y$.
\end{itemize}

\subsubsection{Correlated systematic uncertainties}

The following systematic uncertainties were found to be correlated
bin to bin:
\begin{itemize}
\item $\{\delta_1\}$ electron-energy scale --- the uncertainty in the
  electron-energy scale (as described in \sect{RecoEFinder}) resulted in
  systematic variations in the cross sections of $\pm 2\%$ at high $y$ and in
  negligible uncertainties elsewhere;
\item $\{\delta_2\}$ background estimation --- systematic
  uncertainties arising from the simulation of the PHP background were
  estimated by reducing the cut on $y_e$ to $y_e < 0.9$. The resulting
  changes in the cross sections were typically below $\pm 2\%$; only in
  the highest-$Q^2$ region at low $x$ did the effect increase to
  $\pm 13\%$;
\item $\{\delta_3\}$ variation of selection thresholds (I) --- varying the
  electron-isolation requirement by $\pm2\gev$ caused a negligible
  systematic uncertainty in the cross sections at the lower end of the
  $\qq$ and $y$ range, up to $\pm 2\%$ for the medium $\qq$ and high $y$
  bins and up to $\pm 5\%$ in the highest $\qq$ bins;
\item $\{\delta_4\}$ variation of selection thresholds (II) ---
  varying the $p_\trk$ requirement by $\pm5 \gev$ resulted in
  variations of the cross sections by at most $\pm 2\%$ over nearly
  the full kinematic range, except in a few double-differential bins
  where it became as high as $\pm 8\%$;
\item $\{\delta_5\}$ vertex distribution --- the uncertainty arising
  from the limited knowledge of the shape of the distribution in the
  $Z$ coordinate of the event vertex was obtained by varying the
  contribution of events from the satellite bunches, visible as small
  peaks for $|Z|>50 \cm$ in \fig{CtrlPlts}e), by $+40\%$ and $- 8\%$ in
  the MC simulation, as suggested by comparison with data. The effect
  on the cross sections was at most about $\pm 1\%$;
\item $\{\delta_6\}$ uncertainty in the parton-shower scheme --- a
  comparison of the description of the hadronic energy flow in data
  with the expectations of ARIADNE and the MEPS model of LEPTO was
  made. ARIADNE gave a slightly better description of the data.
  However, small differences with respect to the data were observed
  for both models, particularly in the energy flow between the current
  jet and the target remnant. The effects on the cross sections were
  typically below $\pm 2\%$, reaching as much as $\pm 6\%$ in only a
  few bins.
\end{itemize}

\subsection{Single-differential cross sections}
\label{sec-Single}

\begin{figure}
  \begin{center}
    \includegraphics[width=.9\textwidth]{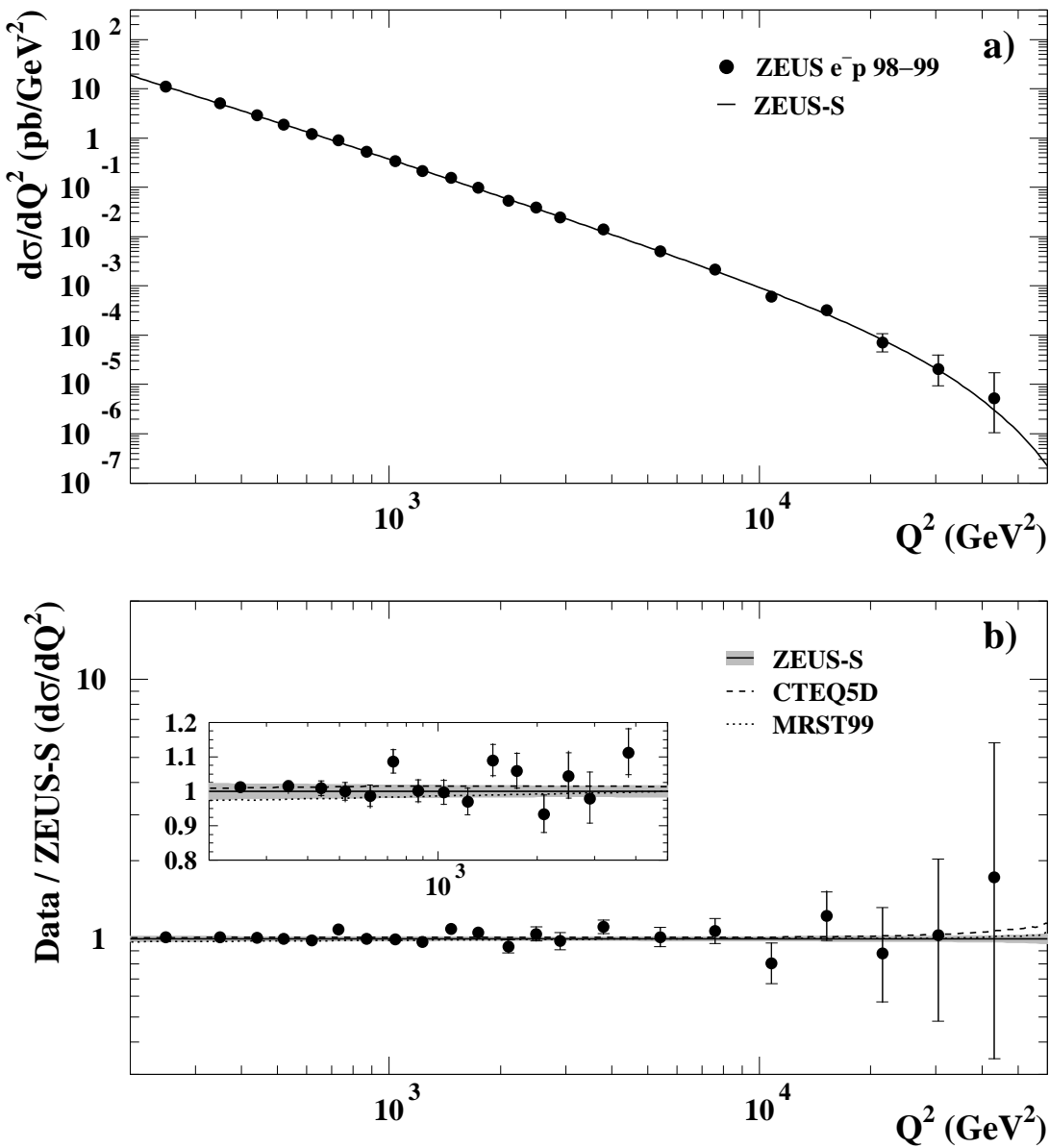}
  \end{center}
  \caption{
    a) The differential $e^-p$ cross-section $d\sigma/dQ^2$ as a
    function of $Q^2$ compared to the SM expectation evaluated using
    the ZEUS-S fit. b) The ratio of the measured cross section to the
    ZEUS-S prediction. Also shown are the ratios of the SM prediction
    using CTEQ5D and MRST99 PDFs to that of ZEUS-S. The inset shows
    the range $200 < Q^2 < 5\,000 \gev^2$ on a linear $y$ scale. The
    shaded band indicates the uncertainty on the calculated cross
    section due to the uncertainty in the ZEUS-S PDFs. The inner error
    bars of the measured points show the statistical uncertainty,
    while the outer ones show the statistical and systematic
    uncertainties added in quadrature.  }
  \label{fig-dsdQ2}
\end{figure}

\begin{figure}
  \begin{center}
    \includegraphics[width=.9\textwidth]{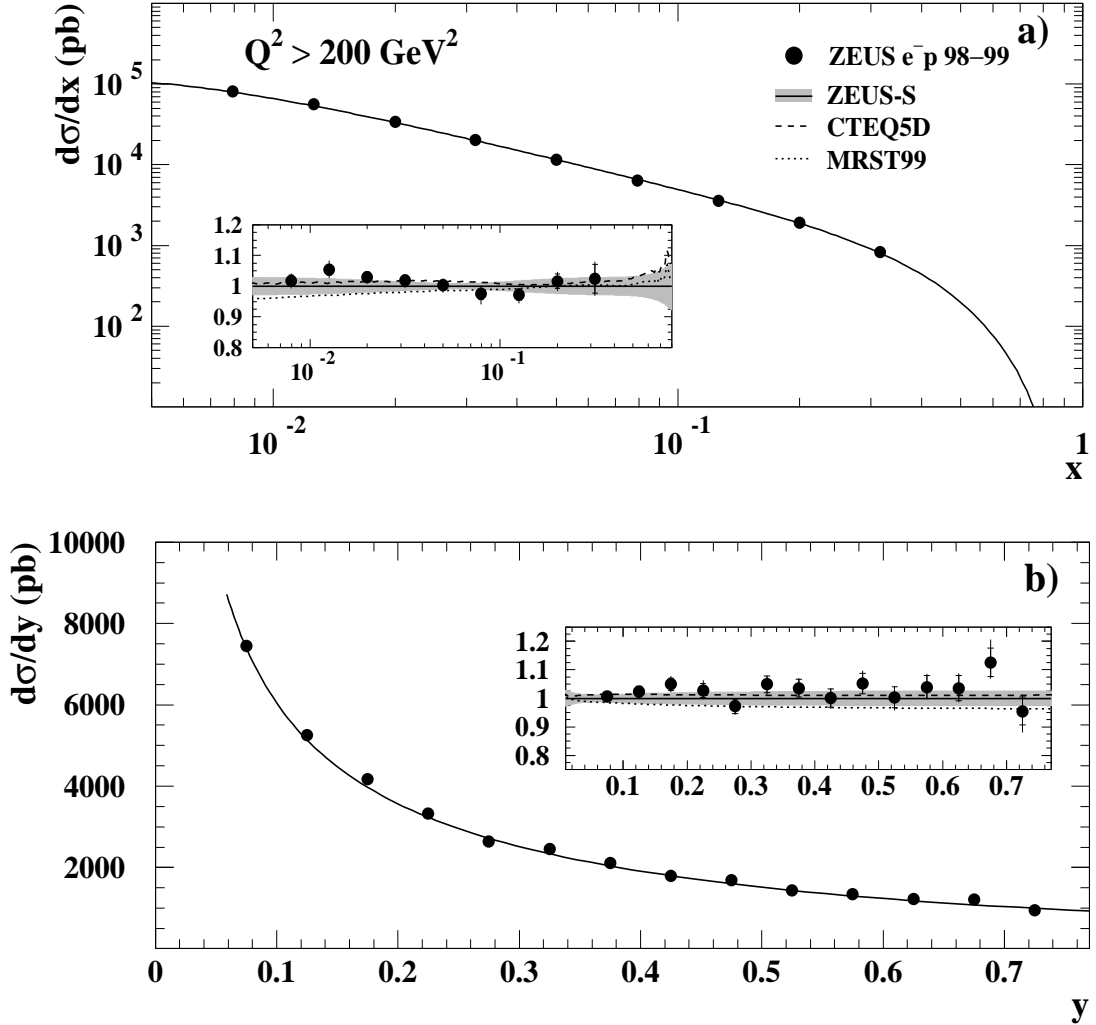}
  \end{center}
  \caption{ a) The differential $e^-p$ cross-section $d\sigma/dx$ for $Q^2 >
    200 \gev^2$ as a function of $x$ compared to the SM expectation
    evaluated using the ZEUS-S fit. b) The differential $e^-p$
    cross-section $d\sigma/dy$ for $Q^2 > 200 \gev^2$ as a function of
    $y$ compared to the SM expectation evaluated using the ZEUS-S fit.
    The insets show ratios of the measured cross sections to the
    ZEUS-S predictions. Also shown are the ratios of SM prediction
    using CTEQ5D and MRST99 PDFs to that of ZEUS-S. The shaded band
    indicates the uncertainty on the calculated cross section due to
    the uncertainty in the ZEUS-S PDFs. The inner error bars of the measured
    points show the statistical uncertainty, while the outer ones show
    the statistical and systematic uncertainties added in quadrature.}
  \label{fig-dsdxdy}
\end{figure}

The single-differential cross-section $d\sigma/ d Q^2$ is shown in
\fig{dsdQ2}a) and tabulated in \tab{dsdQ2}. The systematic
uncertainties are collected in \tab{dsdQ2_c}. The ratio of
$d\sigma/dQ^2$ to ZEUS-S, displayed in \fig{dsdQ2}b), shows that the
SM gives a good description of the data. Note that the ZEUS-S fit did
not use the data presented in this paper. The cross-sections
$d\sigma/dx$ and $d\sigma/dy$ for $Q^2 > 200 \gev^2$ are shown in
\fig{dsdxdy} and are tabulated in \taband{dsdx}{dsdy} (systematic
uncertainties are listed in \taband{dsdx_c}{dsdy_c}). The SM cross
sections, evaluated using the ZEUS-S PDFs, again give a good
description of the data. The plots also contain the SM predictions
using the CTEQ5D \cite{epj:c12:375} and MRST99
\cite{epj:c4:463,*hep-ph-0106075} PDFs.

\begin{figure}
  \begin{center}
    \includegraphics[width=.9\textwidth]{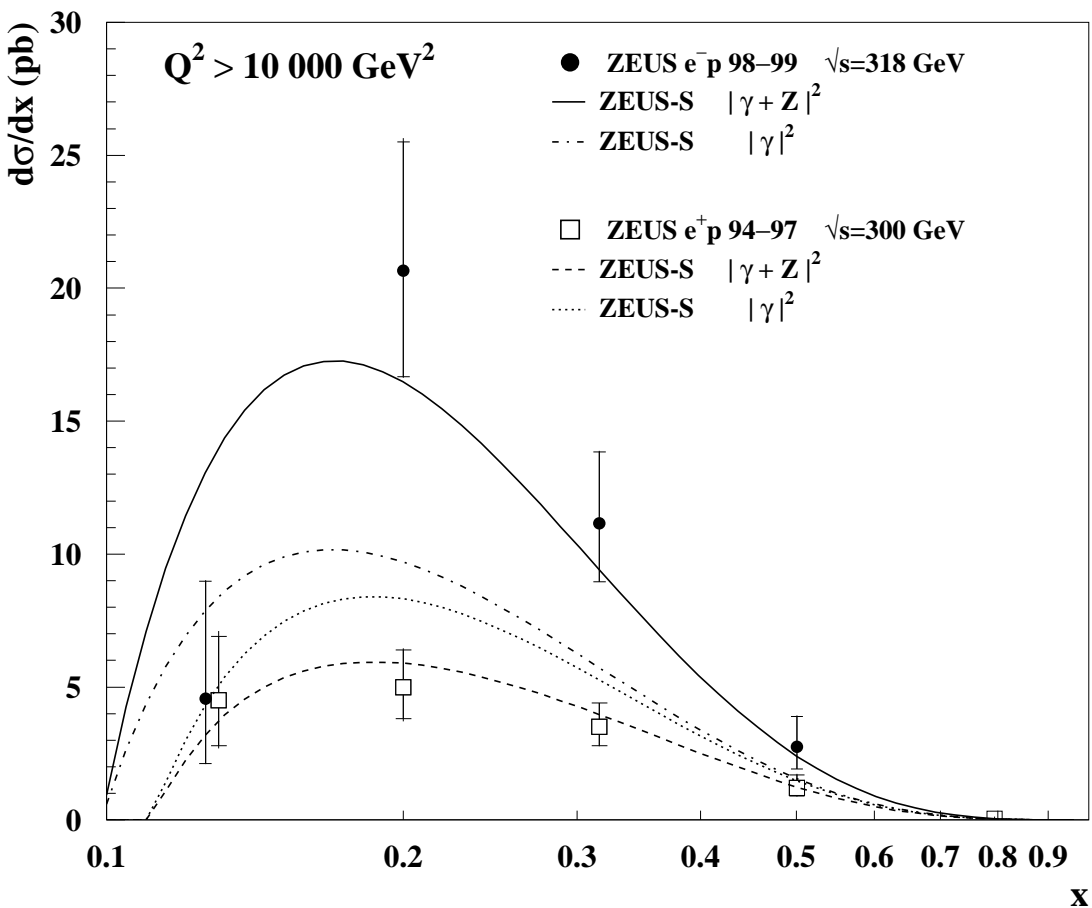}
  \end{center}
  \caption{
    Comparison of measured cross-sections $d\sigma / dx$ for $e^-p$
    (at $\sqrt{s} = 318 \gev$) and $e^+p$ (at $\sqrt{s} = 300 \gev$)
    scattering as a function of $x$ for $Q^2 > 10\,000 \gev^2$. The
    cross sections calculated including the $Z$-exchange contribution
    are shown by the solid and dashed lines. The cross sections
    obtained from photon exchange only are shown by the dash-dotted
    and dotted lines. The lowest $e^-p$ point in $x$ is moved slightly
    to the left for clarity. The inner error bars show the statistical
    error, while the outer ones show the statistical and systematic
    uncertainties added in quadrature.  }
  \label{fig-dsdx10000}
\end{figure}

The $Z$-boson-exchange contribution to NC DIS is clearly seen in
\fig{dsdx10000}, which compares $d\sigma/dx$ for $e^+p$
\cite{epj:c21:443} and $e^-p$ scattering for $\qq > 10\,000
\gev^2$. The $e^-p$ cross section is significantly larger than the
$e^+p$ cross section. This is due to the parity-violating part of the
$Z$-exchange contribution enhancing the $e^-p$ NC DIS cross section
and suppressing the $e^+p$ NC DIS cross section compared to pure
photon exchange. The lines for pure photon exchange are different
because of the different centre-of-mass energies at which the $e^+p$
and $e^-p$ data sets were taken.

\subsection{Reduced cross section}
\label{sec-Double}

\begin{figure}
  \begin{center}
    \includegraphics[width=.9\textwidth]{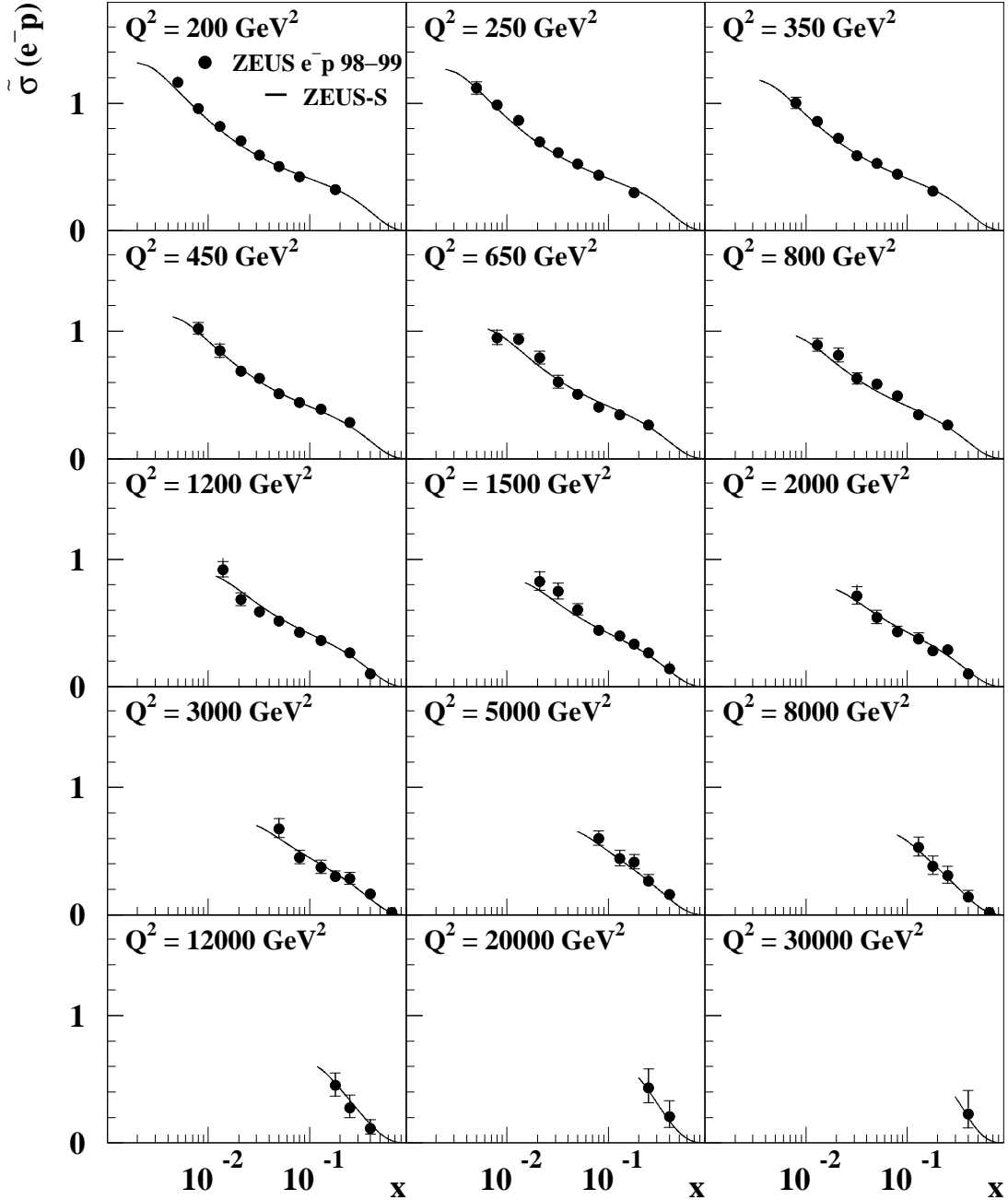}
  \end{center}
  \caption{
    The $e^- p$ reduced cross-section $\tilde{\sigma}(e^- p)$ (solid points)
    plotted as a function of $x$ at fixed $Q^2$ between $200 \gev^2$ and
    $30\,000 \gev^2$. Also shown is the SM expectation evaluated using the
    ZEUS-S PDFs. The inner error bars show the statistical uncertainty, while the
    outer ones show the statistical and systematic uncertainties added in
    quadrature.}
  \label{fig-RedCross1}
\end{figure}

The reduced cross section, $\tilde{\sigma}(e^- p)$, tabulated in
\taband{dsdqdx_1}{dsdqdx_6} (systematic uncertainties are listed in
\taband{dsdxq_c1}{dsdxq_c6}), is shown in \fig{RedCross1} as a
function of $x$ for fixed $Q^2$. The rise of $\tilde{\sigma}(e^-p)$ at
fixed $Q^2$ as $x$ decreases reflects the strong rise of
$F_2$\cite{epj:c21:443}. The SM, evaluated with the ZEUS-S PDFs, gives
a good description of the data. The measurements agree well with those
of the H1 collaboration \cite{epj:c19:269}.

\begin{figure}
  \begin{center}
    \includegraphics[width=.95\textwidth]{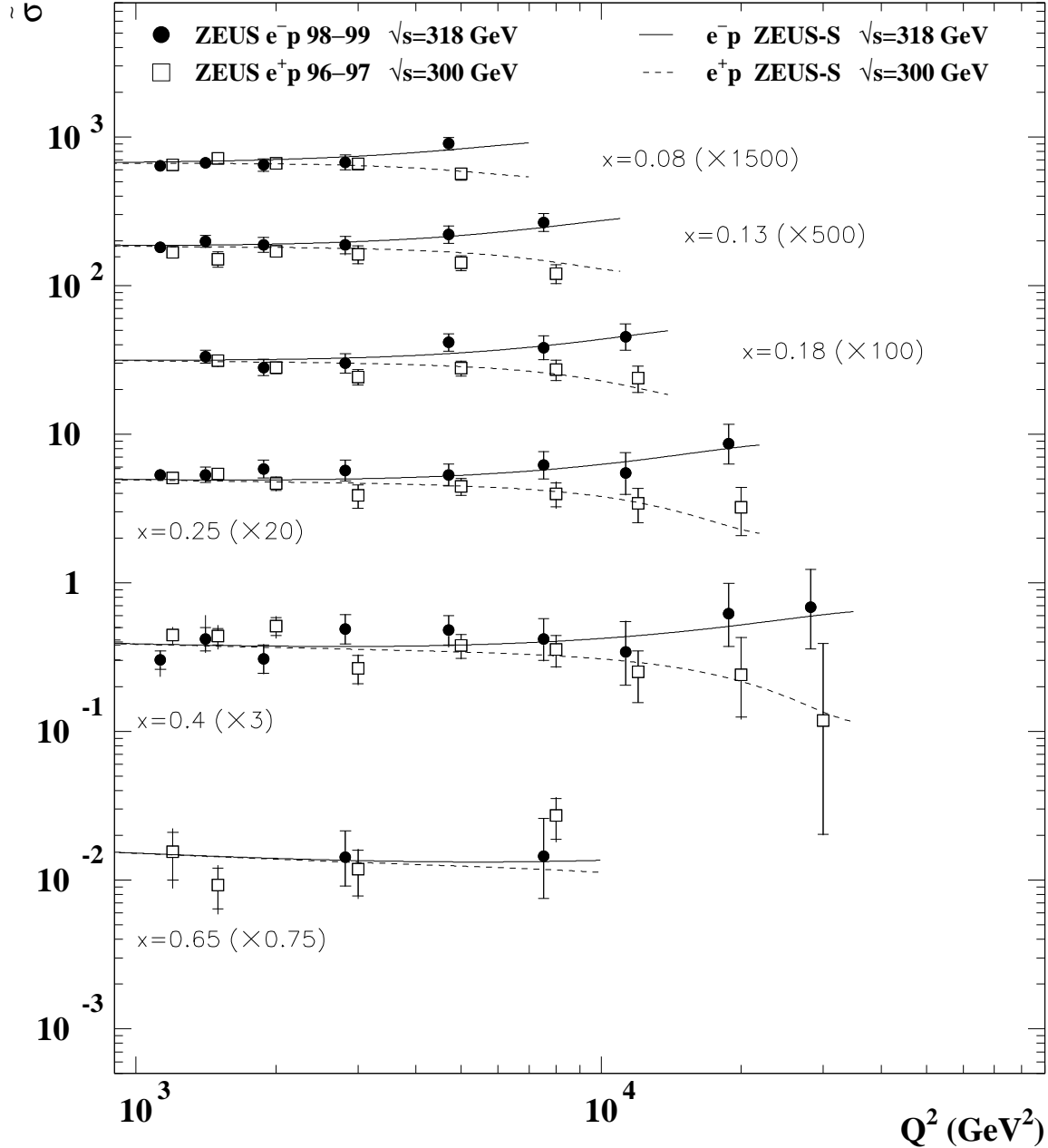}
  \end{center}
  \caption{
    The reduced cross-sections $\tilde{\sigma}$ for $e^-p$ (solid
    points) and $e^+p$ (open squares) scattering as a function of
    $Q^2$ in six different bins of $x$. All $e^-p$ points are moved
    slightly to the left for clarity. The measured values are compared
    to theoretical predictions using the ZEUS-S PDFs. The inner error
    bars show the statistical uncertainty, while the outer ones show
    the statistical and systematic uncertainties added in quadrature.
    The reduced cross sections for a particular $x$ value have been
    scaled by the number shown in parentheses.  }
  \label{fig-RedCross3}
\end{figure}

\Fig{RedCross3} shows the reduced cross section plotted as a function
of $Q^2$ at several values of $x$.  The plot also contains the ZEUS
measurement of $\tilde{\sigma}(e^+p)$ \cite{epj:c21:443}, based on
data collected at $\sqrt{s} = 300 \gev$. For $Q^2$ values below $\sim
3\,000 \gev^2$, the $e^+p$ and $e^-p$ cross sections are nearly
equal. At higher $Q^2$, $\tilde{\sigma}(e^-p)$ is greater than
$\tilde{\sigma}(e^+p)$ as expected from $Z$-boson exchange.

\subsection{\boldmath The $xF_3$ structure function and electroweak analysis}
\label{sec-xF3EW}

The parity-violating structure function, $xF_3$, was obtained, using
\eqsand{Born}{red}, by subtracting the respective reduced cross
sections. Since the $\tilde{\sigma}(e^+p)$ cross section was measured
at a centre-of-mass energy of $300 \gev$ \cite{epj:c21:443}, $xF_3$
was determined by evaluating
\begin{equation}
  xF_3 = \left( 
           \frac{Y_-^{300}}{Y_+^{300}} + \frac{Y_-^{318}}{Y_+^{318}}
           \right)^{-1}
         \left( \tilde{\sigma}(e^-p) - \tilde{\sigma}(e^+p) \right)
         - \Delta_{F_L} \ ,
  \label{eq-xF3} 
\end{equation}
where the superscripts `300' and `318' denote the different
centre-of-mass energies.  The term $\Delta_{F_L}$ in \eq{xF3} is
non-zero because of the different centre-of-mass energies at which the
$e^+p$ and $e^-p$ data were collected.  The relative size of
$\Delta_{F_L}$, computed at NLO in QCD, is less than $1\%$ over most
of the kinematic range in which $xF_3$ is presented and is neglected.

\begin{figure}
  \begin{center}
    \includegraphics[width=.85\textwidth]{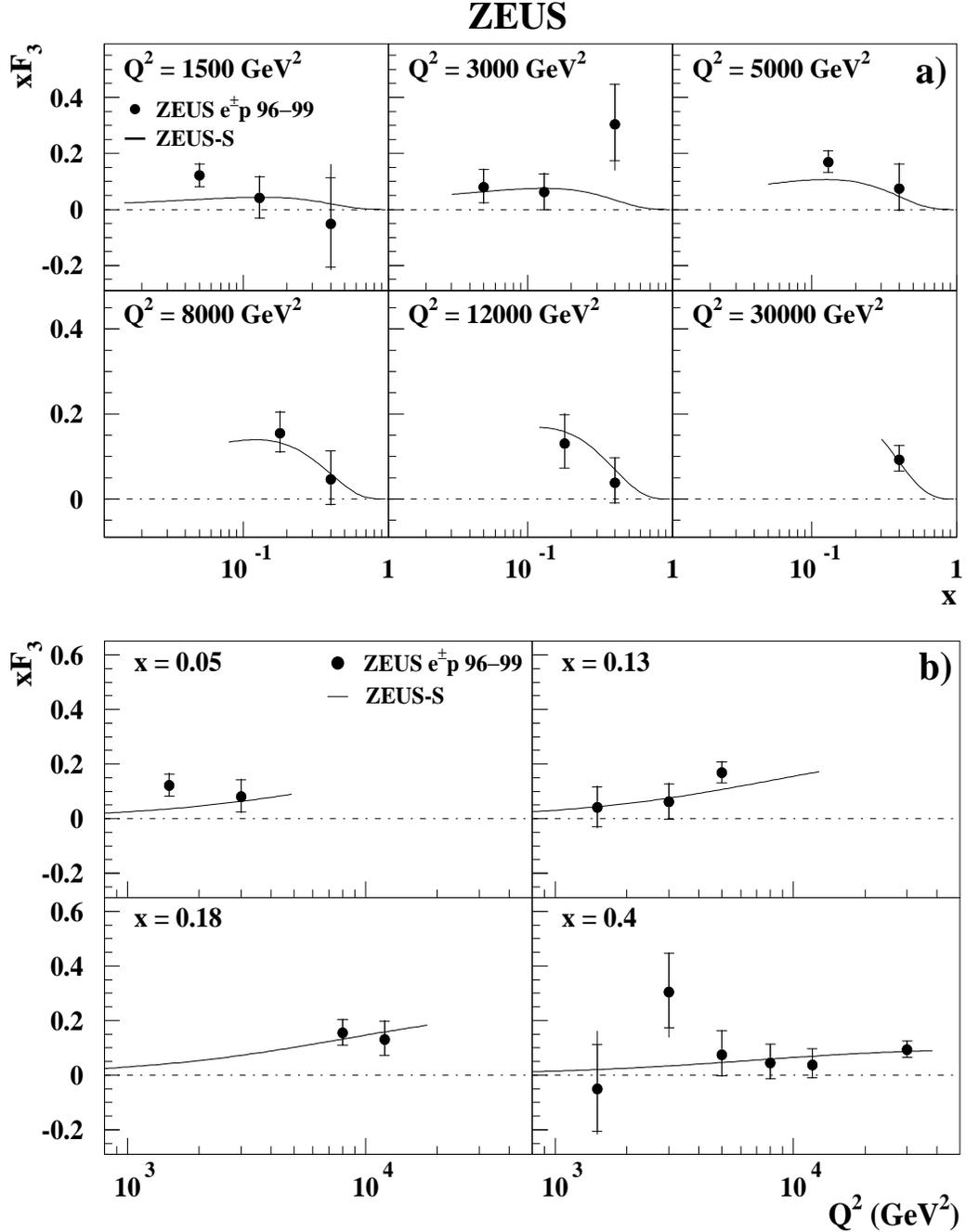}
  \end{center}
  \caption{The structure function $xF_3$ a) as a function of x for
    different \qq\ values and b) as a function of $Q^2$ for different
    $x$ values. The inner error bars show the statistical uncertainty,
    while the outer ones show the statistical and systematic
    uncertainties added in quadrature. Also shown are SM calculations
    using the ZEUS-S PDFs.}
  \label{fig-xF31}
\end{figure}

To reduce statistical fluctuations, several bins used in the
measurement of the double-differential cross section were combined.
\Fig{xF31}a) shows $xF_3$ at fixed values of $Q^2$ as a function of
$x$, whereas \fig{xF31}b) shows $xF_3$ at fixed values of $x$ as a
function of $Q^2$. The measured values are tabulated in \tab{xF3}.
Since the statistical errors dominate the uncertainty, systematic
uncertainties were assumed to be uncorrelated between the $e^+p$ and
the $e^-p$ data sets. The luminosity errors have been included in the
total systematic uncertainty on the most conservative assumption that
they are completely anti-correlated. The expectation of the SM,
evaluated with the ZEUS-S PDFs, gives a good description of the data.
The measurements of the H1 collaboration \cite{epj:c19:269} are in good
agreement with the present results.

To compare the present measurement of $xF_3$ to that obtained at lower
$Q^2$ in fixed-target experiments, it is convenient to use two
structure functions $xG_3(x,Q^2)$ and $xH_3(x,Q^2)$ \cite{zfp:c24:151}
as follows:
\begin{equation*}
  x F_3  =     -a_e \chi_Z\,  xG_3 + 2 v_e a_e \chi_Z^2\, xH_3 \ .
\end{equation*}
The term containing $xG_3$ arises from $\gamma$-$Z$ interference,
while the $xH_3$ term arises purely from $Z$ exchange. The $xH_3$ term
is negligible in comparison to the $xG_3$ term because the coefficient
multiplying $xH_3$ contains the vector coupling of the electron, $v_e
= -0.054$, and $xH_3$ itself is less than half the size of $xG_3$. At
fixed $x$, $xG_3$ depends weakly on $Q^2$
\cite{zfp:c24:151,epjd:cn2:1}. For example, according to the
ZEUS-S PDFs, at $x=0.25$, $xG_3$ varies from $0.46$ at $Q^2 = 100
\gev^2$ to 0.37 at $Q^2 = 10\,000 \gev^2$.

Each value of $xF_3$ was used to obtain an estimate of $xG_3$ by evaluating
\begin{equation*}
  xG_3 \cong \frac{ xF_3}{\left[ -a_e \chi_Z \right]} \ .
\end{equation*}
The weak $Q^2$ dependence of $xG_3$ was accounted for by extrapolating
each $xG_3$ value to $Q^2 = 1\,500 \gev^2$ using ZEUS-S PDFs. In
order to reduce statistical fluctuations, $xG_3$ values from different
bins with the same $x$ were combined by computing the weighted mean of
the individual estimates of $xG_3$. Since the errors are not
symmetric, the mean of the upper and lower statistical error was used
as the weight in this calculation.

\begin{figure}
  \begin{center}
    \includegraphics[width=.93\textwidth]{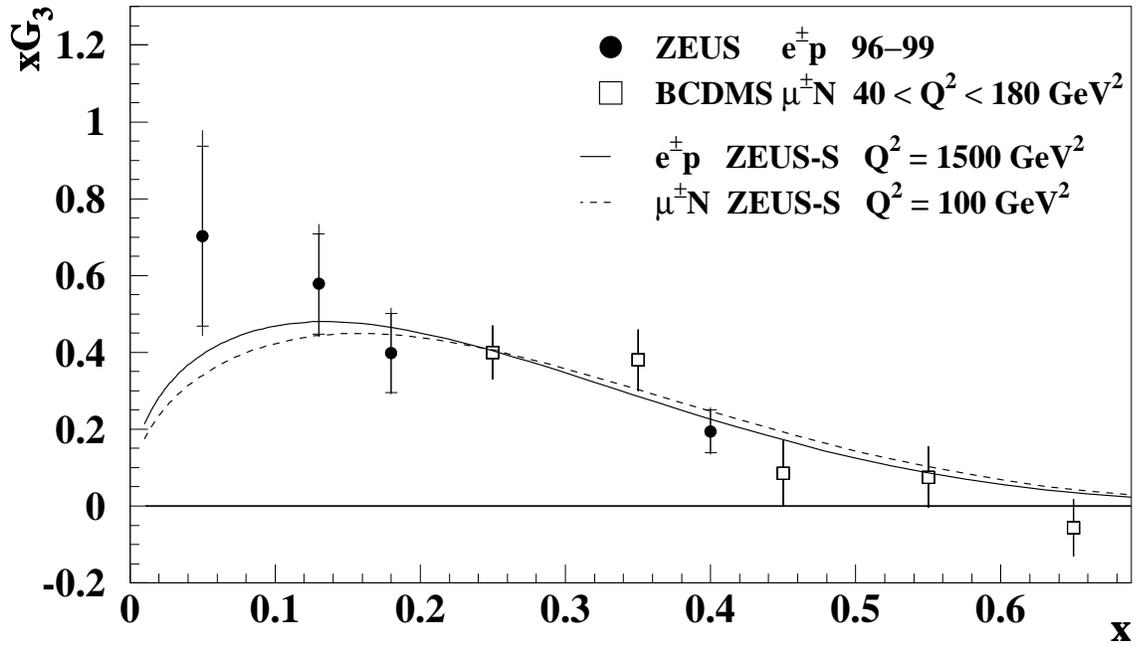}
  \end{center}
  \caption{The structure function $xG_3$ for $e^\pm p$ scattering (solid points)
    compared to that from BCDMS (open squares, total errors only). The
    prediction based on the ZEUS-S PDFs at $Q^2 = 1\,500 \gev^2$ (\,$100
    \gev^2$) is shown as the solid (dashed) line. }
  \label{fig-xG3}
\end{figure}

The result of the above procedure is shown in \fig{xG3}. This figure
also shows the results obtained by the BCDMS collaboration
\cite{pl:b140:142}, which were extracted over the kinematic range $40
< Q^2 < 180 \gev^2$ and $0.2 < x < 0.7$ from NC muon-carbon
scattering. The value of $xG_3$ extracted by the BCDMS collaboration
is therefore the average of $xG_3$ for the proton and the neutron,
since the target nucleus is isoscalar. \Fig{xG3} shows $xG_3$
evaluated for $ep$ scattering at $Q^2 = 1\,500 \gev^2$ and for $\mu N$
scattering at $Q^2 = 100 \gev^2$ using the ZEUS-S PDFs.  The
difference between the theoretical predictions for $ep$ and $\mu N$
scattering, evaluated with ZEUS-S PDFs, is small and the BCDMS data
agree well with the present measurement. The ZEUS data extend the
measurement of $xG_3$ down to $x=0.05$.

%
\section{Summary}
\label{Sect:Summary}

The cross sections for neutral current deep inelastic scattering,
$e^-p \rightarrow e^-X$, have been measured using $15.9\pb^{-1}$ of
data collected with the ZEUS detector during the 1998--99 running
periods. The single-differential cross-sections $d\sigma / dQ^2$,
$d\sigma / dx$ and $d\sigma / dy$ have been measured for $Q^2 > 200
\gev^2$. In order to exhibit the effect of $Z$-boson exchange,
$d\sigma / dx$ has also been measured for $Q^2 > 10\,000 \gev^2$. The
reduced cross section has been measured in the kinematic range $185 <
Q^2 < 50\,000 \,\gevv$ and $0.0037 < x < 0.75$. The Standard Model
predictions, including both $\gamma$ and $Z$ exchange and using
standard parton density functions (ZEUS-S, CTEQ5D and MRST99), are
in good agreement with the data.

The parity-violating structure function, $xF_3$, has been extracted by
combining the data presented here with the published ZEUS measurement
of the reduced cross section for neutral current $e^+p$ deep inelastic
scattering. The structure function $xG_3$ has been extracted from the
$xF_3$ measurement and compared to previous results obtained in
fixed-target muon-carbon scattering by the BCDMS collaboration. The
ZEUS results are in good agreement with the BCDMS measurement and
extend the range of $x$ values covered down to $x=0.05$. The results
are also in good agreement with theoretical predictions and, since
$xF_3$ is non-zero, show the presence of $Z$ exchange in the
space-like $Q^2$ region explored by deep inelastic $ep$
scattering. 

\section{Acknowledgements}

The strong support and encouragement of the DESY Directorate have been
invaluable. The experiment was made possible by the inventiveness and
the diligent efforts of the HERA machine group. The design,
construction and installation of the ZEUS detector have been made
possible by the ingenuity and dedicated efforts of many people from
inside DESY and from home institutes who are not listed as authors.
Their contributions are acknowledged with great appreciation. Special
thanks are due to Hannes Jung who helped to modify RAPGAP so that
diffractive events could be included in the analysis.

\vfill\eject
{
\def\bibname{\Large\bf References}
\def\refname{\Large\bf References}
\pagestyle{plain}
\ifzeusbst
  \bibliographystyle{./BiBTeX/bst/l4z_default}
\fi
\ifzdrftbst
  \bibliographystyle{./BiBTeX/bst/l4z_draft}
\fi
\ifzbstepj
  \bibliographystyle{./BiBTeX/bst/l4z_epj}
\fi
\ifzbstnp
  \bibliographystyle{./BiBTeX/bst/l4z_np}
\fi
\ifzbstpl
  \bibliographystyle{./BiBTeX/bst/l4z_pl}
\fi
{\raggedright
\bibliography{./BiBTeX/user/syn,%
              ./BiBTeX/bib/l4z_articles,%
              ./BiBTeX/bib/l4z_books,%
              ./BiBTeX/bib/l4z_conferences,%
              ./BiBTeX/bib/l4z_h1,%
              ./BiBTeX/bib/l4z_misc,%
              ./BiBTeX/bib/l4z_old,%
              ./BiBTeX/bib/l4z_preprints,%
              ./BiBTeX/bib/l4z_replaced,%
              ./BiBTeX/bib/l4z_temporary,%
              ./BiBTeX/bib/l4z_zeus}}
}
\vfill\eject

\clearpage
\begin{table} [!ht] 
  \begin{center}
    {\footnotesize
\renewcommand{\arraystretch}{1.2}
\begin{tabular}{|r@{ -- }r|r|l@{}l@{$\,$}l@{$\,$}l@{$\,$}l|l|r|r|c|}
\hline
\multicolumn{2}{|c|}{ $Q^2$ range} & 
\multicolumn{1}{c|}{$Q^2_c$} & 
\multicolumn{6}{c|} {$d\sigma / dQ^2\ (\! \pb \, / \gev^2$)} &
{$N_{\rm obs}$} & 
{$N_{\rm bg}$} & 
{$\cal A$} \\
\cline{4-9}
\multicolumn{2}{|c|}{($\Gev^2$)} &
\multicolumn{1}{c|}{($\Gev^2$)} & 
\multicolumn{5}{c|}{measured} &
\multicolumn{1}{c|}{SM} &&& \\
\hline
\hline
$  200.0$ & $  300.0$ & $  250$ && $11.230$ & $\pm 0.105$ & $^{+0.108}_{-0.151}$ & $
$ & $11.220$ &
$15360$ & $3.5$ & $0.81$ \\
$  300.0$ & $  400.0$ & $  350$ && $ 5.040$ & $\pm 0.073$ & $^{+0.038}_{-0.085}$ & $
$ & $ 5.022$ &
$ 6474$ & $2.6$ & $0.77$ \\
$  400.0$ & $  475.7$ & $  440$ && $ 2.879$ & $\pm 0.061$ & $^{+0.026}_{-0.044}$ & $
$ & $ 2.890$ &
$ 2737$ & $0.7$ & $0.74$ \\
$  475.7$ & $  565.7$ & $  520$ && $ 1.899$ & $\pm 0.049$ & $^{+0.021}_{-0.035}$ & $
$ & $ 1.924$ &
$ 1850$ & $1.4$ & $0.65$ \\
$  565.7$ & $  672.7$ & $  620$ && $ 1.217$ & $\pm 0.039$ & $^{+0.013}_{-0.025}$ & $
$ & $ 1.251$ &
$ 1184$ & $0.8$ & $0.54$ \\
$  672.7$ & $  800.0$ & $  730$ && $\left( \right.\!  8.96$ & $\pm  0.28$ & $^{+ 0.10}_{- 0.11}$ & $ \left. \!\!\!\! \right) \cdot 10^{-1}
$ & $  8.36\cdot 10^{-1}$ &
$ 1215$ & $0.9$ & $0.65$ \\
$  800.0$ & $  951.4$ & $  870$ && $\left( \right.\!  5.34$ & $\pm  0.17$ & $^{+ 0.07}_{- 0.10}$ & $ \left. \!\!\!\! \right) \cdot 10^{-1}
$ & $  5.41\cdot 10^{-1}$ &
$ 1171$ & $0.7$ & $0.86$ \\
$  951.4$ & $ 1131.0$ & $ 1040$ && $\left( \right.\!  3.40$ & $\pm  0.12$ & $^{+ 0.06}_{- 0.04}$ & $ \left. \!\!\!\! \right) \cdot 10^{-1}
$ & $  3.47\cdot 10^{-1}$ &
$  973$ & $0.6$ & $0.93$ \\
$ 1131.0$ & $ 1345.0$ & $ 1230$ && $\left( \right.\!  2.17$ & $\pm  0.09$ & $^{+ 0.04}_{- 0.03}$ & $ \left. \!\!\!\! \right) \cdot 10^{-1}
$ & $  2.28\cdot 10^{-1}$ &
$  751$ & $1.4$ & $0.98$ \\
$ 1345.0$ & $ 1600.0$ & $ 1470$ && $\left( \right.\!  1.56$ & $\pm  0.07$ & $^{+ 0.02}_{- 0.03}$ & $ \left. \!\!\!\! \right) \cdot 10^{-1}
$ & $  1.45\cdot 10^{-1}$ &
$  638$ & $0.8$ & $0.98$ \\
$ 1600.0$ & $ 1903.0$ & $ 1740$ && $\left( \right.\!  9.86$ & $\pm  0.48$ & $^{+ 0.16}_{- 0.16}$ & $ \left. \!\!\!\! \right) \cdot 10^{-2}
$ & $  9.46\cdot 10^{-2}$ &
$  488$ & $0.6$ & $0.96$ \\
$ 1903.0$ & $ 2263.0$ & $ 2100$ && $\left( \right.\!  5.39$ & $^{+  0.32}_{- 0.31}$ & $^{+ 0.08}_{- 0.08}$ & $ \left. \!\!\!\! \right) \cdot 10^{-2}
$ & $  5.86\cdot 10^{-2}$ &
$  333$ & $0.6$ & $0.98$ \\
$ 2263.0$ & $ 2691.0$ & $ 2500$ && $\left( \right.\!  3.86$ & $^{+  0.25}_{- 0.24}$ & $^{+ 0.11}_{- 0.07}$ & $ \left. \!\!\!\! \right) \cdot 10^{-2}
$ & $  3.75\cdot 10^{-2}$ &
$  283$ & $1.1$ & $0.97$ \\
$ 2691.0$ & $ 3200.0$ & $ 2900$ && $\left( \right.\!  2.47$ & $^{+  0.19}_{- 0.18}$ & $^{+ 0.04}_{- 0.03}$ & $ \left. \!\!\!\! \right) \cdot 10^{-2}
$ & $  2.56\cdot 10^{-2}$ &
$  196$ & $0.2$ & $0.97$ \\
$ 3200.0$ & $ 4525.0$ & $ 3800$ && $\left( \right.\!  1.40$ & $^{+  0.09}_{- 0.08}$ & $^{+ 0.03}_{- 0.04}$ & $ \left. \!\!\!\! \right) \cdot 10^{-2}
$ & $  1.27\cdot 10^{-2}$ &
$  308$ & $0.4$ & $0.98$ \\
$ 4525.0$ & $ 6400.0$ & $ 5400$ && $\left( \right.\!  5.09$ & $^{+  0.44}_{- 0.41}$ & $^{+ 0.12}_{- 0.13}$ & $ \left. \!\!\!\! \right) \cdot 10^{-3}
$ & $  5.08\cdot 10^{-3}$ &
$  157$ & $0.2$ & $0.96$ \\
$ 6400.0$ & $ 9051.0$ & $ 7600$ && $\left( \right.\!  2.15$ & $^{+  0.25}_{- 0.23}$ & $^{+ 0.03}_{- 0.05}$ & $ \left. \!\!\!\! \right) \cdot 10^{-3}
$ & $  2.03\cdot 10^{-3}$ &
$   91$ & $0.0$ & $0.93$ \\
$ 9051.0$ & $12800.0$ & $10800$ && $\left( \right.\!   6.0$ & $^{+   1.2}_{-  1.0}$ & $^{+  0.3}_{-  0.1}$ & $ \left. \!\!\!\! \right) \cdot 10^{-4}
$ & $   7.6\cdot 10^{-4}$ &
$   35$ & $0.0$ & $0.93$ \\
$12800.0$ & $18100.0$ & $15200$ && $\left( \right.\!   3.2$ & $^{+   0.8}_{-  0.6}$ & $^{+  0.3}_{-  0.1}$ & $ \left. \!\!\!\! \right) \cdot 10^{-4}
$ & $   2.7\cdot 10^{-4}$ &
$   25$ & $0.0$ & $0.90$ \\
$18100.0$ & $25600.0$ & $21500$ && $\left( \right.\!   7.1$ & $^{+   3.6}_{-  2.5}$ & $^{+  0.4}_{-  0.4}$ & $ \left. \!\!\!\! \right) \cdot 10^{-5}
$ & $   8.3\cdot 10^{-5}$ &
$    8$ & $0.0$ & $0.93$ \\
$25600.0$ & $36200.0$ & $30400$ && $\left( \right.\!   2.0$ & $^{+   2.0}_{-  1.1}$ & $^{+  0.3}_{-  0.0}$ & $ \left. \!\!\!\! \right) \cdot 10^{-5}
$ & $   2.0\cdot 10^{-5}$ &
$    3$ & $0.0$ & $0.93$ \\
$36200.0$ & $51200.0$ & $43100$ && $\left( \right.\!   5.3$ & $^{+  12.1}_{-  4.2}$ & $^{+  1.3}_{-  0.2}$ & $ \left. \!\!\!\! \right) \cdot 10^{-6}
$ & $   3.3\cdot 10^{-6}$ &
$    1$ & $0.0$ & $0.93$ \\
\hline
\end{tabular}
}

  \end{center}    
  \caption[this space for rent]{
    The differential cross-section \sigqq\ for the reaction $e^{-} p \rightarrow
    e^{-} X$.  The following quantities are given for each bin: the \qq\ range,
    the value at which the cross section is quoted, \qqc, the measured
    cross-section \sigqq\ corrected to the Born level and the
    corresponding cross 
    section predicted by the SM using CTEQ5D PDFs. The
    first error of the measured cross section gives
    the statistical error and the second is the systematic uncertainty. The last three
    columns contain the number of observed events in data, $N_{\rm obs}$, the number of
    expected background events, $N_{\rm bg}$ and the acceptance, $\cal A$.  }
  \label{tab-dsdQ2}
\end{table}

\begin{table} [!ht] 
  \begin{center}
    {\footnotesize
\renewcommand{\arraystretch}{1.2}
\begin{tabular}{|r|l|c|c||c|c|c|c|c|c|c|}
\hline
\multicolumn{1}{|c|}{$Q^2_c$} & 
\multicolumn{1}{c|}{$d\sigma / dQ^2$} &
stat.  & 
total sys.  & 
uncor. sys. &
$\delta_1$ &
$\delta_2$ &
$\delta_3$ &
$\delta_4$ &
$\delta_5$ &
$\delta_6$ \\
\multicolumn{1}{|c|}{$(\Gev^2)$} &
\multicolumn{1}{c|}{$(\pb \, / \gev^2)$} &
(\%) & 
(\%) &
(\%) &
(\%) &
(\%) &
(\%) &
(\%) &
(\%) &
(\%) \\
\hline \hline
$  250$ & $
11.230$ &
$^{+ 0.9  } _{- 0.9  } $ & $^{+ 1.0  } _{- 1.3  } $ & $^{+ 0.8  } _{- 0.5  } $ & $^{-0.3  }  _{+ 0.4  } $ & $^{}_{+ 0.0  } $ & $^{+ 0.1  } _{+ 0.1  } $ & $^{-0.6  }  _{+ 0.3  } $ & $^{+ 0.3  } _{-1.0  }  $ & $^{+ 0.1  } _{-0.3  }  
$ \\
$  350$ & $
 5.040$ &
$^{+ 1.4  } _{- 1.4  } $ & $^{+ 0.8  } _{- 1.7  } $ & $^{+ 0.5  } _{- 0.8  } $ & $^{-0.4  }  _{+ 0.3  } $ & $^{}_{-0.1  }  $ & $^{-0.1  }  _{-0.1  }  $ & $^{-0.9  }  _{-0.1  }  $ & $^{+ 0.2  } _{-1.0  }  $ & $^{-0.2  }  _{+ 0.5  } 
$ \\
$  440$ & $
 2.879$ &
$^{+ 2.1  } _{- 2.1  } $ & $^{+ 0.9  } _{- 1.5  } $ & $^{+ 0.7  } _{- 0.5  } $ & $^{-0.2  }  _{+ 0.3  } $ & $^{}_{+ 0.1  } $ & $^{+ 0.1  } _{+ 0.2  } $ & $^{-0.5  }  _{+ 0.0  } $ & $^{+ 0.1  } _{-1.0  }  $ & $^{+ 0.4  } _{-0.9  }  
$ \\
$  520$ & $
 1.899$ &
$^{+ 2.6  } _{- 2.5  } $ & $^{+ 1.1  } _{- 1.9  } $ & $^{+ 0.7  } _{- 1.6  } $ & $^{-0.2  }  _{+ 0.3  } $ & $^{}_{- 0.0  } $ & $^{-0.1  }  _{+ 0.0  } $ & $^{-0.4  }  _{+ 0.5  } $ & $^{+ 0.2  } _{-0.9  }  $ & $^{-0.2  }  _{+ 0.5  } 
$ \\
$  620$ & $
 1.217$ &
$^{+ 3.2  } _{- 3.1  } $ & $^{+ 1.1  } _{- 2.0  } $ & $^{+ 1.0  } _{- 1.1  } $ & $^{-0.5  }  _{+ 0.3  } $ & $^{}_{-0.2  }  $ & $^{-0.2  }  _{- 0.0  } $ & $^{-1.1  }  _{+ 0.1  } $ & $^{-0.1  }  _{-1.0  }  $ & $^{+ 0.3  } _{-0.6  }  
$ \\
$  730$ & $
  8.96\cdot 10^{-1}$ &
$^{+ 3.2  } _{- 3.1  } $ & $^{+ 1.1  } _{- 1.2  } $ & $^{+ 0.8  } _{- 0.3  } $ & $^{-0.2  }  _{+ 0.4  } $ & $^{}_{+ 0.4  } $ & $^{+ 0.0  } _{+ 0.0  } $ & $^{-0.6  }  _{+ 0.4  } $ & $^{+ 0.1  } _{-0.9  }  $ & $^{+ 0.2  } _{-0.4  }  
$ \\
$  870$ & $
  5.34\cdot 10^{-1}$ &
$^{+ 3.2  } _{- 3.1  } $ & $^{+ 1.4  } _{- 1.8  } $ & $^{+ 1.2  } _{- 0.6  } $ & $^{+ 0.1  } _{+ 0.2  } $ & $^{}_{-0.1  }  $ & $^{- 0.0  } _{+ 0.0  } $ & $^{-0.9  }  _{-0.1  }  $ & $^{+ 0.3  } _{-0.9  }  $ & $^{+ 0.5  } _{-1.2  }  
$ \\
$ 1040$ & $
  3.40\cdot 10^{-1}$ &
$^{+ 3.5  } _{- 3.4  } $ & $^{+ 1.7  } _{- 1.3  } $ & $^{+ 1.1  } _{- 0.3  } $ & $^{+ 0.1  } _{+ 0.2  } $ & $^{}_{-0.5  }  $ & $^{+ 0.1  } _{+ 0.1  } $ & $^{-0.5  }  _{+ 1.0  } $ & $^{+ 0.4  } _{-1.0  }  $ & $^{-0.2  }  _{+ 0.5  } 
$ \\
$ 1230$ & $
  2.17\cdot 10^{-1}$ &
$^{+ 4.0  } _{- 3.9  } $ & $^{+ 1.7  } _{- 1.2  } $ & $^{+ 1.1  } _{- 0.3  } $ & $^{+ 0.1  } _{+ 0.5  } $ & $^{}_{-0.3  }  $ & $^{+ 0.4  } _{-0.2  }  $ & $^{-0.2  }  _{+ 1.1  } $ & $^{+ 0.5  } _{-1.0  }  $ & $^{+ 0.1  } _{-0.3  }  
$ \\
$ 1470$ & $
  1.56\cdot 10^{-1}$ &
$^{+ 4.3  } _{- 4.1  } $ & $^{+ 1.3  } _{- 1.8  } $ & $^{+ 0.8  } _{- 0.7  } $ & $^{-0.2  }  _{+ 0.1  } $ & $^{}_{-0.5  }  $ & $^{+ 0.3  } _{+ 0.2  } $ & $^{-1.3  }  _{+ 0.4  } $ & $^{+ 0.2  } _{-0.9  }  $ & $^{-0.3  }  _{+ 0.8  } 
$ \\
$ 1740$ & $
  9.86\cdot 10^{-2}$ &
$^{+ 4.9  } _{- 4.7  } $ & $^{+ 1.7  } _{- 1.6  } $ & $^{+ 1.3  } _{- 0.5  } $ & $^{+ 0.1  } _{+ 0.3  } $ & $^{}_{+ 0.3  } $ & $^{+ 0.1  } _{-0.1  }  $ & $^{-0.3  }  _{+ 0.7  } $ & $^{+ 0.2  } _{-0.8  }  $ & $^{+ 0.6  } _{-1.3  }  
$ \\
$ 2100$ & $
  5.39\cdot 10^{-2}$ &
$^{+ 6.0  } _{- 5.7  } $ & $^{+ 1.5  } _{- 1.5  } $ & $^{+ 1.1  } _{- 0.3  } $ & $^{+ 0.0  } _{+ 0.3  } $ & $^{}_{+ 0.2  } $ & $^{+ 0.2  } _{-0.2  }  $ & $^{+ 0.5  } _{+ 0.7  } $ & $^{+ 0.5  } _{-0.8  }  $ & $^{+ 0.5  } _{-1.2  }  
$ \\
$ 2500$ & $
  3.86\cdot 10^{-2}$ &
$^{+ 6.5  } _{- 6.1  } $ & $^{+ 2.8  } _{- 1.9  } $ & $^{+ 1.4  } _{- 0.3  } $ & $^{+ 0.3  } _{+ 0.4  } $ & $^{}_{-0.3  }  $ & $^{- 0.0  } _{+ 0.9  } $ & $^{+ 0.4  } _{+ 1.9  } $ & $^{+ 0.6  } _{-0.7  }  $ & $^{+ 0.7  } _{-1.7  }  
$ \\
$ 2900$ & $
  2.47\cdot 10^{-2}$ &
$^{+ 7.8  } _{- 7.3  } $ & $^{+ 1.5  } _{- 1.3  } $ & $^{+ 1.1  } _{- 0.9  } $ & $^{+ 0.1  } _{+ 0.3  } $ & $^{}_{-0.3  }  $ & $^{- 0.0  } _{+ 0.7  } $ & $^{+ 0.2  } _{+ 0.2  } $ & $^{+ 0.4  } _{-0.8  }  $ & $^{-0.3  }  _{+ 0.6  } 
$ \\
$ 3800$ & $
  1.40\cdot 10^{-2}$ &
$^{+ 6.1  } _{- 5.8  } $ & $^{+ 2.0  } _{- 3.1  } $ & $^{+ 1.7  } _{- 0.7  } $ & $^{+ 0.0  } _{+ 0.3  } $ & $^{}_{-2.0  }  $ & $^{-0.1  }  _{+ 0.3  } $ & $^{-1.6  }  _{+ 0.6  } $ & $^{+ 0.4  } _{-0.8  }  $ & $^{+ 0.6  } _{-1.4  }  
$ \\
$ 5400$ & $
  5.09\cdot 10^{-3}$ &
$^{+ 8.7  } _{- 8.1  } $ & $^{+ 2.4  } _{- 2.5  } $ & $^{+ 1.4  } _{- 0.8  } $ & $^{+ 0.2  } _{+ 0.4  } $ & $^{}_{- 0.0  } $ & $^{+ 1.4  } _{-0.3  }  $ & $^{-2.2  }  _{+ 1.2  } $ & $^{+ 0.5  } _{-0.6  }  $ & $^{+ 0.2  } _{-0.5  }  
$ \\
$ 7600$ & $
  2.15\cdot 10^{-3}$ &
$^{+  12. } _{-  11. } $ & $^{+ 1.4  } _{- 2.3  } $ & $^{+ 0.6  } _{- 1.2  } $ & $^{-0.3  }  _{+ 0.2  } $ & $^{}_{-1.6  }  $ & $^{+ 0.5  } _{+ 1.3  } $ & $^{-0.3  }  _{-0.2  }  $ & $^{+ 0.1  } _{-0.9  }  $ & $^{+ 0.2  } _{-0.4  }  
$ \\
$10800$ & $
   6.0\cdot 10^{-4}$ &
$^{+  20. } _{-  17. } $ & $^{+ 4.3  } _{- 1.6  } $ & $^{+ 3.2  } _{- 0.8  } $ & $^{-0.5  }  _{+ 0.5  } $ & $^{}_{+ 1.6  } $ & $^{-0.6  }  _{+ 1.8  } $ & $^{+ 1.5  } _{-0.3  }  $ & $^{+ 0.3  } _{-1.0  }  $ & $^{+ 0.2  } _{-0.5  }  
$ \\
$15200$ & $
   3.2\cdot 10^{-4}$ &
$^{+  24. } _{-  20. } $ & $^{+ 7.8  } _{- 2.9  } $ & $^{+ 0.8  } _{- 0.4  } $ & $^{-0.6  }  _{+ 0.8  } $ & $^{}_{+ 5.6  } $ & $^{-0.9  }  _{+ 3.5  } $ & $^{-2.4  }  _{+ 3.6  } $ & $^{+ 0.2  } _{-0.9  }  $ & $^{-0.8  }  _{+ 1.9  } 
$ \\
$21500$ & $
   7.1\cdot 10^{-5}$ &
$^{+  50. } _{-  35. } $ & $^{+ 5.0  } _{- 5.1  } $ & $^{+ 1.2  } _{- 0.5  } $ & $^{-0.5  }  _{+ 0.4  } $ & $^{}_{-4.7  }  $ & $^{-1.2  }  _{+ 4.2  } $ & $^{+ 2.3  } _{-0.6  }  $ & $^{+ 0.2  } _{-1.0  }  $ & $^{+ 0.3  } _{-0.7  }  
$ \\
$30400$ & $
   2.0\cdot 10^{-5}$ &
$^{+  97. } _{-  53. } $ & $^{+  13. } _{- 2.4  } $ & $^{+ 0.8  } _{- 0.5  } $ & $^{-0.5  }  _{+ 0.8  } $ & $^{}_{+  11. } $ & $^{-1.9  }  _{+ 5.0  } $ & $^{+ 2.8  } _{-0.3  }  $ & $^{+ 0.2  } _{-0.9  }  $ & $^{-0.7  }  _{+ 1.6  } 
$ \\
$43100$ & $
   5.3\cdot 10^{-6}$ &
$^{+ 230. } _{-  80. } $ & $^{+  24. } _{- 3.9  } $ & $^{+ 1.1  } _{- 1.4  } $ & $^{-1.1  }  _{+ 1.4  } $ & $^{}_{+  23. } $ & $^{-2.5  }  _{+ 5.5  } $ & $^{+ 0.3  } _{+ 0.0  } $ & $^{+ 0.2  } _{-1.0  }  $ & $^{-2.3  }  _{+ 5.3  } 
$ \\
\hline
\end{tabular}
}

  \end{center}    
  \setlength{\localtextwidth}{16.4cm} 
  \caption[this space for rent]{
    Systematic uncertainties with bin-to-bin correlations
    for the differential cross-section \sigqq. The left part of the table
    contains the quoted $Q^2$ value, \qqc, the measured cross section
    \sigqq\ corrected to the Born level, the statistical error and
    the total systematic uncertainty. The right part of the table
    lists the total uncorrelated systematic uncertainty followed by
    the uncertainties $\delta_1$--\,$\delta_6$ (see text) with bin-to-bin
    correlations. For the latter, the upper (lower) numbers refer to 
    positive (negative) variation of e.g. the cut value, whereas the
    signs of the numbers reflect the direction of change in the cross
    sections. 
    }
  \label{tab-dsdQ2_c}
\end{table}

\begin{table} [!ht] 
\begin{center}
  {\footnotesize
\renewcommand{\arraystretch}{1.2}
\begin{tabular}{|c|r@{ -- }l@{$\,$}r|l@{$\,$}r|l@{}l@{$\,$}l@{$\,$}l@{$\,$}l|l|r|r|c|}
\hline
{$Q^2$ cut} & 
\multicolumn{3}{c|}{ $x$ range} & 
\multicolumn{2}{c|}{$x_c$} & 
\multicolumn{6}{c|} {$d\sigma / dx \ $(pb)} &
{$N_{\rm obs}$} & 
{$N_{\rm bg}$} & 
{$\cal A$} \\
\cline{7-12}
{($\Gev^2$)} & 
\multicolumn{3}{c|}{} & 
\multicolumn{2}{c|}{} & 
\multicolumn{5}{c|}{measured} &
\multicolumn{1}{c|}{SM} &&& \\
\hline
\hline
$200$ & $\left( \right.\!0.63$ & $1.00$ & $ \left. \! \right) \cdot 10^{-2}$ & $ 0.794$ & $\,\cdot\,10^{-2}$ &$ \left( \right.$ \hspace{-1mm} & $  8.08$ & $\pm  0.15$ & $^{+ 0.12}_{- 0.12}$ & $ \left. \! \right) \cdot 10^{ 4}
$ & $  8.02\cdot 10^{ 4}$ &
$ 3834$ & $4.3$ & $0.81$ \\
      & $\left( \right.\!0.10$ & $0.16$ & $ \left. \! \right) \cdot 10^{-1}$ & $ 0.126$ & $\,\cdot\,10^{-1}$ &$ \left( \right.$ \hspace{-1mm} & $  5.63$ & $\pm  0.10$ & $^{+ 0.13}_{- 0.12}$ & $ \left. \! \right) \cdot 10^{ 4}
$ & $  5.40\cdot 10^{ 4}$ &
$ 4545$ & $3.2$ & $0.83$ \\
      & $\left( \right.\!0.16$ & $0.25$ & $ \left. \! \right) \cdot 10^{-1}$ & $ 0.200$ & $\,\cdot\,10^{-1}$ &$ \left( \right.$ \hspace{-1mm} & $  3.42$ & $\pm  0.06$ & $^{+ 0.03}_{- 0.04}$ & $ \left. \! \right) \cdot 10^{ 4}
$ & $  3.38\cdot 10^{ 4}$ &
$ 4396$ & $1.9$ & $0.86$ \\
      & $\left( \right.\!0.25$ & $0.40$ & $ \left. \! \right) \cdot 10^{-1}$ & $ 0.316$ & $\,\cdot\,10^{-1}$ &$ \left( \right.$ \hspace{-1mm} & $  2.03$ & $\pm  0.03$ & $^{+ 0.02}_{- 0.03}$ & $ \left. \! \right) \cdot 10^{ 4}
$ & $  2.02\cdot 10^{ 4}$ &
$ 4472$ & $1.7$ & $0.88$ \\
      & $\left( \right.\!0.40$ & $0.63$ & $ \left. \! \right) \cdot 10^{-1}$ & $ 0.501$ & $\,\cdot\,10^{-1}$ &$ \left( \right.$ \hspace{-1mm} & $  1.16$ & $\pm  0.02$ & $^{+ 0.02}_{- 0.02}$ & $ \left. \! \right) \cdot 10^{ 4}
$ & $  1.17\cdot 10^{ 4}$ &
$ 3999$ & $1.1$ & $0.89$ \\
      & $\left( \right.\!0.63$ & $1.00$ & $ \left. \! \right) \cdot 10^{-1}$ & $ 0.794$ & $\,\cdot\,10^{-1}$ &$ \left( \right.$ \hspace{-1mm} & $  6.41$ & $\pm  0.12$ & $^{+ 0.10}_{- 0.19}$ & $ \left. \! \right) \cdot 10^{ 3}
$ & $  6.65\cdot 10^{ 3}$ &
$ 3724$ & $0.3$ & $0.93$ \\
      & $0.10$ & $0.16$ &  & $ 0.126$ &  &$ \left( \right.$ \hspace{-1mm} & $  3.54$ & $\pm  0.07$ & $^{+ 0.05}_{- 0.08}$ & $ \left. \! \right) \cdot 10^{ 3}
$ & $  3.66\cdot 10^{ 3}$ &
$ 3468$ & $0.0$ & $0.98$ \\
      & $0.16$ & $0.25$ &  & $ 0.200$ &  &$ \left( \right.$ \hspace{-1mm} & $  1.91$ & $\pm  0.05$ & $^{+ 0.04}_{- 0.04}$ & $ \left. \! \right) \cdot 10^{ 3}
$ & $  1.89\cdot 10^{ 3}$ &
$ 2218$ & $0.0$ & $0.76$ \\
      & $0.25$ & $0.40$ &  & $ 0.316$ &  &$ \left( \right.$ \hspace{-1mm} & $  8.23$ & $\pm  0.39$ & $^{+ 0.25}_{- 0.24}$ & $ \left. \! \right) \cdot 10^{ 2}
$ & $  8.15\cdot 10^{ 2}$ &
$  550$ & $0.0$ & $0.27$ \\
\hline
$10\,000$ & $0.10$ & $0.16$ &  & $ 0.126$ &  && $  4.56$ & $^{+  4.41}_{- 2.44}$ & $^{+ 0.69}_{- 0.28}$ & $
$ & $ 13.20$ &
$    3$ & $0.0$ & $0.87$ \\
          & $0.16$ & $0.25$ &  & $ 0.200$ &  &$ \left( \right.$ \hspace{-1mm} & $  2.07$ & $^{+  0.49}_{- 0.40}$ & $^{+ 0.12}_{- 0.06}$ & $ \left. \! \right) \cdot 10^{ 1}
$ & $  1.66\cdot 10^{ 1}$ &
$   26$ & $0.0$ & $0.90$ \\
          & $0.25$ & $0.40$ &  & $ 0.316$ &  &$ \left( \right.$ \hspace{-1mm} & $  1.12$ & $^{+  0.27}_{- 0.22}$ & $^{+ 0.04}_{- 0.02}$ & $ \left. \! \right) \cdot 10^{ 1}
$ & $  0.96\cdot 10^{ 1}$ &
$   25$ & $0.0$ & $0.92$ \\
          & $0.40$ & $0.63$ &  & $ 0.501$ &  && $  2.75$ & $^{+  1.14}_{- 0.83}$ & $^{+ 0.32}_{- 0.04}$ & $
$ & $  2.53$ &
$   10$ & $0.0$ & $0.93$ \\
\hline
\end{tabular}
}

\end{center}    
  \setlength{\localtextwidth}{17.5cm} 
  \caption[this space for rent]{
    The differential cross-section \sigx\ for the reaction $e^{-} p \rightarrow
    e^{-} X$. The following quantities are given for each bin: the
    lower $Q^2$ cut, the $x$ range, 
    the value at which the cross section is quoted, $x_c$,  the measured
    \sigx\ corrected to the Born level and the corresponding cross
    section predicted by the SM using CTEQ5D PDFs. The
    first error of the measured cross section gives
    the statistical error and the second is the systematic uncertainty. The last three
    columns contain the number of observed events in data, $N_{\rm obs}$, the
    number of expected background events, $N_{\rm bg}$ and the acceptance, $\cal A$.
  }
  \label{tab-dsdx}
\end{table}

\begin{table} [!ht] 
  \begin{center}
    \hspace*{-5.5mm}{\footnotesize
\renewcommand{\arraystretch}{1.2}
\begin{tabular}{|c|l|l|c|c||c|c|c|c|c|c|c|}
\hline
{$Q^2$ cut} & 
\multicolumn{1}{c|}{$x_c$} & 
\multicolumn{1}{c|}{$d\sigma / dx$} &
stat.  & 
total sys.  & 
uncor. sys. &
$\delta_1$ &
$\delta_2$ &
$\delta_3$ &
$\delta_4$ &
$\delta_5$ &
$\delta_6$ \\
{($\Gev^2$)} & 
$$ &
\multicolumn{1}{c|}{(pb)} &
(\%) & 
(\%) &
(\%) &
(\%) &
(\%) &
(\%) &
(\%) &
(\%) &
(\%) \\
\hline \hline
$200$ & $0.794\,\cdot\,10^{-2}$ & $
  8.08\cdot 10^{ 4}$ &
$^{+ 1.9  } _{- 1.8  } $ & $^{+ 1.4  } _{- 1.5  } $ & $^{+ 1.2  } _{- 0.7  } $ & $^{-0.2  }  _{+ 0.4  } $ & $^{}_{+ 0.1  } $ & $^{+ 0.1  } _{+ 0.0  } $ & $^{-1.1  }  _{+ 0.5  } $ & $^{+ 0.1  } _{-0.9  }  $ & $^{-0.3  }  _{+ 0.8  } 
$ \\
      & $0.126\,\cdot\,10^{-1}$ & $
  5.63\cdot 10^{ 4}$ &
$^{+ 1.7  } _{- 1.7  } $ & $^{+ 2.3  } _{- 2.1  } $ & $^{+ 1.1  } _{- 1.1  } $ & $^{-0.1  }  _{+ 0.1  } $ & $^{}_{-0.3  }  $ & $^{- 0.0  } _{- 0.0  } $ & $^{-1.4  }  _{+ 0.8  } $ & $^{+ 0.2  } _{-1.0  }  $ & $^{-0.8  }  _{+ 1.9  } 
$ \\
      & $0.200\,\cdot\,10^{-1}$ & $
  3.42\cdot 10^{ 4}$ &
$^{+ 1.7  } _{- 1.7  } $ & $^{+ 1.0  } _{- 1.2  } $ & $^{+ 1.1  } _{- 0.8  } $ & $^{- 0.0  } _{+ 0.3  } $ & $^{}_{- 0.0  } $ & $^{+ 0.1  } _{- 0.0  } $ & $^{-0.5  }  _{+ 0.2  } $ & $^{+ 0.3  } _{-0.9  }  $ & $^{+ 0.1  } _{-0.1  }  
$ \\
      & $0.316\,\cdot\,10^{-1}$ & $
  2.03\cdot 10^{ 4}$ &
$^{+ 1.7  } _{- 1.7  } $ & $^{+ 0.8  } _{- 1.7  } $ & $^{+ 0.9  } _{- 1.9  } $ & $^{-0.1  }  _{+ 0.2  } $ & $^{}_{-0.1  }  $ & $^{-0.1  }  _{+ 0.2  } $ & $^{-0.5  }  _{+ 0.4  } $ & $^{+ 0.2  } _{-1.0  }  $ & $^{+ 0.2  } _{-0.6  }  
$ \\
      & $0.501\,\cdot\,10^{-1}$ & $
  1.16\cdot 10^{ 4}$ &
$^{+ 1.8  } _{- 1.8  } $ & $^{+ 1.4  } _{- 1.6  } $ & $^{+ 1.5  } _{- 1.3  } $ & $^{-0.1  }  _{+ 0.3  } $ & $^{}_{+ 0.0  } $ & $^{+ 0.1  } _{+ 0.1  } $ & $^{-0.2  }  _{+ 0.1  } $ & $^{+ 0.3  } _{-0.9  }  $ & $^{+ 0.5  } _{-1.2  }  
$ \\
      & $0.794\,\cdot\,10^{-1}$ & $
  6.41\cdot 10^{ 3}$ &
$^{+ 1.9  } _{- 1.8  } $ & $^{+ 1.5  } _{- 2.9  } $ & $^{+ 1.3  } _{- 1.1  } $ & $^{-0.1  }  _{+ 0.3  } $ & $^{}_{- 0.0  } $ & $^{+ 0.1  } _{+ 0.1  } $ & $^{-0.8  }  _{-0.1  }  $ & $^{+ 0.2  } _{-1.0  }  $ & $^{+ 1.0  } _{-2.4  }  
$ \\
      & $0.126$ & $
  3.54\cdot 10^{ 3}$ &
$^{+ 1.9  } _{- 1.9  } $ & $^{+ 1.3  } _{- 2.2  } $ & $^{+ 1.6  } _{- 0.8  } $ & $^{-0.1  }  _{+ 0.1  } $ & $^{}_{+ 0.0  } $ & $^{- 0.0  } _{- 0.0  } $ & $^{-0.7  }  _{-0.4  }  $ & $^{+ 0.2  } _{-1.0  }  $ & $^{+ 0.7  } _{-1.7  }  
$ \\
      & $0.200$ & $
  1.91\cdot 10^{ 3}$ &
$^{+ 2.4  } _{- 2.4  } $ & $^{+ 1.9  } _{- 2.1  } $ & $^{+ 2.1  } _{- 1.2  } $ & $^{+ 0.0  } _{+ 0.1  } $ & $^{}_{+ 0.1  } $ & $^{+ 0.1  } _{+ 0.1  } $ & $^{-1.1  }  _{+ 0.2  } $ & $^{+ 0.4  } _{-0.9  }  $ & $^{+ 0.4  } _{-0.9  }  
$ \\
      & $0.316$ & $
  8.23\cdot 10^{ 2}$ &
$^{+ 4.7  } _{- 4.5  } $ & $^{+ 3.0  } _{- 2.9  } $ & $^{+ 2.9  } _{- 2.8  } $ & $^{+ 0.1  } _{+ 0.5  } $ & $^{}_{+ 0.4  } $ & $^{+ 0.5  } _{+ 0.2  } $ & $^{+ 0.8  } _{-0.1  }  $ & $^{+ 0.6  } _{-0.5  }  $ & $^{+ 0.3  } _{-0.7  }  
$ \\
\hline
$10\,000$ & $0.126$ & $
  4.56$ &
$^{+  97. } _{-  53. } $ & $^{+  15. } _{- 6.1  } $ & $^{+ 4.3  } _{- 6.5  } $ & $^{-1.9  }  _{+ 2.5  } $ & $^{}_{+  13. } $ & $^{-1.4  }  _{+ 5.0  } $ & $^{+ 3.1  } _{-0.4  }  $ & $^{+ 0.3  } _{-1.1  }  $ & $^{+ 2.3  } _{-5.3  }  
$ \\
          & $0.200$ & $
  2.07\cdot 10^{ 1}$ &
$^{+  24. } _{-  19. } $ & $^{+ 5.7  } _{- 2.7  } $ & $^{+ 3.1  } _{- 1.3  } $ & $^{-0.4  }  _{+ 0.4  } $ & $^{}_{+ 2.5  } $ & $^{-1.1  }  _{+ 3.5  } $ & $^{-2.1  }  _{+ 3.6  } $ & $^{+ 0.2  } _{-1.0  }  $ & $^{-0.2  }  _{+ 0.6  } 
$ \\
          & $0.316$ & $
  1.12\cdot 10^{ 1}$ &
$^{+  24. } _{-  20. } $ & $^{+ 3.9  } _{- 1.8  } $ & $^{+ 1.6  } _{- 1.4  } $ & $^{-0.3  }  _{+ 0.2  } $ & $^{}_{+ 2.0  } $ & $^{-0.7  }  _{+ 2.3  } $ & $^{+ 1.4  } _{-0.4  }  $ & $^{+ 0.3  } _{-1.0  }  $ & $^{-0.8  }  _{+ 1.8  } 
$ \\
          & $0.501$ & $
  2.75$ &
$^{+  42. } _{-  30. } $ & $^{+  12. } _{- 1.5  } $ & $^{+  11. } _{- 2.6  } $ & $^{-0.1  }  _{+ 0.9  } $ & $^{}_{+ 1.4  } $ & $^{-0.3  }  _{+ 2.1  } $ & $^{+ 1.6  } _{+ 0.1  } $ & $^{+ 0.5  } _{-0.5  }  $ & $^{-1.3  }  _{+ 3.2  } 
$ \\
\hline
\end{tabular}
}

  \end{center}    
  \setlength{\localtextwidth}{18.2cm} 
  \caption[this space for rent]{
    Systematic uncertainties with bin-to-bin correlations
    for the differential cross-section \sigx. The left part of the table
    contains the lower $Q^2$ cut, the quoted $x$ value, $x_c$, the
    measured cross-section 
    \sigx\ corrected to the Born level, the statistical error and
    the total systematic uncertainty. The right part of the table
    lists the total uncorrelated systematic uncertainty followed by
    the uncertainties $\delta_1$--\,$\delta_6$ (see text) with bin-to-bin
    correlations. For the latter, the upper (lower) numbers refer to 
    positive (negative) variation of e.g. the cut value, whereas the
    signs of the numbers reflect the direction of change in the cross
    sections. 
    }
  \label{tab-dsdx_c}
\end{table}

\begin{table} [!ht] 
\begin{center}
  {\footnotesize
\renewcommand{\arraystretch}{1.2}
\begin{tabular}{|c|r@{ -- }r|r|l@{}l@{$\,$}l@{$\,$}l@{$\,$}l|l|r|r|c|}
\hline
{$Q^2$ cut} & 
\multicolumn{2}{c|}{ $y$ range} & 
\multicolumn{1}{c|}{$y_c$} & 
\multicolumn{6}{c|} {$d\sigma / dy \ $ (pb)} &
{$N_{\rm obs}$} & 
{$N_{\rm bg}$} & 
{$\cal A$} \\
\cline{5-10}
{($\Gev^2$)} & 
\multicolumn{2}{c|}{} & 
\multicolumn{1}{c|}{} & 
\multicolumn{5}{c|}{measured} &
\multicolumn{1}{c|}{SM} &&& \\
\hline
\hline
$200$ & $0.05$ & $0.10$ & $0.075$ &$ \left( \right.$ \hspace{-1mm} & $  7.45$ & $\pm  0.11$ & $^{+ 0.08}_{- 0.11}$ & $ \left. \! \right) \cdot 10^{ 3}
$ & $  7.50\cdot 10^{ 3}$ &
$ 5709$ & $0.1$ & $0.89$ \\
      & $0.10$ & $0.15$ & $0.125$ &$ \left( \right.$ \hspace{-1mm} & $  5.26$ & $\pm  0.10$ & $^{+ 0.03}_{- 0.08}$ & $ \left. \! \right) \cdot 10^{ 3}
$ & $  5.21\cdot 10^{ 3}$ &
$ 3879$ & $0.1$ & $0.86$ \\
      & $0.15$ & $0.20$ & $0.175$ &$ \left( \right.$ \hspace{-1mm} & $  4.17$ & $\pm  0.09$ & $^{+ 0.07}_{- 0.06}$ & $ \left. \! \right) \cdot 10^{ 3}
$ & $  4.02\cdot 10^{ 3}$ &
$ 2947$ & $0.2$ & $0.84$ \\
      & $0.20$ & $0.25$ & $0.225$ &$ \left( \right.$ \hspace{-1mm} & $  3.33$ & $\pm  0.08$ & $^{+ 0.09}_{- 0.06}$ & $ \left. \! \right) \cdot 10^{ 3}
$ & $  3.28\cdot 10^{ 3}$ &
$ 2327$ & $0.1$ & $0.83$ \\
      & $0.25$ & $0.30$ & $0.275$ &$ \left( \right.$ \hspace{-1mm} & $  2.64$ & $\pm  0.07$ & $^{+ 0.03}_{- 0.05}$ & $ \left. \! \right) \cdot 10^{ 3}
$ & $  2.75\cdot 10^{ 3}$ &
$ 1850$ & $0.0$ & $0.83$ \\
      & $0.30$ & $0.35$ & $0.325$ &$ \left( \right.$ \hspace{-1mm} & $  2.45$ & $\pm  0.07$ & $^{+ 0.02}_{- 0.06}$ & $ \left. \! \right) \cdot 10^{ 3}
$ & $  2.36\cdot 10^{ 3}$ &
$ 1663$ & $0.6$ & $0.82$ \\
      & $0.35$ & $0.40$ & $0.375$ &$ \left( \right.$ \hspace{-1mm} & $  2.11$ & $\pm  0.06$ & $^{+ 0.03}_{- 0.04}$ & $ \left. \! \right) \cdot 10^{ 3}
$ & $  2.06\cdot 10^{ 3}$ &
$ 1453$ & $0.5$ & $0.84$ \\
      & $0.40$ & $0.45$ & $0.425$ &$ \left( \right.$ \hspace{-1mm} & $  1.80$ & $\pm  0.06$ & $^{+ 0.02}_{- 0.03}$ & $ \left. \! \right) \cdot 10^{ 3}
$ & $  1.82\cdot 10^{ 3}$ &
$ 1212$ & $0.8$ & $0.81$ \\
      & $0.45$ & $0.50$ & $0.475$ &$ \left( \right.$ \hspace{-1mm} & $  1.68$ & $\pm  0.06$ & $^{+ 0.05}_{- 0.02}$ & $ \left. \! \right) \cdot 10^{ 3}
$ & $  1.62\cdot 10^{ 3}$ &
$ 1134$ & $0.9$ & $0.81$ \\
      & $0.50$ & $0.55$ & $0.525$ &$ \left( \right.$ \hspace{-1mm} & $  1.44$ & $\pm  0.05$ & $^{+ 0.02}_{- 0.04}$ & $ \left. \! \right) \cdot 10^{ 3}
$ & $  1.45\cdot 10^{ 3}$ &
$  933$ & $1.6$ & $0.77$ \\
      & $0.55$ & $0.60$ & $0.575$ &$ \left( \right.$ \hspace{-1mm} & $  1.35$ & $\pm  0.05$ & $^{+ 0.03}_{- 0.02}$ & $ \left. \! \right) \cdot 10^{ 3}
$ & $  1.31\cdot 10^{ 3}$ &
$  849$ & $1.6$ & $0.76$ \\
      & $0.60$ & $0.65$ & $0.625$ &$ \left( \right.$ \hspace{-1mm} & $  1.23$ & $\pm  0.05$ & $^{+ 0.03}_{- 0.03}$ & $ \left. \! \right) \cdot 10^{ 3}
$ & $  1.20\cdot 10^{ 3}$ &
$  695$ & $1.3$ & $0.69$ \\
      & $0.65$ & $0.70$ & $0.675$ &$ \left( \right.$ \hspace{-1mm} & $  1.22$ & $\pm  0.05$ & $^{+ 0.07}_{- 0.03}$ & $ \left. \! \right) \cdot 10^{ 3}
$ & $  1.10\cdot 10^{ 3}$ &
$  637$ & $1.1$ & $0.63$ \\
      & $0.70$ & $0.75$ & $0.725$ &$ \left( \right.$ \hspace{-1mm} & $  9.52$ & $\pm  0.50$ & $^{+ 0.22}_{- 0.56}$ & $ \left. \! \right) \cdot 10^{ 2}
$ & $ 10.08\cdot 10^{ 2}$ &
$  454$ & $1.7$ & $0.59$ \\
\hline
\end{tabular}
}

\end{center}    
  \caption[this space for rent]{
    The differential cross-section \sigy\ for the reaction $e^{-} p \rightarrow
    e^{-} X$. The following quantities are given for each bin: the lower $Q^2$ cut, the $y$ range,
    the value at which the cross section is quoted, $y_c$, the measured
    cross-section \sigy\ corrected to the  Born level and the corresponding cross
    section predicted by the SM using CTEQ5D fit PDFs. The
    first error of the measured cross section gives
    the statistical error and the second is the systematic uncertainty. The last three
    columns contain the number of observed events in data, $N_{\rm obs}$, the
    number of expected background events, $N_{\rm bg}$ and the acceptance, $\cal
    A$.
  }
\label{tab-dsdy}
\end{table}

\begin{table} [!ht] 
  \begin{center}
    {\footnotesize
\renewcommand{\arraystretch}{1.2}
\begin{tabular}{|c|l|c|c|c||c|c|c|c|c|c|c|}
\hline
{$Q^2$ cut} & 
\multicolumn{1}{c|}{$y_c$} & 
$d\sigma / dy$ &
stat.  & 
total sys.  & 
uncor. sys. &
$\delta_1$ &
$\delta_2$ &
$\delta_3$ &
$\delta_4$ &
$\delta_5$ &
$\delta_6$ \\
{($\Gev^2$)} & 
$$ &
(pb) &
(\%) & 
(\%) &
(\%) &
(\%) &
(\%) &
(\%) &
(\%) &
(\%) &
(\%) \\
\hline \hline
$200$ & $0.075$ & $
  7.45\cdot 10^{ 3}$ &
$^{+ 1.5  } _{- 1.5  } $ & $^{+ 1.0  } _{- 1.5  } $ & $^{+ 4.3  } _{- 6.5  } $ & $^{-0.1  }  _{+ 0.3  } $ & $^{}_{+ 0.1  } $ & $^{+ 0.1  } _{+ 0.2  } $ & $^{-0.3  }  _{+ 0.3  } $ & $^{+ 0.3  } _{-0.9  }  $ & $^{+ 0.4  } _{-1.0  }  
$ \\
      & $0.125$ & $
  5.26\cdot 10^{ 3}$ &
$^{+ 1.8  } _{- 1.8  } $ & $^{+ 0.6  } _{- 1.6  } $ & $^{+ 3.1  } _{- 1.5  } $ & $^{-0.1  }  _{+ 0.1  } $ & $^{}_{- 0.0  } $ & $^{-0.1  }  _{- 0.0  } $ & $^{-0.9  }  _{-0.1  }  $ & $^{+ 0.1  } _{-1.0  }  $ & $^{+ 0.2  } _{-0.5  }  
$ \\
      & $0.175$ & $
  4.17\cdot 10^{ 3}$ &
$^{+ 2.1  } _{- 2.1  } $ & $^{+ 1.8  } _{- 1.5  } $ & $^{+ 1.9  } _{- 1.5  } $ & $^{-0.2  }  _{+ 0.1  } $ & $^{}_{-0.1  }  $ & $^{-0.1  }  _{+ 0.0  } $ & $^{-0.7  }  _{+ 0.1  } $ & $^{+ 0.3  } _{-1.0  }  $ & $^{-0.6  }  _{+ 1.5  } 
$ \\
      & $0.225$ & $
  3.33\cdot 10^{ 3}$ &
$^{+ 2.4  } _{- 2.3  } $ & $^{+ 2.6  } _{- 1.8  } $ & $^{+  11. } _{- 2.7  } $ & $^{-0.1  }  _{+ 0.1  } $ & $^{}_{+ 0.0  } $ & $^{+ 0.1  } _{- 0.0  } $ & $^{-0.7  }  _{+ 0.5  } $ & $^{+ 0.3  } _{-1.0  }  $ & $^{-1.0  }  _{+ 2.5  } 
$ \\
      & $0.275$ & $
  2.64\cdot 10^{ 3}$ &
$^{+ 2.6  } _{- 2.6  } $ & $^{+ 1.1  } _{- 1.7  } $ & $^{+ 2.6  } _{- 2.0  } $ & $^{-0.2  }  _{+ 0.1  } $ & $^{}_{-0.1  }  $ & $^{+ 0.0  } _{+ 0.0  } $ & $^{-0.3  }  _{+ 0.4  } $ & $^{+ 0.0  } _{-1.0  }  $ & $^{-0.1  }  _{+ 0.1  } 
$ \\
      & $0.325$ & $
  2.45\cdot 10^{ 3}$ &
$^{+ 2.8  } _{- 2.8  } $ & $^{+ 1.0  } _{- 2.4  } $ & $^{+ 2.0  } _{- 2.0  } $ & $^{-0.3  }  _{+ 0.0  } $ & $^{}_{-0.1  }  $ & $^{-0.2  }  _{-0.1  }  $ & $^{-1.7  }  _{+ 0.4  } $ & $^{+ 0.1  } _{-1.1  }  $ & $^{+ 0.0  } _{+ 0.1  } 
$ \\
      & $0.375$ & $
  2.11\cdot 10^{ 3}$ &
$^{+ 3.0  } _{- 2.9  } $ & $^{+ 1.5  } _{- 1.8  } $ & $^{+ 3.5  } _{- 5.2  } $ & $^{-0.2  }  _{+ 0.2  } $ & $^{}_{- 0.0  } $ & $^{- 0.0  } _{-0.1  }  $ & $^{-1.3  }  _{+ 0.4  } $ & $^{+ 0.2  } _{-0.8  }  $ & $^{-0.4  }  _{+ 1.1  } 
$ \\
      & $0.425$ & $
  1.80\cdot 10^{ 3}$ &
$^{+ 3.3  } _{- 3.2  } $ & $^{+ 0.9  } _{- 1.9  } $ & $^{+ 2.5  } _{- 2.0  } $ & $^{-0.2  }  _{+ 0.1  } $ & $^{}_{- 0.0  } $ & $^{-0.1  }  _{-0.1  }  $ & $^{-0.8  }  _{-0.1  }  $ & $^{+ 0.2  } _{-1.0  }  $ & $^{-0.1  }  _{+ 0.3  } 
$ \\
      & $0.475$ & $
  1.68\cdot 10^{ 3}$ &
$^{+ 3.4  } _{- 3.3  } $ & $^{+ 2.7  } _{- 1.4  } $ & $^{+ 3.6  } _{- 2.9  } $ & $^{+ 0.0  } _{+ 0.3  } $ & $^{}_{+ 0.1  } $ & $^{+ 0.1  } _{+ 0.3  } $ & $^{-0.8  }  _{+ 0.7  } $ & $^{+ 0.5  } _{-1.0  }  $ & $^{-0.6  }  _{+ 1.4  } 
$ \\
      & $0.525$ & $
  1.44\cdot 10^{ 3}$ &
$^{+ 3.7  } _{- 3.6  } $ & $^{+ 1.3  } _{- 2.7  } $ & $^{+ 1.1  } _{- 2.3  } $ & $^{-0.3  }  _{+ 0.4  } $ & $^{}_{+ 0.1  } $ & $^{+ 0.0  } _{-0.3  }  $ & $^{-1.1  }  _{+ 0.4  } $ & $^{+ 0.1  } _{-1.0  }  $ & $^{-0.5  }  _{+ 1.0  } 
$ \\
      & $0.575$ & $
  1.35\cdot 10^{ 3}$ &
$^{+ 3.9  } _{- 3.8  } $ & $^{+ 2.2  } _{- 1.4  } $ & $^{+ 2.4  } _{- 0.6  } $ & $^{-0.3  }  _{+ 0.7  } $ & $^{}_{+ 0.2  } $ & $^{+ 0.2  } _{-0.3  }  $ & $^{+ 0.5  } _{+ 0.6  } $ & $^{+ 0.2  } _{-0.9  }  $ & $^{+ 0.5  } _{-0.9  }  
$ \\
      & $0.625$ & $
  1.23\cdot 10^{ 3}$ &
$^{+ 4.3  } _{- 4.2  } $ & $^{+ 2.8  } _{- 2.2  } $ & $^{+ 1.8  } _{- 1.4  } $ & $^{-1.1  }  _{+ 1.2  } $ & $^{}_{+ 0.2  } $ & $^{+ 0.3  } _{+ 0.5  } $ & $^{+ 1.8  } _{+ 0.7  } $ & $^{+ 0.4  } _{-1.2  }  $ & $^{-0.3  }  _{+ 0.8  } 
$ \\
      & $0.675$ & $
  1.22\cdot 10^{ 3}$ &
$^{+ 4.5  } _{- 4.4  } $ & $^{+ 5.6  } _{- 2.4  } $ & $^{+ 4.2  } _{- 1.2  } $ & $^{-1.2  }  _{+ 2.2  } $ & $^{}_{+ 0.4  } $ & $^{-0.1  }  _{+ 1.3  } $ & $^{-1.2  }  _{+ 1.5  } $ & $^{+ 0.6  } _{-0.6  }  $ & $^{-1.1  }  _{+ 2.5  } 
$ \\
      & $0.725$ & $
  9.52\cdot 10^{ 2}$ &
$^{+ 5.3  } _{- 5.1  } $ & $^{+ 2.3  } _{- 5.9  } $ & $^{+ 1.5  } _{- 4.7  } $ & $^{-3.0  }  _{+ 1.3  } $ & $^{}_{-0.8  }  $ & $^{- 0.0  } _{-0.2  }  $ & $^{-0.7  }  _{+ 1.2  } $ & $^{-0.8  }  _{-1.6  }  $ & $^{-0.5  }  _{+ 1.1  } 
$ \\
\hline
\end{tabular}
}

  \end{center} 
  \setlength{\localtextwidth}{17.2cm} 
  \caption[this space for rent]{
    Systematic uncertainties with bin-to-bin correlations
    for the differential cross-section \sigy. The left part of the table
    contains the lower $Q^2$ cut, the quoted $y$ value, $y_c$, the
    measured cross-section 
    \sigy\ corrected to the Born level, the statistical error and
    the total systematic uncertainty. The right part of the table
    lists the total uncorrelated systematic uncertainty followed by
    the uncertainties $\delta_1$--\,$\delta_6$ (see text) with bin-to-bin
    correlations. For the latter, the upper (lower) numbers refer to 
    positive (negative) variation of e.g. the cut value, whereas the
    signs of the numbers reflect the direction of change in the cross sections.
    }
  \label{tab-dsdy_c}
\end{table}

\begin{sidewaystable} 
\begin{center}
  {\small
\renewcommand{\arraystretch}{1.2}
\begin{tabular}{|r@{ -- }l|r@{ -- }l@{$\,$}r|r|l|l@{}l@{$\,$}l@{$\,$}l@{$\,$}l|l|r|r|c|}
\hline
\multicolumn{2}{|c|}{ $Q^2$ range} & 
\multicolumn{3}{c|}{ $x$ range} & 
\multicolumn{1}{c|}{$Q^2_c$} & 
\multicolumn{1}{c|}{$x_c$} &
\multicolumn{6}{c|}{$\tilde{\sigma}(e^-p) $} &
{$N_{\rm obs}$} & 
{$N_{\rm bg}$} & 
{$\cal A$} \\
\cline{8-13}
\multicolumn{2}{|c|} {($\Gev^2$)} &
\multicolumn{3}{c|}{} & 
{($\Gev^2$)} && 
\multicolumn{5}{c|}{measured} & 
\multicolumn{1}{c|}{SM} &&& \\ 
\hline
\hline $  185.$ & $  240.$ & $\left( \right.\!0.37$ & $0.60$ & $ \left. \! \right) \cdot 10^{-2}$ & $  200$ & $0.50$$\,\cdot\,10^{-2}$ & $ $ & $ 1.165$ & $\pm 0.033$ & $^{+0.034}_{-0.023}$ & $$ & $ 1.105$ &
$ 1743$ & $1.0$ & $0.90$ \\
\multicolumn{2}{|c|}{} & $\left( \right.\!0.60$ & $1.00$ & $ \left. \! \right) \cdot 10^{-2}$ &  & $0.80$$\,\cdot\,10^{-2}$ & $\left( \right.\! $ \hspace{-0.5mm} & $  9.60$ & $\pm  0.26$ & $^{+ 0.20}_{- 0.16}$ & $ \left. \!\!\!\! \right) \cdot 10^{-1}$ & $  9.44\cdot 10^{-1}$ &
$ 1823$ & $0.1$ & $0.91$ \\
\multicolumn{2}{|c|}{} & $\left( \right.\!0.10$ & $0.17$ & $ \left. \! \right) \cdot 10^{-1}$ &  & $0.13$$\,\cdot\,10^{-1}$ & $\left( \right.\! $ \hspace{-0.5mm} & $  8.18$ & $\pm  0.23$ & $^{+ 0.18}_{- 0.15}$ & $ \left. \!\!\!\! \right) \cdot 10^{-1}$ & $  7.99\cdot 10^{-1}$ &
$ 1791$ & $0.0$ & $0.90$ \\
\multicolumn{2}{|c|}{} & $\left( \right.\!0.17$ & $0.25$ & $ \left. \! \right) \cdot 10^{-1}$ &  & $0.21$$\,\cdot\,10^{-1}$ & $\left( \right.\! $ \hspace{-0.5mm} & $  7.06$ & $\pm  0.24$ & $^{+ 0.13}_{- 0.10}$ & $ \left. \!\!\!\! \right) \cdot 10^{-1}$ & $  6.78\cdot 10^{-1}$ &
$ 1238$ & $0.1$ & $0.92$ \\
\multicolumn{2}{|c|}{} & $\left( \right.\!0.25$ & $0.37$ & $ \left. \! \right) \cdot 10^{-1}$ &  & $0.32$$\,\cdot\,10^{-1}$ & $\left( \right.\! $ \hspace{-0.5mm} & $  5.90$ & $\pm  0.21$ & $^{+ 0.06}_{- 0.13}$ & $ \left. \!\!\!\! \right) \cdot 10^{-1}$ & $  5.90\cdot 10^{-1}$ &
$ 1143$ & $0.0$ & $0.96$ \\
\multicolumn{2}{|c|}{} & $\left( \right.\!0.37$ & $0.60$ & $ \left. \! \right) \cdot 10^{-1}$ &  & $0.50$$\,\cdot\,10^{-1}$ & $\left( \right.\! $ \hspace{-0.5mm} & $  5.02$ & $\pm  0.17$ & $^{+ 0.09}_{- 0.13}$ & $ \left. \!\!\!\! \right) \cdot 10^{-1}$ & $  5.13\cdot 10^{-1}$ &
$ 1277$ & $0.2$ & $0.98$ \\
\multicolumn{2}{|c|}{} & $\left( \right.\!0.60$ & $1.20$ & $ \left. \! \right) \cdot 10^{-1}$ &  & $0.80$$\,\cdot\,10^{-1}$ & $\left( \right.\! $ \hspace{-0.5mm} & $  4.23$ & $\pm  0.12$ & $^{+ 0.09}_{- 0.15}$ & $ \left. \!\!\!\! \right) \cdot 10^{-1}$ & $  4.42\cdot 10^{-1}$ &
$ 1659$ & $0.0$ & $1.09$ \\
\multicolumn{2}{|c|}{} & $0.12$ & $0.25$ & $$ &  & $0.18$$$ & $\left( \right.\! $ \hspace{-0.5mm} & $  3.22$ & $\pm  0.12$ & $^{+ 0.04}_{- 0.13}$ & $ \left. \!\!\!\! \right) \cdot 10^{-1}$ & $  3.28\cdot 10^{-1}$ &
$  948$ & $0.0$ & $0.75$ \\
\hline $  240.$ & $  310.$ & $\left( \right.\!0.37$ & $0.60$ & $ \left. \! \right) \cdot 10^{-2}$ & $  250$ & $0.50$$\,\cdot\,10^{-2}$ & $ $ & $  1.12$ & $\pm  0.05$ & $^{+ 0.03}_{- 0.03}$ & $$ & $  1.13$ &
$  751$ & $0.7$ & $0.59$ \\
\multicolumn{2}{|c|}{} & $\left( \right.\!0.60$ & $1.00$ & $ \left. \! \right) \cdot 10^{-2}$ &  & $0.80$$\,\cdot\,10^{-2}$ & $\left( \right.\! $ \hspace{-0.5mm} & $  9.88$ & $\pm  0.32$ & $^{+ 0.19}_{- 0.24}$ & $ \left. \!\!\!\! \right) \cdot 10^{-1}$ & $  9.64\cdot 10^{-1}$ &
$ 1327$ & $0.6$ & $0.90$ \\
\multicolumn{2}{|c|}{} & $\left( \right.\!0.10$ & $0.17$ & $ \left. \! \right) \cdot 10^{-1}$ &  & $0.13$$\,\cdot\,10^{-1}$ & $\left( \right.\! $ \hspace{-0.5mm} & $  8.64$ & $\pm  0.28$ & $^{+ 0.19}_{- 0.18}$ & $ \left. \!\!\!\! \right) \cdot 10^{-1}$ & $  8.16\cdot 10^{-1}$ &
$ 1354$ & $0.1$ & $0.89$ \\
\multicolumn{2}{|c|}{} & $\left( \right.\!0.17$ & $0.25$ & $ \left. \! \right) \cdot 10^{-1}$ &  & $0.21$$\,\cdot\,10^{-1}$ & $\left( \right.\! $ \hspace{-0.5mm} & $  6.95$ & $\pm  0.27$ & $^{+ 0.07}_{- 0.13}$ & $ \left. \!\!\!\! \right) \cdot 10^{-1}$ & $  6.89\cdot 10^{-1}$ &
$  905$ & $0.0$ & $0.91$ \\
\multicolumn{2}{|c|}{} & $\left( \right.\!0.25$ & $0.37$ & $ \left. \! \right) \cdot 10^{-1}$ &  & $0.32$$\,\cdot\,10^{-1}$ & $\left( \right.\! $ \hspace{-0.5mm} & $  6.11$ & $\pm  0.25$ & $^{+ 0.08}_{- 0.20}$ & $ \left. \!\!\!\! \right) \cdot 10^{-1}$ & $  5.97\cdot 10^{-1}$ &
$  877$ & $0.0$ & $0.95$ \\
\multicolumn{2}{|c|}{} & $\left( \right.\!0.37$ & $0.60$ & $ \left. \! \right) \cdot 10^{-1}$ &  & $0.50$$\,\cdot\,10^{-1}$ & $\left( \right.\! $ \hspace{-0.5mm} & $  5.24$ & $\pm  0.20$ & $^{+ 0.10}_{- 0.14}$ & $ \left. \!\!\!\! \right) \cdot 10^{-1}$ & $  5.17\cdot 10^{-1}$ &
$  960$ & $0.1$ & $0.95$ \\
\multicolumn{2}{|c|}{} & $\left( \right.\!0.60$ & $1.20$ & $ \left. \! \right) \cdot 10^{-1}$ &  & $0.80$$\,\cdot\,10^{-1}$ & $\left( \right.\! $ \hspace{-0.5mm} & $  4.36$ & $\pm  0.15$ & $^{+ 0.07}_{- 0.22}$ & $ \left. \!\!\!\! \right) \cdot 10^{-1}$ & $  4.44\cdot 10^{-1}$ &
$ 1204$ & $0.0$ & $1.00$ \\
\multicolumn{2}{|c|}{} & $0.12$ & $0.25$ & $$ &  & $0.18$$$ & $\left( \right.\! $ \hspace{-0.5mm} & $  2.98$ & $\pm  0.12$ & $^{+ 0.20}_{- 0.08}$ & $ \left. \!\!\!\! \right) \cdot 10^{-1}$ & $  3.26\cdot 10^{-1}$ &
$  817$ & $0.0$ & $0.90$ \\
\hline
\end{tabular}
}

\end{center}    
  \setlength{\localtextwidth}{21.3cm} 
  \caption[this space for rent]{
    The reduced cross-section $\tilde{\sigma}(e^- p)$ for the reaction
    $e^{-} p \rightarrow e^{-} X$.  The following quantities are given
    for each bin: the $Q^2$ and $x$ ranges, the values at which the
    cross section is quoted, \qqc\ and $x_c$, the measured reduced
    cross-section, $\tilde{\sigma}(e^- p)$, corrected to Born level
    and the corresponding cross section predicted by the SM using
    CTEQ5D PDFs. The first error of the measured cross section gives
    the statistical error and the second is the systematic
    uncertainty. The last three columns contain the number of observed
    events in data, $N_{\rm obs}$, the number of expected background
    events, $N_{\rm bg}$ and the acceptance, $\cal A$.  }
  \label{tab-dsdqdx_1}
\end{sidewaystable}
\begin{sidewaystable}
  \begin{center}
    {\small
\renewcommand{\arraystretch}{1.2}
\begin{tabular}{|r@{ -- }l|r@{ -- }l@{$\,$}r|r|l|l@{}l@{$\,$}l@{$\,$}l@{$\,$}l|l|r|r|c|}
\hline
\multicolumn{2}{|c|}{ $Q^2$ range} & 
\multicolumn{3}{c|}{ $x$ range} & 
\multicolumn{1}{c|}{$Q^2_c$} & 
\multicolumn{1}{c|}{$x_c$} &
\multicolumn{6}{c|}{$\tilde{\sigma}(e^-p) $} &
{$N_{\rm obs}$} & 
{$N_{\rm bg}$} & 
{$\cal A$} \\
\cline{8-13}
\multicolumn{2}{|c|} {($\Gev^2$)} &
\multicolumn{3}{c|}{} & 
{($\Gev^2$)} && 
\multicolumn{5}{c|}{measured} & 
\multicolumn{1}{c|}{SM} &&& \\ 
\hline
\hline $  310.$ & $  410.$ & $\left( \right.\!0.60$ & $1.00$ & $ \left. \! \right) \cdot 10^{-2}$ & $  350$ & $0.80$$\,\cdot\,10^{-2}$ & $ $ & $  1.00$ & $\pm  0.05$ & $^{+ 0.02}_{- 0.02}$ & $$ & $  0.99$ &
$  668$ & $1.3$ & $0.59$ \\
\multicolumn{2}{|c|}{} & $\left( \right.\!0.10$ & $0.17$ & $ \left. \! \right) \cdot 10^{-1}$ &  & $0.13$$\,\cdot\,10^{-1}$ & $\left( \right.\! $ \hspace{-0.5mm} & $  8.57$ & $\pm  0.32$ & $^{+ 0.27}_{- 0.19}$ & $ \left. \! \right) \cdot 10^{-1}$ & $  8.39\cdot 10^{-1}$ &
$  980$ & $0.1$ & $0.84$ \\
\multicolumn{2}{|c|}{} & $\left( \right.\!0.17$ & $0.25$ & $ \left. \! \right) \cdot 10^{-1}$ &  & $0.21$$\,\cdot\,10^{-1}$ & $\left( \right.\! $ \hspace{-0.5mm} & $  7.25$ & $\pm  0.31$ & $^{+ 0.06}_{- 0.20}$ & $ \left. \! \right) \cdot 10^{-1}$ & $  7.06\cdot 10^{-1}$ &
$  745$ & $0.0$ & $0.91$ \\
\multicolumn{2}{|c|}{} & $\left( \right.\!0.25$ & $0.37$ & $ \left. \! \right) \cdot 10^{-1}$ &  & $0.32$$\,\cdot\,10^{-1}$ & $\left( \right.\! $ \hspace{-0.5mm} & $  5.86$ & $\pm  0.27$ & $^{+ 0.06}_{- 0.13}$ & $ \left. \! \right) \cdot 10^{-1}$ & $  6.08\cdot 10^{-1}$ &
$  675$ & $0.0$ & $0.93$ \\
\multicolumn{2}{|c|}{} & $\left( \right.\!0.37$ & $0.60$ & $ \left. \! \right) \cdot 10^{-1}$ &  & $0.50$$\,\cdot\,10^{-1}$ & $\left( \right.\! $ \hspace{-0.5mm} & $  5.27$ & $\pm  0.22$ & $^{+ 0.09}_{- 0.06}$ & $ \left. \! \right) \cdot 10^{-1}$ & $  5.23\cdot 10^{-1}$ &
$  776$ & $0.0$ & $0.93$ \\
\multicolumn{2}{|c|}{} & $\left( \right.\!0.60$ & $1.20$ & $ \left. \! \right) \cdot 10^{-1}$ &  & $0.80$$\,\cdot\,10^{-1}$ & $\left( \right.\! $ \hspace{-0.5mm} & $  4.41$ & $\pm  0.17$ & $^{+ 0.05}_{- 0.10}$ & $ \left. \! \right) \cdot 10^{-1}$ & $  4.46\cdot 10^{-1}$ &
$  997$ & $0.3$ & $1.01$ \\
\multicolumn{2}{|c|}{} & $0.12$ & $0.25$ & $$ &  & $0.18$$$ & $\left( \right.\! $ \hspace{-0.5mm} & $  3.10$ & $\pm  0.13$ & $^{+ 0.09}_{- 0.08}$ & $ \left. \! \right) \cdot 10^{-1}$ & $  3.23\cdot 10^{-1}$ &
$  782$ & $0.0$ & $0.98$ \\
\hline $  410.$ & $  530.$ & $\left( \right.\!0.60$ & $1.00$ & $ \left. \! \right) \cdot 10^{-2}$ & $  450$ & $0.80$$\,\cdot\,10^{-2}$ & $ $ & $  1.02$ & $\pm  0.05$ & $^{+ 0.02}_{- 0.03}$ & $$ & $  1.00$ &
$  588$ & $1.5$ & $0.81$ \\
\multicolumn{2}{|c|}{} & $\left( \right.\!0.10$ & $0.17$ & $ \left. \! \right) \cdot 10^{-1}$ &  & $0.13$$\,\cdot\,10^{-1}$ & $\left( \right.\! $ \hspace{-0.5mm} & $  8.47$ & $\pm  0.53$ & $^{+ 0.37}_{- 0.35}$ & $ \left. \! \right) \cdot 10^{-1}$ & $  8.55\cdot 10^{-1}$ &
$  332$ & $0.1$ & $0.45$ \\
\multicolumn{2}{|c|}{} & $\left( \right.\!0.17$ & $0.25$ & $ \left. \! \right) \cdot 10^{-1}$ &  & $0.21$$\,\cdot\,10^{-1}$ & $\left( \right.\! $ \hspace{-0.5mm} & $  6.85$ & $\pm  0.42$ & $^{+ 0.17}_{- 0.08}$ & $ \left. \! \right) \cdot 10^{-1}$ & $  7.18\cdot 10^{-1}$ &
$  348$ & $0.0$ & $0.67$ \\
\multicolumn{2}{|c|}{} & $\left( \right.\!0.25$ & $0.37$ & $ \left. \! \right) \cdot 10^{-1}$ &  & $0.32$$\,\cdot\,10^{-1}$ & $\left( \right.\! $ \hspace{-0.5mm} & $  6.28$ & $\pm  0.35$ & $^{+ 0.11}_{- 0.11}$ & $ \left. \! \right) \cdot 10^{-1}$ & $  6.17\cdot 10^{-1}$ &
$  422$ & $0.0$ & $0.81$ \\
\multicolumn{2}{|c|}{} & $\left( \right.\!0.37$ & $0.60$ & $ \left. \! \right) \cdot 10^{-1}$ &  & $0.50$$\,\cdot\,10^{-1}$ & $\left( \right.\! $ \hspace{-0.5mm} & $  5.08$ & $\pm  0.25$ & $^{+ 0.09}_{- 0.15}$ & $ \left. \! \right) \cdot 10^{-1}$ & $  5.28\cdot 10^{-1}$ &
$  519$ & $0.0$ & $0.92$ \\
\multicolumn{2}{|c|}{} & $\left( \right.\!0.60$ & $1.00$ & $ \left. \! \right) \cdot 10^{-1}$ &  & $0.80$$\,\cdot\,10^{-1}$ & $\left( \right.\! $ \hspace{-0.5mm} & $  4.41$ & $\pm  0.22$ & $^{+ 0.07}_{- 0.09}$ & $ \left. \! \right) \cdot 10^{-1}$ & $  4.48\cdot 10^{-1}$ &
$  500$ & $0.0$ & $0.96$ \\
\multicolumn{2}{|c|}{} & $0.10$ & $0.17$ & $$ &  & $0.13$$$ & $\left( \right.\! $ \hspace{-0.5mm} & $  3.90$ & $\pm  0.21$ & $^{+ 0.07}_{- 0.07}$ & $ \left. \! \right) \cdot 10^{-1}$ & $  3.73\cdot 10^{-1}$ &
$  458$ & $0.0$ & $0.94$ \\
\multicolumn{2}{|c|}{} & $0.17$ & $0.30$ & $$ &  & $0.25$$$ & $\left( \right.\! $ \hspace{-0.5mm} & $  2.83$ & $\pm  0.17$ & $^{+ 0.06}_{- 0.16}$ & $ \left. \! \right) \cdot 10^{-1}$ & $  2.59\cdot 10^{-1}$ &
$  365$ & $0.0$ & $0.87$ \\
\hline
\end{tabular}
}

  \end{center}    
  \setlength{\localtextwidth}{21.3cm} 
  \caption[this space for rent]{
    The reduced cross-section $\tilde{\sigma}(e^- p)$ for the reaction
    $e^{-} p \rightarrow e^{-} X$.  The following quantities are given
    for each bin: the $Q^2$ and $x$ ranges, the values at which the
    cross section is quoted, \qqc\ and $x_c$, the measured reduced
    cross-section, $\tilde{\sigma}(e^- p)$, corrected to Born level
    and the corresponding cross section predicted by the SM using
    CTEQ5D PDFs. The first error of the measured cross section gives
    the statistical error and the second is the systematic
    uncertainty. The last three columns contain the number of observed
    events in data, $N_{\rm obs}$, the number of expected background
    events, $N_{\rm bg}$ and the acceptance, $\cal A$.  }
  \label{tab-dsdqdx_2}
\end{sidewaystable}
\begin{sidewaystable}
  \begin{center}
    {\small
\renewcommand{\arraystretch}{1.2}
\begin{tabular}{|r@{ -- }l|r@{ -- }l@{$\,$}r|r|l|l@{}l@{$\,$}l@{$\,$}l@{$\,$}l|l|r|r|c|}
\hline
\multicolumn{2}{|c|}{ $Q^2$ range} & 
\multicolumn{3}{c|}{ $x$ range} & 
\multicolumn{1}{c|}{$Q^2_c$} & 
\multicolumn{1}{c|}{$x_c$} &
\multicolumn{6}{c|}{$\tilde{\sigma}(e^-p) $} &
{$N_{\rm obs}$} & 
{$N_{\rm bg}$} & 
{$\cal A$} \\
\cline{8-13}
\multicolumn{2}{|c|} {($\Gev^2$)} &
\multicolumn{3}{c|}{} & 
{($\Gev^2$)} && 
\multicolumn{5}{c|}{measured} & 
\multicolumn{1}{c|}{SM} &&& \\ 
\hline
\hline $  530.$ & $  710.$ & $\left( \right.\!0.60$ & $1.00$ & $ \left. \! \right) \cdot 10^{-2}$ & $  650$ & $0.80$$\,\cdot\,10^{-2}$ & $\left( \right.\! $ \hspace{-0.5mm} & $  9.50$ & $^{+  0.59}_{- 0.56}$ & $^{+ 0.29}_{- 0.26}$ & $ \left. \! \right) \cdot 10^{-1}$ & $ 10.00\cdot 10^{-1}$ &
$  328$ & $0.8$ & $0.73$ \\
\multicolumn{2}{|c|}{} & $\left( \right.\!0.10$ & $0.17$ & $ \left. \! \right) \cdot 10^{-1}$ &  & $0.13$$\,\cdot\,10^{-1}$ & $\left( \right.\! $ \hspace{-0.5mm} & $  9.38$ & $\pm  0.43$ & $^{+ 0.13}_{- 0.30}$ & $ \left. \! \right) \cdot 10^{-1}$ & $  8.74\cdot 10^{-1}$ &
$  593$ & $0.6$ & $0.87$ \\
\multicolumn{2}{|c|}{} & $\left( \right.\!0.17$ & $0.25$ & $ \left. \! \right) \cdot 10^{-1}$ &  & $0.21$$\,\cdot\,10^{-1}$ & $\left( \right.\! $ \hspace{-0.5mm} & $  7.92$ & $^{+  0.53}_{- 0.50}$ & $^{+ 0.06}_{- 0.19}$ & $ \left. \! \right) \cdot 10^{-1}$ & $  7.36\cdot 10^{-1}$ &
$  289$ & $0.0$ & $0.58$ \\
\multicolumn{2}{|c|}{} & $\left( \right.\!0.25$ & $0.37$ & $ \left. \! \right) \cdot 10^{-1}$ &  & $0.32$$\,\cdot\,10^{-1}$ & $\left( \right.\! $ \hspace{-0.5mm} & $  6.03$ & $^{+  0.52}_{- 0.49}$ & $^{+ 0.18}_{- 0.06}$ & $ \left. \! \right) \cdot 10^{-1}$ & $  6.29\cdot 10^{-1}$ &
$  177$ & $0.0$ & $0.42$ \\
\multicolumn{2}{|c|}{} & $\left( \right.\!0.37$ & $0.60$ & $ \left. \! \right) \cdot 10^{-1}$ &  & $0.50$$\,\cdot\,10^{-1}$ & $\left( \right.\! $ \hspace{-0.5mm} & $  5.05$ & $^{+  0.42}_{- 0.40}$ & $^{+ 0.14}_{- 0.18}$ & $ \left. \! \right) \cdot 10^{-1}$ & $  5.35\cdot 10^{-1}$ &
$  186$ & $0.0$ & $0.41$ \\
\multicolumn{2}{|c|}{} & $\left( \right.\!0.60$ & $1.00$ & $ \left. \! \right) \cdot 10^{-1}$ &  & $0.80$$\,\cdot\,10^{-1}$ & $\left( \right.\! $ \hspace{-0.5mm} & $  4.03$ & $^{+  0.35}_{- 0.32}$ & $^{+ 0.08}_{- 0.09}$ & $ \left. \! \right) \cdot 10^{-1}$ & $  4.51\cdot 10^{-1}$ &
$  177$ & $0.0$ & $0.45$ \\
\multicolumn{2}{|c|}{} & $0.10$ & $0.17$ & $$ &  & $0.13$$$ & $\left( \right.\! $ \hspace{-0.5mm} & $  3.45$ & $^{+  0.28}_{- 0.27}$ & $^{+ 0.06}_{- 0.13}$ & $ \left. \! \right) \cdot 10^{-1}$ & $  3.73\cdot 10^{-1}$ &
$  196$ & $0.0$ & $0.51$ \\
\multicolumn{2}{|c|}{} & $0.17$ & $0.30$ & $$ &  & $0.25$$$ & $\left( \right.\! $ \hspace{-0.5mm} & $  2.63$ & $^{+  0.22}_{- 0.21}$ & $^{+ 0.15}_{- 0.10}$ & $ \left. \! \right) \cdot 10^{-1}$ & $  2.55\cdot 10^{-1}$ &
$  193$ & $0.0$ & $0.58$ \\
\hline $  710.$ & $  900.$ & $\left( \right.\!0.90$ & $1.70$ & $ \left. \! \right) \cdot 10^{-2}$ & $  800$ & $1.30$$\,\cdot\,10^{-2}$ & $\left( \right.\! $ \hspace{-0.5mm} & $  8.93$ & $\pm  0.50$ & $^{+ 0.26}_{- 0.10}$ & $ \left. \! \right) \cdot 10^{-1}$ & $  8.81\cdot 10^{-1}$ &
$  406$ & $0.8$ & $0.95$ \\
\multicolumn{2}{|c|}{} & $\left( \right.\!0.17$ & $0.25$ & $ \left. \! \right) \cdot 10^{-1}$ &  & $0.21$$\,\cdot\,10^{-1}$ & $\left( \right.\! $ \hspace{-0.5mm} & $  8.11$ & $^{+  0.56}_{- 0.53}$ & $^{+ 0.14}_{- 0.20}$ & $ \left. \! \right) \cdot 10^{-1}$ & $  7.46\cdot 10^{-1}$ &
$  272$ & $0.0$ & $0.97$ \\
\multicolumn{2}{|c|}{} & $\left( \right.\!0.25$ & $0.37$ & $ \left. \! \right) \cdot 10^{-1}$ &  & $0.32$$\,\cdot\,10^{-1}$ & $\left( \right.\! $ \hspace{-0.5mm} & $  6.30$ & $^{+  0.46}_{- 0.44}$ & $^{+ 0.11}_{- 0.27}$ & $ \left. \! \right) \cdot 10^{-1}$ & $  6.37\cdot 10^{-1}$ &
$  241$ & $0.0$ & $0.92$ \\
\multicolumn{2}{|c|}{} & $\left( \right.\!0.37$ & $0.60$ & $ \left. \! \right) \cdot 10^{-1}$ &  & $0.50$$\,\cdot\,10^{-1}$ & $\left( \right.\! $ \hspace{-0.5mm} & $  5.85$ & $^{+  0.41}_{- 0.39}$ & $^{+ 0.10}_{- 0.14}$ & $ \left. \! \right) \cdot 10^{-1}$ & $  5.40\cdot 10^{-1}$ &
$  266$ & $0.0$ & $0.80$ \\
\multicolumn{2}{|c|}{} & $\left( \right.\!0.60$ & $1.00$ & $ \left. \! \right) \cdot 10^{-1}$ &  & $0.80$$\,\cdot\,10^{-1}$ & $\left( \right.\! $ \hspace{-0.5mm} & $  4.93$ & $^{+  0.40}_{- 0.37}$ & $^{+ 0.07}_{- 0.17}$ & $ \left. \! \right) \cdot 10^{-1}$ & $  4.53\cdot 10^{-1}$ &
$  206$ & $0.0$ & $0.70$ \\
\multicolumn{2}{|c|}{} & $0.10$ & $0.17$ & $$ &  & $0.13$$$ & $\left( \right.\! $ \hspace{-0.5mm} & $  3.43$ & $^{+  0.33}_{- 0.31}$ & $^{+ 0.10}_{- 0.23}$ & $ \left. \! \right) \cdot 10^{-1}$ & $  3.73\cdot 10^{-1}$ &
$  143$ & $0.0$ & $0.61$ \\
\multicolumn{2}{|c|}{} & $0.17$ & $0.30$ & $$ &  & $0.25$$$ & $\left( \right.\! $ \hspace{-0.5mm} & $  2.64$ & $^{+  0.31}_{- 0.29}$ & $^{+ 0.18}_{- 0.06}$ & $ \left. \! \right) \cdot 10^{-1}$ & $  2.53\cdot 10^{-1}$ &
$   99$ & $0.0$ & $0.52$ \\
\hline
\end{tabular}
}

  \end{center}    
  \setlength{\localtextwidth}{21.3cm} 
  \caption[this space for rent]{
    The reduced cross-section $\tilde{\sigma}(e^- p)$ for the reaction
    $e^{-} p \rightarrow e^{-} X$.  The following quantities are given
    for each bin: the $Q^2$ and $x$ ranges, the values at which the
    cross section is quoted, \qqc\ and $x_c$, the measured reduced
    cross-section, $\tilde{\sigma}(e^- p)$, corrected to Born level
    and the corresponding cross section predicted by the SM using
    CTEQ5D PDFs. The first error of the measured cross section gives
    the statistical error and the second is the systematic
    uncertainty. The last three columns contain the number of observed
    events in data, $N_{\rm obs}$, the number of expected background
    events, $N_{\rm bg}$ and the acceptance, $\cal A$.  }
  \label{tab-dsdqdx_3}
\end{sidewaystable}
\clearpage
\begin{sidewaystable}
  \begin{center}
    {\small
\renewcommand{\arraystretch}{1.2}
\begin{tabular}{|r@{ -- }l|r@{ -- }l@{$\,$}r|r|l|l@{}l@{$\,$}l@{$\,$}l@{$\,$}l|l|r|r|c|}
\hline
\multicolumn{2}{|c|}{ $Q^2$ range} & 
\multicolumn{3}{c|}{ $x$ range} & 
\multicolumn{1}{c|}{$Q^2_c$} & 
\multicolumn{1}{c|}{$x_c$} &
\multicolumn{6}{c|}{$\tilde{\sigma}(e^-p) $} &
{$N_{\rm obs}$} & 
{$N_{\rm bg}$} & 
{$\cal A$} \\
\cline{8-13}
\multicolumn{2}{|c|} {($\Gev^2$)} &
\multicolumn{3}{c|}{} & 
{($\Gev^2$)} && 
\multicolumn{5}{c|}{measured} & 
\multicolumn{1}{c|}{SM} &&& \\ 
\hline
\hline $  900.$ & $ 1300.$ & $\left( \right.\!0.10$ & $0.17$ & $ \left. \! \right) \cdot 10^{-1}$ & $ 1200$ & $0.14$$\,\cdot\,10^{-1}$ & $\left( \right.\! $ \hspace{-0.5mm} & $  9.20$ & $^{+  0.61}_{- 0.58}$ & $^{+ 0.67}_{- 0.27}$ & $ \left. \! \right) \cdot 10^{-1}$ & $  8.61\cdot 10^{-1}$ &
$  284$ & $1.3$ & $0.95$ \\
\multicolumn{2}{|c|}{} & $\left( \right.\!0.17$ & $0.25$ & $ \left. \! \right) \cdot 10^{-1}$ &  & $0.21$$\,\cdot\,10^{-1}$ & $\left( \right.\! $ \hspace{-0.5mm} & $  6.86$ & $^{+  0.50}_{- 0.47}$ & $^{+ 0.12}_{- 0.12}$ & $ \left. \! \right) \cdot 10^{-1}$ & $  7.64\cdot 10^{-1}$ &
$  239$ & $0.2$ & $0.97$ \\
\multicolumn{2}{|c|}{} & $\left( \right.\!0.25$ & $0.37$ & $ \left. \! \right) \cdot 10^{-1}$ &  & $0.32$$\,\cdot\,10^{-1}$ & $\left( \right.\! $ \hspace{-0.5mm} & $  5.86$ & $^{+  0.42}_{- 0.40}$ & $^{+ 0.06}_{- 0.18}$ & $ \left. \! \right) \cdot 10^{-1}$ & $  6.54\cdot 10^{-1}$ &
$  242$ & $0.0$ & $0.97$ \\
\multicolumn{2}{|c|}{} & $\left( \right.\!0.37$ & $0.60$ & $ \left. \! \right) \cdot 10^{-1}$ &  & $0.50$$\,\cdot\,10^{-1}$ & $\left( \right.\! $ \hspace{-0.5mm} & $  5.15$ & $^{+  0.33}_{- 0.31}$ & $^{+ 0.12}_{- 0.06}$ & $ \left. \! \right) \cdot 10^{-1}$ & $  5.52\cdot 10^{-1}$ &
$  311$ & $0.2$ & $0.98$ \\
\multicolumn{2}{|c|}{} & $\left( \right.\!0.60$ & $1.00$ & $ \left. \! \right) \cdot 10^{-1}$ &  & $0.80$$\,\cdot\,10^{-1}$ & $\left( \right.\! $ \hspace{-0.5mm} & $  4.28$ & $^{+  0.28}_{- 0.27}$ & $^{+ 0.07}_{- 0.11}$ & $ \left. \! \right) \cdot 10^{-1}$ & $  4.59\cdot 10^{-1}$ &
$  291$ & $0.0$ & $0.98$ \\
\multicolumn{2}{|c|}{} & $0.10$ & $0.17$ & $$ &  & $0.13$$$ & $\left( \right.\! $ \hspace{-0.5mm} & $  3.62$ & $^{+  0.25}_{- 0.24}$ & $^{+ 0.12}_{- 0.05}$ & $ \left. \! \right) \cdot 10^{-1}$ & $  3.75\cdot 10^{-1}$ &
$  264$ & $0.0$ & $0.97$ \\
\multicolumn{2}{|c|}{} & $0.17$ & $0.30$ & $$ &  & $0.25$$$ & $\left( \right.\! $ \hspace{-0.5mm} & $  2.65$ & $^{+  0.21}_{- 0.19}$ & $^{+ 0.04}_{- 0.05}$ & $ \left. \! \right) \cdot 10^{-1}$ & $  2.50\cdot 10^{-1}$ &
$  217$ & $0.0$ & $0.93$ \\
\multicolumn{2}{|c|}{} & $0.30$ & $0.53$ & $$ &  & $0.40$$$ & $\left( \right.\! $ \hspace{-0.5mm} & $  1.01$ & $^{+  0.15}_{- 0.14}$ & $^{+ 0.04}_{- 0.18}$ & $ \left. \! \right) \cdot 10^{-1}$ & $  1.31\cdot 10^{-1}$ &
$   61$ & $0.0$ & $0.66$ \\
\hline $ 1300.$ & $ 1800.$ & $\left( \right.\!0.17$ & $0.25$ & $ \left. \! \right) \cdot 10^{-1}$ & $ 1500$ & $0.21$$\,\cdot\,10^{-1}$ & $\left( \right.\! $ \hspace{-0.5mm} & $  8.26$ & $^{+  0.75}_{- 0.70}$ & $^{+ 0.35}_{- 0.17}$ & $ \left. \! \right) \cdot 10^{-1}$ & $  7.72\cdot 10^{-1}$ &
$  152$ & $0.5$ & $1.00$ \\
\multicolumn{2}{|c|}{} & $\left( \right.\!0.25$ & $0.37$ & $ \left. \! \right) \cdot 10^{-1}$ &  & $0.32$$\,\cdot\,10^{-1}$ & $\left( \right.\! $ \hspace{-0.5mm} & $  7.50$ & $^{+  0.64}_{- 0.59}$ & $^{+ 0.20}_{- 0.14}$ & $ \left. \! \right) \cdot 10^{-1}$ & $  6.66\cdot 10^{-1}$ &
$  173$ & $0.4$ & $0.97$ \\
\multicolumn{2}{|c|}{} & $\left( \right.\!0.37$ & $0.60$ & $ \left. \! \right) \cdot 10^{-1}$ &  & $0.50$$\,\cdot\,10^{-1}$ & $\left( \right.\! $ \hspace{-0.5mm} & $  6.07$ & $^{+  0.46}_{- 0.44}$ & $^{+ 0.06}_{- 0.15}$ & $ \left. \! \right) \cdot 10^{-1}$ & $  5.61\cdot 10^{-1}$ &
$  210$ & $0.0$ & $1.00$ \\
\multicolumn{2}{|c|}{} & $\left( \right.\!0.60$ & $1.00$ & $ \left. \! \right) \cdot 10^{-1}$ &  & $0.80$$\,\cdot\,10^{-1}$ & $\left( \right.\! $ \hspace{-0.5mm} & $  4.45$ & $^{+  0.37}_{- 0.35}$ & $^{+ 0.05}_{- 0.12}$ & $ \left. \! \right) \cdot 10^{-1}$ & $  4.65\cdot 10^{-1}$ &
$  176$ & $0.0$ & $0.96$ \\
\multicolumn{2}{|c|}{} & $0.10$ & $0.15$ & $$ &  & $0.13$$$ & $\left( \right.\! $ \hspace{-0.5mm} & $  3.97$ & $^{+  0.39}_{- 0.36}$ & $^{+ 0.12}_{- 0.12}$ & $ \left. \! \right) \cdot 10^{-1}$ & $  3.77\cdot 10^{-1}$ &
$  131$ & $0.0$ & $0.98$ \\
\multicolumn{2}{|c|}{} & $0.15$ & $0.23$ & $$ &  & $0.18$$$ & $\left( \right.\! $ \hspace{-0.5mm} & $  3.33$ & $^{+  0.34}_{- 0.31}$ & $^{+ 0.09}_{- 0.05}$ & $ \left. \! \right) \cdot 10^{-1}$ & $  3.18\cdot 10^{-1}$ &
$  124$ & $0.0$ & $1.01$ \\
\multicolumn{2}{|c|}{} & $0.23$ & $0.35$ & $$ &  & $0.25$$$ & $\left( \right.\! $ \hspace{-0.5mm} & $  2.66$ & $^{+  0.33}_{- 0.30}$ & $^{+ 0.06}_{- 0.02}$ & $ \left. \! \right) \cdot 10^{-1}$ & $  2.50\cdot 10^{-1}$ &
$   84$ & $0.0$ & $0.94$ \\
\multicolumn{2}{|c|}{} & $0.35$ & $0.53$ & $$ &  & $0.40$$$ & $\left( \right.\! $ \hspace{-0.5mm} & $  1.40$ & $^{+  0.27}_{- 0.24}$ & $^{+ 0.57}_{- 0.14}$ & $ \left. \! \right) \cdot 10^{-1}$ & $  1.30\cdot 10^{-1}$ &
$   38$ & $0.0$ & $0.89$ \\
\hline
\end{tabular}
}

  \end{center}    
  \setlength{\localtextwidth}{21.3cm} 
  \caption[this space for rent]{
    The reduced cross-section $\tilde{\sigma}(e^- p)$ for the reaction
    $e^{-} p \rightarrow e^{-} X$.  The following quantities are given
    for each bin: the $Q^2$ and $x$ ranges, the values at which the
    cross section is quoted, \qqc\ and $x_c$, the measured reduced
    cross-section, $\tilde{\sigma}(e^- p)$, corrected to Born level
    and the corresponding cross section predicted by the SM using
    CTEQ5D PDFs. The first error of the measured cross section gives
    the statistical error and the second is the systematic
    uncertainty. The last three columns contain the number of observed
    events in data, $N_{\rm obs}$, the number of expected background
    events, $N_{\rm bg}$ and the acceptance, $\cal A$.  }
  \label{tab-dsdqdx_4}
\end{sidewaystable}

\begin{sidewaystable}
  \begin{center}
    {\small
\renewcommand{\arraystretch}{1.2}
\begin{tabular}{|r@{ -- }l|r@{ -- }l@{$\,$}r|r|l|l@{}l@{$\,$}l@{$\,$}l@{$\,$}l|l|r|r|c|}
\hline
\multicolumn{2}{|c|}{ $Q^2$ range} & 
\multicolumn{3}{c|}{ $x$ range} & 
\multicolumn{1}{c|}{$Q^2_c$} & 
\multicolumn{1}{c|}{$x_c$} &
\multicolumn{6}{c|}{$\tilde{\sigma}(e^-p) $} &
{$N_{\rm obs}$} & 
{$N_{\rm bg}$} & 
{$\cal A$} \\
\cline{8-13}
\multicolumn{2}{|c|} {($\Gev^2$)} &
\multicolumn{3}{c|}{} & 
{($\Gev^2$)} && 
\multicolumn{5}{c|}{measured} & 
\multicolumn{1}{c|}{SM} &&& \\ 
\hline
\hline $ 1800.$ & $ 2500.$ & $\left( \right.\!0.23$ & $0.37$ & $ \left. \! \right) \cdot 10^{-1}$ & $ 2000$ & $0.32$$\,\cdot\,10^{-1}$ & $\left( \right.\! $ \hspace{-0.5mm} & $  7.15$ & $^{+  0.73}_{- 0.67}$ & $^{+ 0.45}_{- 0.12}$ & $ \left. \! \right) \cdot 10^{-1}$ & $  6.83\cdot 10^{-1}$ &
$  123$ & $1.2$ & $0.98$ \\
\multicolumn{2}{|c|}{} & $\left( \right.\!0.37$ & $0.60$ & $ \left. \! \right) \cdot 10^{-1}$ &  & $0.50$$\,\cdot\,10^{-1}$ & $\left( \right.\! $ \hspace{-0.5mm} & $  5.45$ & $^{+  0.55}_{- 0.51}$ & $^{+ 0.09}_{- 0.16}$ & $ \left. \! \right) \cdot 10^{-1}$ & $  5.77\cdot 10^{-1}$ &
$  123$ & $0.1$ & $0.97$ \\
\multicolumn{2}{|c|}{} & $\left( \right.\!0.60$ & $1.00$ & $ \left. \! \right) \cdot 10^{-1}$ &  & $0.80$$\,\cdot\,10^{-1}$ & $\left( \right.\! $ \hspace{-0.5mm} & $  4.31$ & $^{+  0.44}_{- 0.40}$ & $^{+ 0.10}_{- 0.20}$ & $ \left. \! \right) \cdot 10^{-1}$ & $  4.75\cdot 10^{-1}$ &
$  122$ & $0.0$ & $0.99$ \\
\multicolumn{2}{|c|}{} & $0.10$ & $0.15$ & $$ &  & $0.13$$$ & $\left( \right.\! $ \hspace{-0.5mm} & $  3.77$ & $^{+  0.45}_{- 0.41}$ & $^{+ 0.17}_{- 0.07}$ & $ \left. \! \right) \cdot 10^{-1}$ & $  3.82\cdot 10^{-1}$ &
$   90$ & $0.0$ & $0.96$ \\
\multicolumn{2}{|c|}{} & $0.15$ & $0.23$ & $$ &  & $0.18$$$ & $\left( \right.\! $ \hspace{-0.5mm} & $  2.82$ & $^{+  0.37}_{- 0.33}$ & $^{+ 0.08}_{- 0.08}$ & $ \left. \! \right) \cdot 10^{-1}$ & $  3.20\cdot 10^{-1}$ &
$   76$ & $0.0$ & $0.99$ \\
\multicolumn{2}{|c|}{} & $0.23$ & $0.35$ & $$ &  & $0.25$$$ & $\left( \right.\! $ \hspace{-0.5mm} & $  2.92$ & $^{+  0.42}_{- 0.38}$ & $^{+ 0.07}_{- 0.08}$ & $ \left. \! \right) \cdot 10^{-1}$ & $  2.50\cdot 10^{-1}$ &
$   65$ & $0.0$ & $0.96$ \\
\multicolumn{2}{|c|}{} & $0.35$ & $0.53$ & $$ &  & $0.40$$$ & $\left( \right.\! $ \hspace{-0.5mm} & $  1.03$ & $^{+  0.25}_{- 0.20}$ & $^{+ 0.10}_{- 0.08}$ & $ \left. \! \right) \cdot 10^{-1}$ & $  1.28\cdot 10^{-1}$ &
$   26$ & $0.0$ & $1.00$ \\
\hline $ 2500.$ & $ 3500.$ & $\left( \right.\!0.37$ & $0.60$ & $ \left. \! \right) \cdot 10^{-1}$ & $ 3000$ & $0.50$$\,\cdot\,10^{-1}$ & $\left( \right.\! $ \hspace{-0.5mm} & $  6.77$ & $^{+  0.78}_{- 0.71}$ & $^{+ 0.21}_{- 0.30}$ & $ \left. \! \right) \cdot 10^{-1}$ & $  6.11\cdot 10^{-1}$ &
$   95$ & $0.0$ & $0.99$ \\
\multicolumn{2}{|c|}{} & $\left( \right.\!0.60$ & $1.00$ & $ \left. \! \right) \cdot 10^{-1}$ &  & $0.80$$\,\cdot\,10^{-1}$ & $\left( \right.\! $ \hspace{-0.5mm} & $  4.50$ & $^{+  0.55}_{- 0.50}$ & $^{+ 0.10}_{- 0.14}$ & $ \left. \! \right) \cdot 10^{-1}$ & $  5.01\cdot 10^{-1}$ &
$   84$ & $0.0$ & $0.99$ \\
\multicolumn{2}{|c|}{} & $0.10$ & $0.15$ & $$ &  & $0.13$$$ & $\left( \right.\! $ \hspace{-0.5mm} & $  3.75$ & $^{+  0.55}_{- 0.48}$ & $^{+ 0.11}_{- 0.28}$ & $ \left. \! \right) \cdot 10^{-1}$ & $  3.97\cdot 10^{-1}$ &
$   62$ & $0.0$ & $0.96$ \\
\multicolumn{2}{|c|}{} & $0.15$ & $0.23$ & $$ &  & $0.18$$$ & $\left( \right.\! $ \hspace{-0.5mm} & $  3.00$ & $^{+  0.47}_{- 0.41}$ & $^{+ 0.10}_{- 0.07}$ & $ \left. \! \right) \cdot 10^{-1}$ & $  3.29\cdot 10^{-1}$ &
$   54$ & $0.0$ & $0.98$ \\
\multicolumn{2}{|c|}{} & $0.23$ & $0.35$ & $$ &  & $0.25$$$ & $\left( \right.\! $ \hspace{-0.5mm} & $  2.85$ & $^{+  0.50}_{- 0.43}$ & $^{+ 0.12}_{- 0.13}$ & $ \left. \! \right) \cdot 10^{-1}$ & $  2.54\cdot 10^{-1}$ &
$   45$ & $0.0$ & $0.92$ \\
\multicolumn{2}{|c|}{} & $0.35$ & $0.53$ & $$ &  & $0.40$$$ & $\left( \right.\! $ \hspace{-0.5mm} & $  1.63$ & $^{+  0.41}_{- 0.34}$ & $^{+ 0.10}_{- 0.08}$ & $ \left. \! \right) \cdot 10^{-1}$ & $  1.28\cdot 10^{-1}$ &
$   24$ & $0.0$ & $0.93$ \\
\multicolumn{2}{|c|}{} & $0.53$ & $0.75$ & $$ &  & $0.65$$$ & $\left( \right.\! $ \hspace{-0.5mm} & $  1.90$ & $^{+  0.96}_{- 0.68}$ & $^{+ 0.17}_{- 0.10}$ & $ \left. \! \right) \cdot 10^{-2}$ & $  1.94\cdot 10^{-2}$ &
$    8$ & $0.0$ & $0.95$ \\
\hline
\end{tabular}
}

  \end{center}    
  \setlength{\localtextwidth}{21.3cm} 
  \caption[this space for rent]{
    The reduced cross-section $\tilde{\sigma}(e^- p)$ for the reaction
    $e^{-} p \rightarrow e^{-} X$.  The following quantities are given
    for each bin: the $Q^2$ and $x$ ranges, the values at which the
    cross section is quoted, \qqc\ and $x_c$, the measured reduced
    cross-section, $\tilde{\sigma}(e^- p)$, corrected to Born level
    and the corresponding cross section predicted by the SM using
    CTEQ5D PDFs. The first error of the measured cross section gives
    the statistical error and the second is the systematic
    uncertainty. The last three columns contain the number of observed
    events in data, $N_{\rm obs}$, the number of expected background
    events, $N_{\rm bg}$ and the acceptance, $\cal A$.  }
  \label{tab-dsdqdx_5}
\end{sidewaystable}
\begin{sidewaystable}
  \begin{center}
    {\small
\renewcommand{\arraystretch}{1.2}
\begin{tabular}{|r@{ -- }l|r@{ -- }l@{$\,$}r|r|l|l@{}l@{$\,$}l@{$\,$}l@{$\,$}l|l|r|r|c|}
\hline
\multicolumn{2}{|c|}{ $Q^2$ range} & 
\multicolumn{3}{c|}{ $x$ range} & 
\multicolumn{1}{c|}{$Q^2_c$} & 
\multicolumn{1}{c|}{$x_c$} &
\multicolumn{6}{c|}{$\tilde{\sigma}(e^-p) $} &
{$N_{\rm obs}$} & 
{$N_{\rm bg}$} & 
{$\cal A$} \\
\cline{8-13}
\multicolumn{2}{|c|} {($\Gev^2$)} &
\multicolumn{3}{c|}{} & 
{($\Gev^2$)} && 
\multicolumn{5}{c|}{measured} & 
\multicolumn{1}{c|}{SM} &&& \\ 
\hline
\hline $ 3500.$ & $ 5600.$ & $\left( \right.\!0.40$ & $1.00$ & $ \left. \! \right) \cdot 10^{-1}$ & $ 5000$ & $0.80$$\,\cdot\,10^{-1}$ & $\left( \right.\! $ \hspace{-0.5mm} & $  6.01$ & $^{+  0.59}_{- 0.54}$ & $^{+ 0.20}_{- 0.06}$ & $ \left. \! \right) \cdot 10^{-1}$ & $  5.62\cdot 10^{-1}$ &
$  126$ & $0.6$ & $0.96$ \\
\multicolumn{2}{|c|}{} & $0.10$ & $0.15$ & $$ &  & $0.13$$$ & $\left( \right.\! $ \hspace{-0.5mm} & $  4.40$ & $^{+  0.64}_{- 0.57}$ & $^{+ 0.12}_{- 0.08}$ & $ \left. \! \right) \cdot 10^{-1}$ & $  4.37\cdot 10^{-1}$ &
$   62$ & $0.0$ & $0.98$ \\
\multicolumn{2}{|c|}{} & $0.15$ & $0.23$ & $$ &  & $0.18$$$ & $\left( \right.\! $ \hspace{-0.5mm} & $  4.15$ & $^{+  0.59}_{- 0.52}$ & $^{+ 0.14}_{- 0.14}$ & $ \left. \! \right) \cdot 10^{-1}$ & $  3.56\cdot 10^{-1}$ &
$   65$ & $0.0$ & $0.95$ \\
\multicolumn{2}{|c|}{} & $0.23$ & $0.35$ & $$ &  & $0.25$$$ & $\left( \right.\! $ \hspace{-0.5mm} & $  2.67$ & $^{+  0.49}_{- 0.42}$ & $^{+ 0.04}_{- 0.07}$ & $ \left. \! \right) \cdot 10^{-1}$ & $  2.68\cdot 10^{-1}$ &
$   41$ & $0.0$ & $1.02$ \\
\multicolumn{2}{|c|}{} & $0.35$ & $0.53$ & $$ &  & $0.40$$$ & $\left( \right.\! $ \hspace{-0.5mm} & $  1.60$ & $^{+  0.40}_{- 0.33}$ & $^{+ 0.06}_{- 0.17}$ & $ \left. \! \right) \cdot 10^{-1}$ & $  1.31\cdot 10^{-1}$ &
$   24$ & $0.0$ & $0.99$ \\
\hline $ 5600.$ & $ 9000.$ & $\left( \right.\!0.70$ & $1.50$ & $ \left. \! \right) \cdot 10^{-1}$ & $ 8000$ & $1.30$$\,\cdot\,10^{-1}$ & $\left( \right.\! $ \hspace{-0.5mm} & $  5.32$ & $^{+  0.78}_{- 0.69}$ & $^{+ 0.16}_{- 0.27}$ & $ \left. \! \right) \cdot 10^{-1}$ & $  5.08\cdot 10^{-1}$ &
$   60$ & $0.0$ & $0.91$ \\
\multicolumn{2}{|c|}{} & $0.15$ & $0.23$ & $$ &  & $0.18$$$ & $\left( \right.\! $ \hspace{-0.5mm} & $  3.83$ & $^{+  0.77}_{- 0.65}$ & $^{+ 0.15}_{- 0.16}$ & $ \left. \! \right) \cdot 10^{-1}$ & $  4.04\cdot 10^{-1}$ &
$   34$ & $0.0$ & $0.97$ \\
\multicolumn{2}{|c|}{} & $0.23$ & $0.35$ & $$ &  & $0.25$$$ & $\left( \right.\! $ \hspace{-0.5mm} & $  3.10$ & $^{+  0.73}_{- 0.60}$ & $^{+ 0.06}_{- 0.04}$ & $ \left. \! \right) \cdot 10^{-1}$ & $  2.97\cdot 10^{-1}$ &
$   26$ & $0.0$ & $1.02$ \\
\multicolumn{2}{|c|}{} & $0.35$ & $0.53$ & $$ &  & $0.40$$$ & $\left( \right.\! $ \hspace{-0.5mm} & $  1.39$ & $^{+  0.52}_{- 0.39}$ & $^{+ 0.08}_{- 0.12}$ & $ \left. \! \right) \cdot 10^{-1}$ & $  1.39\cdot 10^{-1}$ &
$   12$ & $0.0$ & $1.00$ \\
\multicolumn{2}{|c|}{} & $0.53$ & $0.75$ & $$ &  & $0.65$$$ & $\left( \right.\! $ \hspace{-0.5mm} & $   1.9$ & $^{+   1.5}_{-  0.9}$ & $^{+  0.1}_{-  0.0}$ & $ \left. \! \right) \cdot 10^{-2}$ & $   1.9\cdot 10^{-2}$ &
$    4$ & $0.0$ & $0.94$ \\
\hline $ 9000.$ & $15000.$ & $0.11$ & $0.23$ & $$ & $12000$ & $0.18$$$ & $\left( \right.\! $ \hspace{-0.5mm} & $   4.5$ & $^{+   1.0}_{-  0.8}$ & $^{+  0.3}_{-  0.1}$ & $ \left. \! \right) \cdot 10^{-1}$ & $   4.7\cdot 10^{-1}$ &
$   29$ & $0.0$ & $0.90$ \\
\multicolumn{2}{|c|}{} & $0.23$ & $0.35$ & $$ &  & $0.25$$$ & $\left( \right.\! $ \hspace{-0.5mm} & $   2.7$ & $^{+   1.0}_{-  0.8}$ & $^{+  0.1}_{-  0.2}$ & $ \left. \! \right) \cdot 10^{-1}$ & $   3.4\cdot 10^{-1}$ &
$   12$ & $0.0$ & $0.94$ \\
\multicolumn{2}{|c|}{} & $0.35$ & $0.53$ & $$ &  & $0.40$$$ & $\left( \right.\! $ \hspace{-0.5mm} & $   1.1$ & $^{+   0.7}_{-  0.5}$ & $^{+  0.2}_{-  0.0}$ & $ \left. \! \right) \cdot 10^{-1}$ & $   1.5\cdot 10^{-1}$ &
$    6$ & $0.0$ & $0.99$ \\
\hline $15000.$ & $25000.$ & $0.18$ & $0.35$ & $$ & $20000$ & $0.25$$$ & $\left( \right.\! $ \hspace{-0.5mm} & $   4.3$ & $^{+   1.5}_{-  1.1}$ & $^{+  0.2}_{-  0.1}$ & $ \left. \! \right) \cdot 10^{-1}$ & $   4.2\cdot 10^{-1}$ &
$   13$ & $0.0$ & $0.92$ \\
\multicolumn{2}{|c|}{} & $0.35$ & $0.75$ & $$ &  & $0.40$$$ & $\left( \right.\! $ \hspace{-0.5mm} & $   2.1$ & $^{+   1.2}_{-  0.8}$ & $^{+  0.1}_{-  0.0}$ & $ \left. \! \right) \cdot 10^{-1}$ & $   1.8\cdot 10^{-1}$ &
$    6$ & $0.0$ & $0.90$ \\
\hline $25000.$ & $50000.$ & $0.30$ & $0.75$ & $$ & $30000$ & $0.40$$$ & $\left( \right.\! $ \hspace{-0.5mm} & $   2.3$ & $^{+   1.8}_{-  1.1}$ & $^{+  0.3}_{-  0.1}$ & $ \left. \! \right) \cdot 10^{-1}$ & $   2.1\cdot 10^{-1}$ &
$    4$ & $0.0$ & $0.92$ \\
\hline
\end{tabular}
}

  \end{center}    
  \setlength{\localtextwidth}{21.5cm} 
  \caption[this space for rent]{
    The reduced cross-section $\tilde{\sigma}(e^- p)$ for the reaction
    $e^{-} p \rightarrow e^{-} X$.  The following quantities are given
    for each bin: the $Q^2$ and $x$ ranges, the values at which the
    cross section is quoted, \qqc\ and $x_c$, the measured reduced
    cross-section, $\tilde{\sigma}(e^- p)$, corrected to Born level
    and the corresponding cross section predicted by the SM using
    CTEQ5D PDFs. The first error of the measured cross section gives
    the statistical error and the second is the systematic
    uncertainty. The last three columns contain the number of observed
    events in data, $N_{\rm obs}$, the number of expected background
    events, $N_{\rm bg}$ and the acceptance, $\cal A$.  }
  \label{tab-dsdqdx_6}
\end{sidewaystable}

\clearpage

\begin{sidewaystable}
  \begin{center}
    {\small
\renewcommand{\arraystretch}{1.2}
\begin{tabular}{|c|l|c|c|c||c|c|c|c|c|c|c|}
\hline
{$Q^2_c$} & 
\multicolumn{1}{c|}{$x_c$} & 
 $\tilde{\sigma}(e^-p) $&
stat.  & 
total sys.  & 
uncor. sys. &
$\delta_1$ &
$\delta_2$ &
$\delta_3$ &
$\delta_4$ &
$\delta_5$ &
$\delta_6$ \\
{($\gev^2$)} & 
$$ &
 &
(\%) & 
(\%) &
(\%) &
(\%) &
(\%) &
(\%) &
(\%) &
(\%) &
(\%) \\
\hline
\hline
  200      & $0.50\,\cdot\,10^{-2}$ & $
  1.16$ &
$^{+ 2.8  } _{- 2.8  } $ & $^{+ 2.9  } _{- 2.0  } $ & $^{+ 1.0  } _{- 1.1  } $ & $^{-0.4  }  _{+ 0.2  } $ & $^{}_{- 0.0  } $ & $^{+ 0.0  } _{- 0.0  } $ & $^{-0.8  }  _{+ 1.1  } $ & $^{+ 0.3  } _{-0.9  }  $ & $^{-1.1  }  _{+ 2.5  } 
$ \\
           & $0.80\,\cdot\,10^{-2}$ & $
  0.96$ &
$^{+ 2.7  } _{- 2.7  } $ & $^{+ 2.1  } _{- 1.6  } $ & $^{+ 1.2  } _{- 0.6  } $ & $^{-0.1  }  _{+ 0.3  } $ & $^{}_{+ 0.1  } $ & $^{+ 0.1  } _{+ 0.0  } $ & $^{-1.1  }  _{+ 0.4  } $ & $^{+ 0.2  } _{-0.8  }  $ & $^{-0.6  }  _{+ 1.5  } 
$ \\
           & $0.13\,\cdot\,10^{-1}$ & $
  0.82$ &
$^{+ 2.8  } _{- 2.7  } $ & $^{+ 2.2  } _{- 1.9  } $ & $^{+ 0.9  } _{- 1.2  } $ & $^{-0.1  }  _{+ 0.1  } $ & $^{}_{- 0.0  } $ & $^{- 0.0  } _{+ 0.1  } $ & $^{-0.8  }  _{-0.1  }  $ & $^{+ 0.2  } _{-0.8  }  $ & $^{-0.9  }  _{+ 2.0  } 
$ \\
           & $0.21\,\cdot\,10^{-1}$ & $
  0.71$ &
$^{+ 3.4  } _{- 3.3  } $ & $^{+ 1.8  } _{- 1.5  } $ & $^{+ 1.4  } _{- 1.1  } $ & $^{-0.1  }  _{+ 0.3  } $ & $^{}_{-0.1  }  $ & $^{+ 0.2  } _{+ 0.0  } $ & $^{+ 0.1  } _{-0.1  }  $ & $^{+ 0.2  } _{-0.8  }  $ & $^{-0.5  }  _{+ 1.1  } 
$ \\
           & $0.32\,\cdot\,10^{-1}$ & $
  0.59$ &
$^{+ 3.5  } _{- 3.4  } $ & $^{+ 1.1  } _{- 2.2  } $ & $^{+ 0.9  } _{- 1.9  } $ & $^{-0.3  }  _{+ 0.3  } $ & $^{}_{+ 0.0  } $ & $^{+ 0.1  } _{- 0.0  } $ & $^{-0.3  }  _{+ 0.2  } $ & $^{+ 0.2  } _{-1.0  }  $ & $^{-0.1  }  _{+ 0.4  } 
$ \\
           & $0.50\,\cdot\,10^{-1}$ & $
  0.50$ &
$^{+ 3.3  } _{- 3.2  } $ & $^{+ 1.7  } _{- 2.5  } $ & $^{+ 1.3  } _{- 0.7  } $ & $^{-0.1  }  _{+ 0.3  } $ & $^{}_{- 0.0  } $ & $^{+ 0.1  } _{+ 0.1  } $ & $^{-0.1  }  _{-0.5  }  $ & $^{+ 0.4  } _{-1.1  }  $ & $^{+ 0.9  } _{-2.1  }  
$ \\
           & $0.80\,\cdot\,10^{-1}$ & $
  0.42$ &
$^{+ 2.9  } _{- 2.8  } $ & $^{+ 2.0  } _{- 3.5  } $ & $^{+ 1.4  } _{- 0.7  } $ & $^{-0.2  }  _{+ 0.3  } $ & $^{}_{+ 0.0  } $ & $^{-0.1  }  _{+ 0.1  } $ & $^{+ 0.0  } _{+ 0.4  } $ & $^{+ 0.3  } _{-1.2  }  $ & $^{+ 1.3  } _{-3.2  }  
$ \\
           & $0.18$ & $
  0.32$ &
$^{+ 3.9  } _{- 3.8  } $ & $^{+ 1.1  } _{- 3.9  } $ & $^{+ 0.7  } _{- 3.3  } $ & $^{-0.3  }  _{+ 0.0  } $ & $^{}_{- 0.0  } $ & $^{-0.1  }  _{+ 0.0  } $ & $^{-0.2  }  _{-0.7  }  $ & $^{+ 0.1  } _{-1.2  }  $ & $^{+ 0.8  } _{-1.6  }  
$ \\
\hline
  250      & $0.50\,\cdot\,10^{-2}$ & $
  1.12$ &
$^{+ 4.3  } _{- 4.2  } $ & $^{+ 2.5  } _{- 3.0  } $ & $^{+ 1.2  } _{- 2.1  } $ & $^{-1.3  }  _{+ 0.7  } $ & $^{}_{-0.3  }  $ & $^{-0.1  }  _{-0.6  }  $ & $^{+ 0.2  } _{+ 0.6  } $ & $^{-0.2  }  _{-1.3  }  $ & $^{-0.8  }  _{+ 1.9  } 
$ \\
           & $0.80\,\cdot\,10^{-2}$ & $
  0.99$ &
$^{+ 3.2  } _{- 3.2  } $ & $^{+ 1.9  } _{- 2.4  } $ & $^{+ 1.1  } _{- 1.5  } $ & $^{-0.3  }  _{+ 0.2  } $ & $^{}_{-0.1  }  $ & $^{+ 0.0  } _{-0.1  }  $ & $^{-1.4  }  _{+ 0.8  } $ & $^{+ 0.1  } _{-1.1  }  $ & $^{-0.5  }  _{+ 1.3  } 
$ \\
           & $0.13\,\cdot\,10^{-1}$ & $
  0.86$ &
$^{+ 3.2  } _{- 3.2  } $ & $^{+ 2.3  } _{- 2.1  } $ & $^{+ 1.0  } _{- 1.0  } $ & $^{-0.1  }  _{+ 0.1  } $ & $^{}_{-0.1  }  $ & $^{- 0.0  } _{- 0.0  } $ & $^{-1.2  }  _{+ 0.9  } $ & $^{+ 0.4  } _{-1.1  }  $ & $^{-0.7  }  _{+ 1.8  } 
$ \\
           & $0.21\,\cdot\,10^{-1}$ & $
  0.69$ &
$^{+ 3.9  } _{- 3.8  } $ & $^{+ 1.0  } _{- 1.9  } $ & $^{+ 1.0  } _{- 1.3  } $ & $^{-0.1  }  _{-0.1  }  $ & $^{}_{-0.2  }  $ & $^{-0.1  }  _{-0.1  }  $ & $^{-0.6  }  _{-0.7  }  $ & $^{+ 0.2  } _{-1.1  }  $ & $^{+ 0.1  } _{+ 0.0  } 
$ \\
           & $0.32\,\cdot\,10^{-1}$ & $
  0.61$ &
$^{+ 4.0  } _{- 3.9  } $ & $^{+ 1.2  } _{- 3.3  } $ & $^{+ 0.6  } _{- 1.4  } $ & $^{-0.1  }  _{+ 0.1  } $ & $^{}_{+ 0.0  } $ & $^{+ 0.0  } _{+ 0.1  } $ & $^{-1.0  }  _{+ 0.3  } $ & $^{+ 0.0  } _{-1.0  }  $ & $^{+ 1.0  } _{-2.6  }  
$ \\
           & $0.50\,\cdot\,10^{-1}$ & $
  0.52$ &
$^{+ 3.8  } _{- 3.7  } $ & $^{+ 1.9  } _{- 2.7  } $ & $^{+ 1.6  } _{- 0.8  } $ & $^{-0.3  }  _{+ 0.2  } $ & $^{}_{+ 0.1  } $ & $^{+ 0.1  } _{- 0.0  } $ & $^{-1.2  }  _{+ 0.2  } $ & $^{+ 0.2  } _{-1.0  }  $ & $^{+ 0.9  } _{-2.1  }  
$ \\
           & $0.80\,\cdot\,10^{-1}$ & $
  0.44$ &
$^{+ 3.4  } _{- 3.3  } $ & $^{+ 1.5  } _{- 4.9  } $ & $^{+ 0.5  } _{- 3.0  } $ & $^{-0.3  }  _{+ 0.2  } $ & $^{}_{-0.1  }  $ & $^{+ 0.1  } _{-0.2  }  $ & $^{-1.8  }  _{-0.4  }  $ & $^{+ 0.0  } _{-1.0  }  $ & $^{+ 1.4  } _{-3.4  }  
$ \\
           & $0.18$ & $
  0.30$ &
$^{+ 4.1  } _{- 4.0  } $ & $^{+ 6.8  } _{- 2.8  } $ & $^{+ 6.8  } _{- 2.1  } $ & $^{-0.1  }  _{- 0.0  } $ & $^{}_{- 0.0  } $ & $^{+ 0.0  } _{-0.2  }  $ & $^{-0.3  }  _{-0.2  }  $ & $^{+ 0.1  } _{-1.1  }  $ & $^{+ 0.6  } _{-1.4  }  
$ \\
\hline
\end{tabular}
}

  \end{center}    
  \setlength{\localtextwidth}{18.7cm} 
  \caption[this space for rent]{
    Systematic uncertainties with bin-to-bin correlations
    for the reduced cross-section $\tilde{\sigma}(e^- p)$. The left
    part of the table contains the quoted $Q^2$ and $x$ values,
    $Q^2_c$ and $x_c$, the measured cross-section 
    $\tilde{\sigma}(e^- p)$ corrected to the Born level, the
    statistical error and the total systematic uncertainty. The right
    part of the table lists the total uncorrelated systematic
    uncertainty followed by the uncertainties $\delta_1$--\,$\delta_6$ (see text)
    with bin-to-bin correlations. For the latter, the upper (lower)
    numbers refer to positive (negative) variation of e.g. the cut
    value, whereas the signs of the numbers reflect the direction of
    change in the cross sections. 
    }
  \label{tab-dsdxq_c1}
\end{sidewaystable}

\begin{sidewaystable}
  \begin{center}
    {\small
\renewcommand{\arraystretch}{1.2}
\begin{tabular}{|c|l|c|c|c||c|c|c|c|c|c|c|}
\hline
{$Q^2_c$} & 
\multicolumn{1}{c|}{$x_c$} & 
 $\tilde{\sigma}(e^-p) $&
stat.  & 
total sys.  & 
uncor. sys. &
$\delta_1$ &
$\delta_2$ &
$\delta_3$ &
$\delta_4$ &
$\delta_5$ &
$\delta_6$ \\
{($\gev^2$)} & 
$$ &
 &
(\%) & 
(\%) &
(\%) &
(\%) &
(\%) &
(\%) &
(\%) &
(\%) &
(\%) \\
\hline
  350      & $0.80\,\cdot\,10^{-2}$ & $
  1.00$ &
$^{+ 4.6  } _{- 4.4  } $ & $^{+ 1.8  } _{- 1.6  } $ & $^{+ 1.3  } _{- 1.4  } $ & $^{+ 0.1  } _{+ 0.2  } $ & $^{}_{+ 0.1  } $ & $^{+ 0.0  } _{-0.2  }  $ & $^{-0.2  }  _{+ 0.5  } $ & $^{+ 0.2  } _{-0.6  }  $ & $^{-0.4  }  _{+ 1.1  } 
$ \\
           & $0.13\,\cdot\,10^{-1}$ & $
  0.86$ &
$^{+ 3.8  } _{- 3.7  } $ & $^{+ 3.1  } _{- 2.2  } $ & $^{+ 1.1  } _{- 1.2  } $ & $^{-0.3  }  _{-0.1  }  $ & $^{}_{-0.2  }  $ & $^{-0.2  }  _{-0.2  }  $ & $^{-0.5  }  _{- 0.0  } $ & $^{+ 0.1  } _{-1.2  }  $ & $^{-1.2  }  _{+ 2.9  } 
$ \\
           & $0.21\,\cdot\,10^{-1}$ & $
  0.72$ &
$^{+ 4.3  } _{- 4.2  } $ & $^{+ 0.9  } _{- 2.7  } $ & $^{+ 0.7  } _{- 1.3  } $ & $^{-0.3  }  _{+ 0.1  } $ & $^{}_{-0.1  }  $ & $^{-0.2  }  _{-0.1  }  $ & $^{-2.1  }  _{+ 0.4  } $ & $^{+ 0.2  } _{-1.1  }  $ & $^{-0.1  }  _{+ 0.2  } 
$ \\
           & $0.32\,\cdot\,10^{-1}$ & $
  0.59$ &
$^{+ 4.5  } _{- 4.4  } $ & $^{+ 1.1  } _{- 2.3  } $ & $^{+ 1.0  } _{- 1.7  } $ & $^{-0.4  }  _{+ 0.0  } $ & $^{}_{-0.1  }  $ & $^{-0.4  }  _{-0.1  }  $ & $^{-0.7  }  _{-0.1  }  $ & $^{+ 0.0  } _{-1.3  }  $ & $^{-0.2  }  _{+ 0.4  } 
$ \\
           & $0.50\,\cdot\,10^{-1}$ & $
  0.53$ &
$^{+ 4.3  } _{- 4.1  } $ & $^{+ 1.7  } _{- 1.1  } $ & $^{+ 1.3  } _{- 0.9  } $ & $^{+ 0.0  } _{+ 0.5  } $ & $^{}_{+ 0.1  } $ & $^{+ 0.4  } _{+ 0.4  } $ & $^{+ 0.6  } _{+ 0.3  } $ & $^{+ 0.4  } _{-0.6  }  $ & $^{-0.2  }  _{+ 0.6  } 
$ \\
           & $0.80\,\cdot\,10^{-1}$ & $
  0.44$ &
$^{+ 3.8  } _{- 3.7  } $ & $^{+ 1.2  } _{- 2.2  } $ & $^{+ 0.8  } _{- 0.9  } $ & $^{-0.1  }  _{+ 0.4  } $ & $^{}_{- 0.0  } $ & $^{+ 0.0  } _{+ 0.1  } $ & $^{-0.6  }  _{-0.3  }  $ & $^{+ 0.2  } _{-0.8  }  $ & $^{+ 0.8  } _{-1.8  }  
$ \\
           & $0.18$ & $
  0.31$ &
$^{+ 4.2  } _{- 4.1  } $ & $^{+ 2.8  } _{- 2.7  } $ & $^{+ 2.8  } _{- 1.7  } $ & $^{-0.1  }  _{+ 0.2  } $ & $^{}_{- 0.0  } $ & $^{+ 0.2  } _{+ 0.2  } $ & $^{-1.5  }  _{-0.1  }  $ & $^{+ 0.5  } _{-0.8  }  $ & $^{+ 0.4  } _{-1.2  }  
$ \\
\hline
  450      & $0.80\,\cdot\,10^{-2}$ & $
  1.02$ &
$^{+ 4.6  } _{- 4.5  } $ & $^{+ 2.1  } _{- 3.0  } $ & $^{+ 1.4  } _{- 1.2  } $ & $^{+ 0.1  } _{+ 0.6  } $ & $^{}_{- 0.0  } $ & $^{+ 0.5  } _{+ 0.4  } $ & $^{-1.0  }  _{+ 1.0  } $ & $^{-0.2  }  _{-1.1  }  $ & $^{+ 0.9  } _{-2.3  }  
$ \\
           & $0.13\,\cdot\,10^{-1}$ & $
  0.85$ &
$^{+ 6.2  } _{- 5.9  } $ & $^{+ 4.4  } _{- 4.2  } $ & $^{+ 3.3  } _{- 2.5  } $ & $^{-0.3  }  _{-0.4  }  $ & $^{}_{-0.3  }  $ & $^{-0.5  }  _{+ 0.2  } $ & $^{-2.9  }  _{-0.9  }  $ & $^{-0.3  }  _{-0.9  }  $ & $^{-1.2  }  _{+ 2.9  } 
$ \\
           & $0.21\,\cdot\,10^{-1}$ & $
  0.69$ &
$^{+ 6.1  } _{- 5.8  } $ & $^{+ 2.6  } _{- 1.1  } $ & $^{+ 2.1  } _{- 0.7  } $ & $^{+ 0.1  } _{+ 0.6  } $ & $^{}_{+ 0.3  } $ & $^{+ 0.5  } _{+ 0.4  } $ & $^{+ 1.1  } _{+ 0.3  } $ & $^{+ 0.5  } _{-0.4  }  $ & $^{+ 0.2  } _{-0.8  }  
$ \\
           & $0.32\,\cdot\,10^{-1}$ & $
  0.63$ &
$^{+ 5.5  } _{- 5.3  } $ & $^{+ 1.8  } _{- 1.8  } $ & $^{+ 1.8  } _{- 1.3  } $ & $^{-0.1  }  _{+ 0.0  } $ & $^{}_{+ 0.2  } $ & $^{-0.2  }  _{+ 0.2  } $ & $^{-1.1  }  _{+ 0.4  } $ & $^{+ 0.1  } _{-0.7  }  $ & $^{+ 0.0  } _{-0.2  }  
$ \\
           & $0.50\,\cdot\,10^{-1}$ & $
  0.51$ &
$^{+ 4.9  } _{- 4.8  } $ & $^{+ 1.8  } _{- 2.9  } $ & $^{+ 1.1  } _{- 0.8  } $ & $^{-0.1  }  _{+ 0.4  } $ & $^{}_{+ 0.1  } $ & $^{+ 0.0  } _{+ 0.3  } $ & $^{+ 0.5  } _{-0.3  }  $ & $^{+ 0.5  } _{-0.7  }  $ & $^{+ 1.2  } _{-2.7  }  
$ \\
           & $0.80\,\cdot\,10^{-1}$ & $
  0.44$ &
$^{+ 5.1  } _{- 4.9  } $ & $^{+ 1.6  } _{- 2.0  } $ & $^{+ 1.5  } _{- 1.4  } $ & $^{-0.5  }  _{+ 0.2  } $ & $^{}_{+ 0.0  } $ & $^{-0.1  }  _{+ 0.0  } $ & $^{+ 0.3  } _{-0.6  }  $ & $^{+ 0.1  } _{-1.1  }  $ & $^{+ 0.2  } _{-0.5  }  
$ \\
           & $0.13$ & $
  0.39$ &
$^{+ 5.3  } _{- 5.1  } $ & $^{+ 1.7  } _{- 1.7  } $ & $^{+ 1.5  } _{- 0.8  } $ & $^{+ 0.2  } _{+ 0.3  } $ & $^{}_{+ 0.2  } $ & $^{+ 0.2  } _{+ 0.2  } $ & $^{-0.4  }  _{-0.1  }  $ & $^{+ 0.4  } _{-0.8  }  $ & $^{+ 0.5  } _{-1.2  }  
$ \\
           & $0.25$ & $
  0.28$ &
$^{+ 6.0  } _{- 5.7  } $ & $^{+ 2.1  } _{- 5.6  } $ & $^{+ 1.7  } _{- 5.5  } $ & $^{+ 0.1  } _{+ 0.5  } $ & $^{}_{+ 0.3  } $ & $^{+ 0.4  } _{+ 0.2  } $ & $^{-0.8  }  _{+ 0.8  } $ & $^{+ 0.5  } _{-0.8  }  $ & $^{-0.1  }  _{+ 0.4  } 
$ \\
\hline
\end{tabular}
}

  \end{center}    
  \setlength{\localtextwidth}{18.7cm} 
  \caption[this space for rent]{
    Systematic uncertainties with bin-to-bin correlations
    for the reduced cross-section $\tilde{\sigma}(e^- p)$. The left
    part of the table contains the quoted $Q^2$ and $x$ values,
    $Q^2_c$ and $x_c$, the measured cross-section 
    $\tilde{\sigma}(e^- p)$ corrected to the Born level, the
    statistical error and the total systematic uncertainty. The right
    part of the table lists the total uncorrelated systematic
    uncertainty followed by the uncertainties $\delta_1$--\,$\delta_6$ (see text)
    with bin-to-bin correlations. For the latter, the upper (lower)
    numbers refer to positive (negative) variation of e.g. the cut
    value, whereas the signs of the numbers reflect the direction of
    change in the cross sections. 
    }
  \label{tab-dsdxq_c2}
\end{sidewaystable}

\begin{sidewaystable}
  \begin{center}
    {\small
\renewcommand{\arraystretch}{1.2}
\begin{tabular}{|c|l|c|c|c||c|c|c|c|c|c|c|}
\hline
{$Q^2_c$} & 
\multicolumn{1}{c|}{$x_c$} & 
 $\tilde{\sigma}(e^-p) $&
stat.  & 
total sys.  & 
uncor. sys. &
$\delta_1$ &
$\delta_2$ &
$\delta_3$ &
$\delta_4$ &
$\delta_5$ &
$\delta_6$ \\
{($\gev^2$)} & 
$$ &
 &
(\%) & 
(\%) &
(\%) &
(\%) &
(\%) &
(\%) &
(\%) &
(\%) &
(\%) \\
\hline
\hline
  650      & $0.80\,\cdot\,10^{-2}$ & $
  0.95$ &
$^{+ 6.2  } _{- 5.9  } $ & $^{+ 3.0  } _{- 2.8  } $ & $^{+ 2.7  } _{- 1.6  } $ & $^{-1.6  }  _{+ 1.3  } $ & $^{}_{+ 0.2  } $ & $^{+ 0.1  } _{+ 0.2  } $ & $^{-0.7  }  _{-0.1  }  $ & $^{- 0.0  } _{-1.3  }  $ & $^{-0.3  }  _{+ 0.6  } 
$ \\
           & $0.13\,\cdot\,10^{-1}$ & $
  0.94$ &
$^{+ 4.6  } _{- 4.4  } $ & $^{+ 1.3  } _{- 3.2  } $ & $^{+ 0.6  } _{- 2.0  } $ & $^{-0.4  }  _{-0.1  }  $ & $^{}_{-0.2  }  $ & $^{-0.4  }  _{-0.5  }  $ & $^{-1.9  }  _{+ 0.9  } $ & $^{-0.3  }  _{-1.4  }  $ & $^{-0.3  }  _{+ 0.7  } 
$ \\
           & $0.21\,\cdot\,10^{-1}$ & $
  0.79$ &
$^{+ 6.7  } _{- 6.3  } $ & $^{+ 0.8  } _{- 2.5  } $ & $^{+ 0.6  } _{- 2.0  } $ & $^{-0.1  }  _{+ 0.2  } $ & $^{}_{+ 0.1  } $ & $^{+ 0.1  } _{-0.2  }  $ & $^{-0.6  }  _{-0.5  }  $ & $^{+ 0.2  } _{-1.1  }  $ & $^{+ 0.4  } _{-0.7  }  
$ \\
           & $0.32\,\cdot\,10^{-1}$ & $
  0.60$ &
$^{+ 8.6  } _{- 8.1  } $ & $^{+ 2.9  } _{- 1.0  } $ & $^{+ 2.7  } _{- 0.6  } $ & $^{-0.1  }  _{+ 0.1  } $ & $^{}_{+ 0.1  } $ & $^{-0.2  }  _{+ 0.3  } $ & $^{+ 0.2  } _{+ 0.0  } $ & $^{+ 0.2  } _{-0.6  }  $ & $^{-0.5  }  _{+ 1.1  } 
$ \\
           & $0.50\,\cdot\,10^{-1}$ & $
  0.51$ &
$^{+ 8.4  } _{- 7.9  } $ & $^{+ 2.8  } _{- 3.5  } $ & $^{+ 1.6  } _{- 3.5  } $ & $^{+ 0.2  } _{+ 0.0  } $ & $^{}_{+ 0.1  } $ & $^{+ 0.4  } _{+ 0.2  } $ & $^{+ 0.0  } _{+ 2.1  } $ & $^{+ 0.0  } _{-0.2  }  $ & $^{-0.3  }  _{+ 0.8  } 
$ \\
           & $0.80\,\cdot\,10^{-1}$ & $
  0.40$ &
$^{+ 8.6  } _{- 8.0  } $ & $^{+ 2.0  } _{- 2.3  } $ & $^{+ 1.9  } _{- 2.0  } $ & $^{+ 0.1  } _{+ 0.2  } $ & $^{}_{-0.2  }  $ & $^{+ 0.1  } _{+ 0.1  } $ & $^{-0.5  }  _{+ 0.3  } $ & $^{+ 0.0  } _{-0.3  }  $ & $^{+ 0.4  } _{-1.0  }  
$ \\
           & $0.13$ & $
  0.35$ &
$^{+ 8.2  } _{- 7.7  } $ & $^{+ 1.8  } _{- 3.7  } $ & $^{+ 1.7  } _{- 3.1  } $ & $^{-0.6  }  _{+ 0.1  } $ & $^{}_{-0.1  }  $ & $^{-0.5  }  _{-0.4  }  $ & $^{-0.2  }  _{+ 0.1  } $ & $^{+ 0.3  } _{-1.1  }  $ & $^{+ 0.6  } _{-1.5  }  
$ \\
           & $0.25$ & $
  0.26$ &
$^{+ 8.3  } _{- 7.8  } $ & $^{+ 5.7  } _{- 3.9  } $ & $^{+ 5.7  } _{- 2.9  } $ & $^{-0.1  }  _{+ 0.2  } $ & $^{}_{-0.1  }  $ & $^{-0.2  }  _{+ 0.0  } $ & $^{-0.6  }  _{-0.5  }  $ & $^{+ 0.2  } _{-0.8  }  $ & $^{+ 1.0  } _{-2.4  }  
$ \\
\hline
  800      & $0.13\,\cdot\,10^{-1}$ & $
  0.89$ &
$^{+ 5.6  } _{- 5.3  } $ & $^{+ 3.0  } _{- 1.1  } $ & $^{+ 2.1  } _{- 1.0  } $ & $^{+ 0.6  } _{+ 0.7  } $ & $^{}_{+ 0.4  } $ & $^{+ 0.0  } _{+ 0.4  } $ & $^{+ 0.4  } _{+ 1.8  } $ & $^{+ 0.3  } _{-0.5  }  $ & $^{+ 0.2  } _{-0.4  }  
$ \\
           & $0.21\,\cdot\,10^{-1}$ & $
  0.81$ &
$^{+ 6.9  } _{- 6.5  } $ & $^{+ 1.7  } _{- 2.5  } $ & $^{+ 1.6  } _{- 1.6  } $ & $^{+ 0.1  } _{+ 0.2  } $ & $^{}_{+ 0.1  } $ & $^{-0.1  }  _{+ 0.1  } $ & $^{-0.7  }  _{-0.9  }  $ & $^{+ 0.1  } _{-1.2  }  $ & $^{+ 0.5  } _{-1.1  }  
$ \\
           & $0.32\,\cdot\,10^{-1}$ & $
  0.63$ &
$^{+ 7.3  } _{- 6.9  } $ & $^{+ 1.8  } _{- 4.2  } $ & $^{+ 0.7  } _{- 1.2  } $ & $^{+ 0.1  } _{+ 0.1  } $ & $^{}_{+ 0.1  } $ & $^{+ 0.1  } _{-0.2  }  $ & $^{-0.6  }  _{+ 0.1  } $ & $^{+ 0.2  } _{-1.0  }  $ & $^{+ 1.7  } _{-3.9  }  
$ \\
           & $0.50\,\cdot\,10^{-1}$ & $
  0.59$ &
$^{+ 7.0  } _{- 6.6  } $ & $^{+ 1.6  } _{- 2.3  } $ & $^{+ 1.4  } _{- 0.9  } $ & $^{-0.3  }  _{+ 0.1  } $ & $^{}_{+ 0.1  } $ & $^{-0.2  }  _{-0.3  }  $ & $^{-0.3  }  _{+ 0.0  } $ & $^{+ 0.3  } _{-0.8  }  $ & $^{+ 0.8  } _{-1.9  }  
$ \\
           & $0.80\,\cdot\,10^{-1}$ & $
  0.49$ &
$^{+ 8.0  } _{- 7.6  } $ & $^{+ 1.3  } _{- 3.4  } $ & $^{+ 0.6  } _{- 2.7  } $ & $^{-0.3  }  _{-0.3  }  $ & $^{}_{-0.3  }  $ & $^{-0.4  }  _{-0.3  }  $ & $^{-1.4  }  _{-1.2  }  $ & $^{-0.3  }  _{-1.2  }  $ & $^{-0.4  }  _{+ 1.2  } 
$ \\
           & $0.13$ & $
  0.34$ &
$^{+ 9.7  } _{- 9.0  } $ & $^{+ 3.0  } _{- 6.7  } $ & $^{+ 1.6  } _{- 2.4  } $ & $^{-0.4  }  _{-0.2  }  $ & $^{}_{-0.3  }  $ & $^{-0.4  }  _{-0.6  }  $ & $^{-2.4  }  _{-1.4  }  $ & $^{-0.4  }  _{-1.0  }  $ & $^{+ 2.6  } _{-5.7  }  
$ \\
           & $0.25$ & $
  0.26$ &
$^{+  12. } _{-  11. } $ & $^{+ 6.7  } _{- 2.2  } $ & $^{+ 4.7  } _{- 0.6  } $ & $^{+ 0.4  } _{+ 1.1  } $ & $^{}_{+ 0.7  } $ & $^{-0.2  }  _{+ 0.4  } $ & $^{-0.9  }  _{+ 0.5  } $ & $^{+ 1.3  } _{- 0.0  } $ & $^{-1.9  }  _{+ 4.3  } 
$ \\
\hline
\end{tabular}
}

  \end{center}    
  \setlength{\localtextwidth}{18.7cm} 
  \caption[this space for rent]{
    Systematic uncertainties with bin-to-bin correlations
    for the reduced cross-section $\tilde{\sigma}(e^- p)$. The left
    part of the table contains the quoted $Q^2$ and $x$ values,
    $Q^2_c$ and $x_c$, the measured cross-section 
    $\tilde{\sigma}(e^- p)$ corrected to the Born level, the
    statistical error and the total systematic uncertainty. The right
    part of the table lists the total uncorrelated systematic
    uncertainty followed by the uncertainties $\delta_1$--\,$\delta_6$ (see text)
    with bin-to-bin correlations. For the latter, the upper (lower)
    numbers refer to positive (negative) variation of e.g. the cut
    value, whereas the signs of the numbers reflect the direction of
    change in the cross sections. 
    }
  \label{tab-dsdxq_c3}
\end{sidewaystable}

\begin{sidewaystable}
  \begin{center}
    {\small
\renewcommand{\arraystretch}{1.2}
\begin{tabular}{|c|l|c|c|c||c|c|c|c|c|c|c|}
\hline
{$Q^2_c$} & 
\multicolumn{1}{c|}{$x_c$} & 
 $\tilde{\sigma}(e^-p) $&
stat.  & 
total sys.  & 
uncor. sys. &
$\delta_1$ &
$\delta_2$ &
$\delta_3$ &
$\delta_4$ &
$\delta_5$ &
$\delta_6$ \\
{($\gev^2$)} & 
$$ &
 &
(\%) & 
(\%) &
(\%) &
(\%) &
(\%) &
(\%) &
(\%) &
(\%) &
(\%) \\
\hline
\hline
 1200      & $0.14\,\cdot\,10^{-1}$ & $
  0.92$ &
$^{+ 6.7  } _{- 6.3  } $ & $^{+ 7.3  } _{- 2.9  } $ & $^{+ 3.2  } _{- 0.7  } $ & $^{- 0.0  } _{+ 0.5  } $ & $^{}_{-2.3  }  $ & $^{+ 1.3  } _{-0.2  }  $ & $^{-0.8  }  _{+ 5.5  } $ & $^{+ 1.1  } _{-0.6  }  $ & $^{-1.3  }  _{+ 3.0  } 
$ \\
           & $0.21\,\cdot\,10^{-1}$ & $
  0.69$ &
$^{+ 7.3  } _{- 6.9  } $ & $^{+ 1.7  } _{- 1.8  } $ & $^{+ 1.5  } _{- 1.2  } $ & $^{+ 0.1  } _{+ 0.4  } $ & $^{}_{+ 0.2  } $ & $^{+ 0.4  } _{-0.4  }  $ & $^{-0.3  }  _{-0.2  }  $ & $^{+ 0.2  } _{-1.0  }  $ & $^{+ 0.3  } _{-0.8  }  
$ \\
           & $0.32\,\cdot\,10^{-1}$ & $
  0.59$ &
$^{+ 7.2  } _{- 6.8  } $ & $^{+ 1.0  } _{- 3.1  } $ & $^{+ 0.9  } _{- 2.2  } $ & $^{-0.2  }  _{-0.2  }  $ & $^{}_{-0.2  }  $ & $^{-0.2  }  _{-0.1  }  $ & $^{-1.8  }  _{-0.4  }  $ & $^{+ 0.2  } _{-1.3  }  $ & $^{-0.1  }  _{+ 0.1  } 
$ \\
           & $0.50\,\cdot\,10^{-1}$ & $
  0.52$ &
$^{+ 6.4  } _{- 6.1  } $ & $^{+ 2.4  } _{- 1.1  } $ & $^{+ 1.9  } _{- 0.8  } $ & $^{+ 0.2  } _{+ 0.5  } $ & $^{}_{+ 0.6  } $ & $^{+ 0.3  } _{+ 0.3  } $ & $^{+ 1.0  } _{+ 0.9  } $ & $^{+ 0.4  } _{-0.8  }  $ & $^{- 0.0  } _{+ 0.2  } 
$ \\
           & $0.80\,\cdot\,10^{-1}$ & $
  0.43$ &
$^{+ 6.6  } _{- 6.3  } $ & $^{+ 1.7  } _{- 2.5  } $ & $^{+ 1.5  } _{- 1.0  } $ & $^{-0.1  }  _{+ 0.1  } $ & $^{}_{-0.1  }  $ & $^{+ 0.0  } _{-0.3  }  $ & $^{-1.4  }  _{-0.4  }  $ & $^{+ 0.5  } _{-1.3  }  $ & $^{+ 0.5  } _{-1.2  }  
$ \\
           & $0.13$ & $
  0.36$ &
$^{+ 7.0  } _{- 6.6  } $ & $^{+ 3.3  } _{- 1.5  } $ & $^{+ 3.1  } _{- 0.3  } $ & $^{+ 0.3  } _{+ 0.1  } $ & $^{}_{+ 0.1  } $ & $^{+ 0.2  } _{+ 0.3  } $ & $^{-0.8  }  _{+ 0.8  } $ & $^{+ 0.5  } _{-1.0  }  $ & $^{+ 0.2  } _{-0.6  }  
$ \\
           & $0.25$ & $
  0.27$ &
$^{+ 7.8  } _{- 7.3  } $ & $^{+ 1.4  } _{- 1.7  } $ & $^{+ 1.2  } _{- 1.2  } $ & $^{+ 0.2  } _{+ 0.0  } $ & $^{}_{+ 0.0  } $ & $^{+ 0.2  } _{+ 0.0  } $ & $^{+ 0.0  } _{-0.4  }  $ & $^{+ 0.7  } _{-1.1  }  $ & $^{+ 0.3  } _{-0.2  }  
$ \\
           & $0.40$ & $
  0.10$ &
$^{+  15. } _{-  13. } $ & $^{+ 4.1  } _{-  18. } $ & $^{+ 1.7  } _{-  18. } $ & $^{-0.2  }  _{+ 0.8  } $ & $^{}_{+ 0.3  } $ & $^{+ 0.8  } _{+ 0.3  } $ & $^{+ 3.1  } _{-0.5  }  $ & $^{+ 0.7  } _{-0.6  }  $ & $^{+ 1.6  } _{-3.7  }  
$ \\
\hline
 1500      & $0.21\,\cdot\,10^{-1}$ & $
  0.83$ &
$^{+ 9.1  } _{- 8.4  } $ & $^{+ 4.2  } _{- 2.0  } $ & $^{+ 2.7  } _{- 1.6  } $ & $^{+ 0.3  } _{+ 0.7  } $ & $^{}_{+ 1.7  } $ & $^{+ 0.4  } _{-0.5  }  $ & $^{-0.7  }  _{+ 2.5  } $ & $^{+ 0.4  } _{-0.9  }  $ & $^{-0.3  }  _{+ 0.8  } 
$ \\
           & $0.32\,\cdot\,10^{-1}$ & $
  0.75$ &
$^{+ 8.5  } _{- 7.9  } $ & $^{+ 2.7  } _{- 1.9  } $ & $^{+ 1.0  } _{- 0.7  } $ & $^{-0.1  }  _{+ 0.3  } $ & $^{}_{+ 0.0  } $ & $^{-0.4  }  _{+ 1.1  } $ & $^{+ 1.9  } _{-1.0  }  $ & $^{+ 0.0  } _{-1.3  }  $ & $^{-0.4  }  _{+ 1.0  } 
$ \\
           & $0.50\,\cdot\,10^{-1}$ & $
  0.61$ &
$^{+ 7.7  } _{- 7.2  } $ & $^{+ 1.0  } _{- 2.5  } $ & $^{+ 0.7  } _{- 1.2  } $ & $^{-0.2  }  _{+ 0.0  } $ & $^{}_{-0.1  }  $ & $^{+ 0.1  } _{-0.2  }  $ & $^{-1.7  }  _{- 0.0  } $ & $^{+ 0.4  } _{-0.8  }  $ & $^{+ 0.5  } _{-1.1  }  
$ \\
           & $0.80\,\cdot\,10^{-1}$ & $
  0.45$ &
$^{+ 8.4  } _{- 7.8  } $ & $^{+ 1.1  } _{- 2.6  } $ & $^{+ 1.0  } _{- 1.5  } $ & $^{-0.2  }  _{-0.3  }  $ & $^{}_{-0.3  }  $ & $^{+ 0.0  } _{-0.2  }  $ & $^{-1.4  }  _{+ 0.4  } $ & $^{-0.1  }  _{-1.2  }  $ & $^{+ 0.4  } _{-1.0  }  
$ \\
           & $0.13$ & $
  0.40$ &
$^{+ 9.9  } _{- 9.1  } $ & $^{+ 3.0  } _{- 3.0  } $ & $^{+ 0.8  } _{- 2.4  } $ & $^{-0.3  }  _{-0.1  }  $ & $^{}_{-0.1  }  $ & $^{-0.3  }  _{+ 0.1  } $ & $^{+ 0.1  } _{-0.8  }  $ & $^{-0.1  }  _{-1.0  }  $ & $^{-1.2  }  _{+ 2.9  } 
$ \\
           & $0.18$ & $
  0.33$ &
$^{+  10. } _{- 9.4  } $ & $^{+ 2.7  } _{- 1.4  } $ & $^{+ 2.5  } _{- 0.6  } $ & $^{+ 0.3  } _{+ 0.4  } $ & $^{}_{+ 0.2  } $ & $^{-0.1  }  _{+ 0.2  } $ & $^{-1.1  }  _{+ 0.5  } $ & $^{+ 0.5  } _{-0.7  }  $ & $^{-0.3  }  _{+ 0.5  } 
$ \\
           & $0.25$ & $
  0.27$ &
$^{+  13. } _{-  11. } $ & $^{+ 2.3  } _{- 0.9  } $ & $^{+ 2.1  } _{- 0.2  } $ & $^{+ 0.0  } _{+ 0.6  } $ & $^{}_{+ 0.2  } $ & $^{+ 0.6  } _{-0.1  }  $ & $^{-0.1  }  _{-0.1  }  $ & $^{+ 0.1  } _{-0.7  }  $ & $^{+ 0.2  } _{-0.5  }  
$ \\
           & $0.40$ & $
  0.14$ &
$^{+  20. } _{-  17. } $ & $^{+  40. } _{- 9.7  } $ & $^{+  40. } _{- 9.6  } $ & $^{-0.9  }  _{-0.2  }  $ & $^{}_{+ 0.7  } $ & $^{+ 0.2  } _{-0.2  }  $ & $^{+ 0.0  } _{-0.2  }  $ & $^{+ 0.0  } _{-0.7  }  $ & $^{-0.8  }  _{+ 1.5  } 
$ \\
\hline
\end{tabular}
}

  \end{center}    
  \setlength{\localtextwidth}{18.7cm} 
  \caption[this space for rent]{
    Systematic uncertainties with bin-to-bin correlations
    for the reduced cross-section $\tilde{\sigma}(e^- p)$. The left
    part of the table contains the quoted $Q^2$ and $x$ values,
    $Q^2_c$ and $x_c$, the measured cross-section 
    $\tilde{\sigma}(e^- p)$ corrected to the Born level, the
    statistical error and the total systematic uncertainty. The right
    part of the table lists the total uncorrelated systematic
    uncertainty followed by the uncertainties $\delta_1$--\,$\delta_6$ (see text)
    with bin-to-bin correlations. For the latter, the upper (lower)
    numbers refer to positive (negative) variation of e.g. the cut
    value, whereas the signs of the numbers reflect the direction of
    change in the cross sections. 
    }
  \label{tab-dsdxq_c4}
\end{sidewaystable}

\begin{sidewaystable}
  \begin{center}
    {\small
\renewcommand{\arraystretch}{1.2}
\begin{tabular}{|c|l|c|c|c||c|c|c|c|c|c|c|}
\hline
{$Q^2_c$} & 
\multicolumn{1}{c|}{$x_c$} & 
 $\tilde{\sigma}(e^-p) $&
stat.  & 
total sys.  & 
uncor. sys. &
$\delta_1$ &
$\delta_2$ &
$\delta_3$ &
$\delta_4$ &
$\delta_5$ &
$\delta_6$ \\
{($\gev^2$)} & 
$$ &
 &
(\%) & 
(\%) &
(\%) &
(\%) &
(\%) &
(\%) &
(\%) &
(\%) &
(\%) \\
\hline
\hline
 2000      & $0.32\,\cdot\,10^{-1}$ & $
  0.72$ &
$^{+  10. } _{- 9.3  } $ & $^{+ 6.3  } _{- 1.7  } $ & $^{+ 5.2  } _{- 1.0  } $ & $^{+ 0.8  } _{+ 1.0  } $ & $^{}_{-1.4  }  $ & $^{+ 1.7  } _{+ 0.7  } $ & $^{+ 1.0  } _{+ 2.5  } $ & $^{+ 1.4  } _{-0.1  }  $ & $^{- 0.0  } _{- 0.0  } 
$ \\
           & $0.50\,\cdot\,10^{-1}$ & $
  0.54$ &
$^{+  10. } _{- 9.3  } $ & $^{+ 1.7  } _{- 3.0  } $ & $^{+ 1.1  } _{- 1.5  } $ & $^{- 0.0  } _{+ 0.1  } $ & $^{}_{+ 0.2  } $ & $^{-0.1  }  _{-0.1  }  $ & $^{+ 0.7  } _{+ 0.1  } $ & $^{+ 0.3  } _{-0.8  }  $ & $^{+ 1.0  } _{-2.5  }  
$ \\
           & $0.80\,\cdot\,10^{-1}$ & $
  0.43$ &
$^{+  10. } _{- 9.3  } $ & $^{+ 2.4  } _{- 4.7  } $ & $^{+ 0.8  } _{- 2.7  } $ & $^{- 0.0  } _{- 0.0  } $ & $^{}_{+ 0.2  } $ & $^{+ 0.2  } _{+ 0.3  } $ & $^{+ 1.6  } _{+ 1.6  } $ & $^{+ 0.2  } _{-1.2  }  $ & $^{+ 1.6  } _{-3.6  }  
$ \\
           & $0.13$ & $
  0.38$ &
$^{+  12. } _{-  11. } $ & $^{+ 4.6  } _{- 1.9  } $ & $^{+ 4.5  } _{- 1.5  } $ & $^{+ 0.2  } _{+ 0.0  } $ & $^{}_{-0.2  }  $ & $^{-0.4  }  _{+ 0.0  } $ & $^{-0.3  }  _{+ 0.3  } $ & $^{+ 0.3  } _{-0.9  }  $ & $^{+ 0.3  } _{-0.6  }  
$ \\
           & $0.18$ & $
  0.28$ &
$^{+  13. } _{-  12. } $ & $^{+ 3.0  } _{- 2.8  } $ & $^{+ 2.8  } _{- 0.8  } $ & $^{-0.2  }  _{+ 0.1  } $ & $^{}_{-0.2  }  $ & $^{+ 0.1  } _{-0.5  }  $ & $^{-1.0  }  _{-0.5  }  $ & $^{+ 0.2  } _{-1.0  }  $ & $^{+ 1.0  } _{-2.3  }  
$ \\
           & $0.25$ & $
  0.29$ &
$^{+  14. } _{-  13. } $ & $^{+ 2.5  } _{- 2.6  } $ & $^{+ 1.7  } _{- 2.0  } $ & $^{+ 0.5  } _{+ 0.5  } $ & $^{}_{+ 0.3  } $ & $^{+ 0.4  } _{+ 1.0  } $ & $^{-0.1  }  _{+ 0.3  } $ & $^{+ 1.2  } _{-0.5  }  $ & $^{+ 0.7  } _{-1.6  }  
$ \\
           & $0.40$ & $
  0.10$ &
$^{+  24. } _{-  20. } $ & $^{+ 9.9  } _{- 7.5  } $ & $^{+ 8.9  } _{- 3.8  } $ & $^{-1.0  }  _{-0.5  }  $ & $^{}_{-1.0  }  $ & $^{-0.7  }  _{-0.2  }  $ & $^{-1.0  }  _{+ 3.6  } $ & $^{+ 0.0  } _{-1.7  }  $ & $^{+ 2.6  } _{-5.9  }  
$ \\
\hline
 3000      & $0.50\,\cdot\,10^{-1}$ & $
  0.68$ &
$^{+  12. } _{-  10. } $ & $^{+ 3.1  } _{- 4.4  } $ & $^{+ 1.5  } _{- 1.2  } $ & $^{+ 0.1  } _{+ 0.4  } $ & $^{}_{+ 0.9  } $ & $^{-0.5  }  _{+ 1.1  } $ & $^{-4.0  }  _{+ 2.2  } $ & $^{+ 0.5  } _{-0.9  }  $ & $^{+ 0.3  } _{-0.9  }  
$ \\
           & $0.80\,\cdot\,10^{-1}$ & $
  0.45$ &
$^{+  12. } _{-  11. } $ & $^{+ 2.3  } _{- 3.1  } $ & $^{+ 1.4  } _{- 0.6  } $ & $^{- 0.0  } _{+ 0.0  } $ & $^{}_{+ 0.3  } $ & $^{+ 0.0  } _{+ 0.3  } $ & $^{+ 1.2  } _{+ 0.9  } $ & $^{+ 0.4  } _{-0.8  }  $ & $^{+ 1.2  } _{-2.9  }  
$ \\
           & $0.13$ & $
  0.38$ &
$^{+  15. } _{-  13. } $ & $^{+ 2.9  } _{- 7.5  } $ & $^{+ 2.2  } _{- 7.4  } $ & $^{+ 0.4  } _{+ 0.5  } $ & $^{}_{- 0.0  } $ & $^{+ 0.5  } _{+ 0.4  } $ & $^{+ 1.7  } _{+ 0.0  } $ & $^{+ 0.5  } _{-0.6  }  $ & $^{+ 0.2  } _{-0.4  }  
$ \\
           & $0.18$ & $
  0.30$ &
$^{+  16. } _{-  14. } $ & $^{+ 3.2  } _{- 2.4  } $ & $^{+ 2.8  } _{- 2.2  } $ & $^{- 0.0  } _{+ 0.1  } $ & $^{}_{-0.2  }  $ & $^{+ 0.1  } _{+ 0.3  } $ & $^{+ 0.8  } _{-0.4  }  $ & $^{+ 0.3  } _{-0.7  }  $ & $^{-0.4  }  _{+ 1.2  } 
$ \\
           & $0.25$ & $
  0.28$ &
$^{+  17. } _{-  15. } $ & $^{+ 4.4  } _{- 4.5  } $ & $^{+ 0.8  } _{- 3.6  } $ & $^{-0.1  }  _{+ 0.0  } $ & $^{}_{+ 0.0  } $ & $^{+ 0.1  } _{+ 0.2  } $ & $^{+ 0.7  } _{-1.7  }  $ & $^{+ 0.1  } _{-0.9  }  $ & $^{-1.8  }  _{+ 4.2  } 
$ \\
           & $0.40$ & $
  0.16$ &
$^{+  25. } _{-  21. } $ & $^{+ 6.1  } _{- 5.0  } $ & $^{+ 4.6  } _{- 4.6  } $ & $^{-0.3  }  _{-0.3  }  $ & $^{}_{+ 1.0  } $ & $^{+ 0.2  } _{+ 0.7  } $ & $^{+ 0.9  } _{-0.3  }  $ & $^{-1.1  }  _{-1.3  }  $ & $^{-1.6  }  _{+ 3.7  } 
$ \\
           & $0.65$ & $
  0.02$ &
$^{+  51. } _{-  36. } $ & $^{+ 9.0  } _{- 5.1  } $ & $^{+ 8.4  } _{- 3.5  } $ & $^{+ 0.0  } _{+ 2.1  } $ & $^{}_{+ 0.0  } $ & $^{+ 1.5  } _{-1.0  }  $ & $^{+ 0.2  } _{+ 0.0  } $ & $^{+ 0.8  } _{-0.6  }  $ & $^{+ 1.5  } _{-3.5  }  
$ \\
\hline
\end{tabular}
}

  \end{center}    
  \setlength{\localtextwidth}{18.7cm} 
  \caption[this space for rent]{
    Systematic uncertainties with bin-to-bin correlations
    for the reduced cross-section $\tilde{\sigma}(e^- p)$. The left
    part of the table contains the quoted $Q^2$ and $x$ values,
    $Q^2_c$ and $x_c$, the measured cross-section 
    $\tilde{\sigma}(e^- p)$ corrected to the Born level, the
    statistical error and the total systematic uncertainty. The right
    part of the table lists the total uncorrelated systematic
    uncertainty followed by the uncertainties $\delta_1$--\,$\delta_6$ (see text)
    with bin-to-bin correlations. For the latter, the upper (lower)
    numbers refer to positive (negative) variation of e.g. the cut
    value, whereas the signs of the numbers reflect the direction of
    change in the cross sections. 
    }
  \label{tab-dsdxq_c5}
\end{sidewaystable}

\begin{sidewaystable}
  \begin{center}
    {\small
\renewcommand{\arraystretch}{1.2}
\begin{tabular}{|c|l|c|c|c||c|c|c|c|c|c|c|}
\hline
{$Q^2_c$} & 
\multicolumn{1}{c|}{$x_c$} & 
 $\tilde{\sigma}(e^-p) $&
stat.  & 
total sys.  & 
uncor. sys. &
$\delta_1$ &
$\delta_2$ &
$\delta_3$ &
$\delta_4$ &
$\delta_5$ &
$\delta_6$ \\
{($\gev^2$)} & 
$$ &
 &
(\%) & 
(\%) &
(\%) &
(\%) &
(\%) &
(\%) &
(\%) &
(\%) &
(\%) \\
\hline
\hline
 5000      & $0.80\,\cdot\,10^{-1}$ & $
  0.60$ &
$^{+ 9.9  } _{- 9.1  } $ & $^{+ 3.3  } _{- 0.9  } $ & $^{+ 2.5  } _{- 0.6  } $ & $^{+ 0.4  } _{+ 0.6  } $ & $^{}_{-0.5  }  $ & $^{+ 0.9  } _{+ 0.3  } $ & $^{+ 0.1  } _{+ 1.8  } $ & $^{+ 0.7  } _{-0.4  }  $ & $^{+ 0.1  } _{-0.2  }  
$ \\
           & $0.13$ & $
  0.44$ &
$^{+  15. } _{-  13. } $ & $^{+ 2.7  } _{- 1.8  } $ & $^{+ 2.7  } _{- 0.4  } $ & $^{+ 0.2  } _{+ 0.3  } $ & $^{}_{-0.2  }  $ & $^{-0.1  }  _{-1.2  }  $ & $^{-0.9  }  _{-0.3  }  $ & $^{+ 0.1  } _{-0.9  }  $ & $^{-0.1  }  _{+ 0.1  } 
$ \\
           & $0.18$ & $
  0.41$ &
$^{+  14. } _{-  13. } $ & $^{+ 3.3  } _{- 3.3  } $ & $^{+ 2.8  } _{- 1.3  } $ & $^{-0.1  }  _{+ 0.2  } $ & $^{}_{+ 0.2  } $ & $^{+ 0.2  } _{+ 0.5  } $ & $^{+ 0.8  } _{-0.2  }  $ & $^{+ 0.5  } _{-0.6  }  $ & $^{+ 1.3  } _{-3.0  }  
$ \\
           & $0.25$ & $
  0.27$ &
$^{+  18. } _{-  16. } $ & $^{+ 1.6  } _{- 2.6  } $ & $^{+ 1.2  } _{- 0.4  } $ & $^{+ 0.0  } _{-0.3  }  $ & $^{}_{+ 0.1  } $ & $^{+ 0.0  } _{+ 0.2  } $ & $^{-0.1  }  _{-0.2  }  $ & $^{+ 0.3  } _{-0.6  }  $ & $^{+ 1.1  } _{-2.5  }  
$ \\
           & $0.40$ & $
  0.16$ &
$^{+  25. } _{-  20. } $ & $^{+ 4.0  } _{-  11. } $ & $^{+ 3.7  } _{- 6.8  } $ & $^{-0.4  }  _{+ 0.9  } $ & $^{}_{+ 0.4  } $ & $^{+ 0.7  } _{-3.5  }  $ & $^{-7.5  }  _{+ 0.2  } $ & $^{+ 0.7  } _{-0.9  }  $ & $^{-0.1  }  _{+ 0.3  } 
$ \\
\hline
 8000      & $0.13$ & $
  0.53$ &
$^{+  15. } _{-  13. } $ & $^{+ 3.0  } _{- 5.0  } $ & $^{+ 0.7  } _{- 3.7  } $ & $^{-0.5  }  _{+ 0.3  } $ & $^{}_{-2.2  }  $ & $^{+ 2.5  } _{+ 2.5  } $ & $^{-2.2  }  _{+ 1.3  } $ & $^{+ 0.3  } _{-1.0  }  $ & $^{+ 0.3  } _{-0.6  }  
$ \\
           & $0.18$ & $
  0.38$ &
$^{+  20. } _{-  17. } $ & $^{+ 4.0  } _{- 4.1  } $ & $^{+ 2.5  } _{- 3.2  } $ & $^{-0.1  }  _{+ 0.1  } $ & $^{}_{-0.3  }  $ & $^{-0.2  }  _{+ 0.0  } $ & $^{+ 0.2  } _{+ 3.0  } $ & $^{-0.1  }  _{-1.0  }  $ & $^{+ 0.9  } _{-2.2  }  
$ \\
           & $0.25$ & $
  0.31$ &
$^{+  24. } _{-  19. } $ & $^{+ 2.1  } _{- 1.4  } $ & $^{+ 1.7  } _{- 0.5  } $ & $^{+ 0.1  } _{+ 0.1  } $ & $^{}_{+ 0.2  } $ & $^{+ 0.3  } _{+ 0.2  } $ & $^{+ 1.0  } _{-0.2  }  $ & $^{+ 0.4  } _{-0.5  }  $ & $^{+ 0.5  } _{-1.1  }  
$ \\
           & $0.40$ & $
  0.14$ &
$^{+  37. } _{-  28. } $ & $^{+ 6.1  } _{- 8.5  } $ & $^{+ 3.1  } _{- 1.0  } $ & $^{-0.2  }  _{+ 0.3  } $ & $^{}_{-0.5  }  $ & $^{+ 0.4  } _{-0.7  }  $ & $^{-8.1  }  _{-0.8  }  $ & $^{+ 0.7  } _{-0.4  }  $ & $^{-2.2  }  _{+ 5.1  } 
$ \\
           & $0.65$ & $
  0.02$ &
$^{+  80. } _{-  48. } $ & $^{+ 6.7  } _{- 1.6  } $ & $^{+ 6.5  } _{- 1.4  } $ & $^{+ 0.1  } _{+ 0.0  } $ & $^{}_{+ 0.0  } $ & $^{+ 0.4  } _{+ 0.4  } $ & $^{+ 1.0  } _{+ 0.0  } $ & $^{-0.4  }  _{-0.2  }  $ & $^{-0.6  }  _{+ 1.1  } 
$ \\
\hline
12000      & $0.18$ & $
  0.45$ &
$^{+  22. } _{-  18. } $ & $^{+ 6.3  } _{- 2.9  } $ & $^{+ 3.2  } _{- 0.9  } $ & $^{-0.7  }  _{+ 0.6  } $ & $^{}_{+ 3.4  } $ & $^{-1.1  }  _{+ 3.0  } $ & $^{-1.7  }  _{+ 2.9  } $ & $^{+ 0.1  } _{-1.2  }  $ & $^{+ 0.6  } _{-1.3  }  
$ \\
           & $0.25$ & $
  0.27$ &
$^{+  37. } _{-  28. } $ & $^{+ 3.1  } _{- 8.5  } $ & $^{+ 1.1  } _{- 8.3  } $ & $^{-0.1  }  _{+ 0.1  } $ & $^{}_{-0.1  }  $ & $^{-0.5  }  _{+ 0.7  } $ & $^{+ 0.2  } _{-0.2  }  $ & $^{+ 0.4  } _{-1.1  }  $ & $^{-1.2  }  _{+ 2.8  } 
$ \\
           & $0.40$ & $
  0.11$ &
$^{+  60. } _{-  40. } $ & $^{+  18. } _{- 0.6  } $ & $^{+  18. } _{- 0.1  } $ & $^{+ 0.2  } _{+ 0.9  } $ & $^{}_{+ 0.2  } $ & $^{+ 0.6  } _{+ 0.7  } $ & $^{+ 0.9  } _{+ 0.2  } $ & $^{+ 0.7  } _{-0.4  }  $ & $^{-0.5  }  _{+ 1.3  } 
$ \\
\hline
20000      & $0.25$ & $
  0.43$ &
$^{+  35. } _{-  27. } $ & $^{+ 5.7  } _{- 2.0  } $ & $^{+ 1.2  } _{- 0.8  } $ & $^{-0.6  }  _{+ 0.3  } $ & $^{}_{+ 0.0  } $ & $^{-1.2  }  _{+ 4.6  } $ & $^{+ 3.1  } _{-0.6  }  $ & $^{+ 0.2  } _{-1.0  }  $ & $^{+ 0.1  } _{-0.1  }  
$ \\
           & $0.40$ & $
  0.21$ &
$^{+  60. } _{-  40. } $ & $^{+ 3.6  } _{- 2.0  } $ & $^{+ 2.5  } _{- 1.1  } $ & $^{-0.4  }  _{+ 0.2  } $ & $^{}_{- 0.0  } $ & $^{-1.0  }  _{+ 2.3  } $ & $^{+ 1.0  } _{-0.3  }  $ & $^{+ 0.1  } _{-1.2  }  $ & $^{-0.1  }  _{+ 0.3  } 
$ \\
\hline
30000      & $0.40$ & $
  0.23$ &
$^{+  80. } _{-  48. } $ & $^{+  11. } _{- 2.4  } $ & $^{+ 0.8  } _{- 0.7  } $ & $^{-0.6  }  _{+ 0.8  } $ & $^{}_{+ 9.5  } $ & $^{-1.8  }  _{+ 4.7  } $ & $^{+ 2.4  } _{-0.3  }  $ & $^{+ 0.2  } _{-0.9  }  $ & $^{-1.0  }  _{+ 2.2  } 
$ \\
\hline
\end{tabular}
}

  \end{center}    
  \setlength{\localtextwidth}{18.7cm} 
  \caption[this space for rent]{
    Systematic uncertainties with bin-to-bin correlations
    for the reduced cross-section $\tilde{\sigma}(e^- p)$. The left
    part of the table contains the quoted $Q^2$ and $x$ values,
    $Q^2_c$ and $x_c$, the measured cross-section 
    $\tilde{\sigma}(e^- p)$ corrected to the Born level, the
    statistical error and the total systematic uncertainty. The right
    part of the table lists the total uncorrelated systematic
    uncertainty followed by the uncertainties $\delta_1$--\,$\delta_6$ (see text)
    with bin-to-bin correlations. For the latter, the upper (lower)
    numbers refer to positive (negative) variation of e.g. the cut
    value, whereas the signs of the numbers reflect the direction of
    change in the cross sections. 
    }
  \label{tab-dsdxq_c6}
\end{sidewaystable}

\clearpage

\begin{sidewaystable} 
  \begin{center}
    {\small
\renewcommand{\arraystretch}{1.2}
\begin{tabular}{|r@{ -- }l|r@{ -- }l@{$\,$}r|r|l|l@{}r@{$\,$}l@{$\,$}l@{$\,$}l|r|r|r|}
\hline
\multicolumn{2}{|c|}{ $Q^2$ range} & 
\multicolumn{3}{c|}{ $x$ range} & 
\multicolumn{1}{c|}{$Q^2_c$} & 
\multicolumn{1}{c|}{$x_c$} &
\multicolumn{6}{c|}{$xF_3$} &
{$N_{e^-p}$} & 
{$N_{e^+p}$} \\
\cline{8-13}
\multicolumn{2}{|c|} {($\Gev^2$)} &
\multicolumn{3}{c|}{} & 
{($\Gev^2$)} && 
\multicolumn{5}{c|}{measured} & 
\multicolumn{1}{c|}{SM} && \\ 
\hline
\hline $  1300$ & $  2500$ & $0.017$ & $0.060$ & $$ & $ 1500$ & $0.050$$$ & $\left( \right.\! \! $ & $  12.2$ & $\pm   4.1$ & $^{+  2.2}_{-  1.3}$ & $ \left. \! \right) \cdot 10^{-2}$ & $   3.7\cdot 10^{-2}$ &
$  783$ & $ 1112$ \\
\multicolumn{2}{|c|}{} & $0.060$ & $0.230$ & $$ & $$ & $0.130$$$ & $\left( \right.\! \! $ & $   4.2$ & $\pm   7.4$ & $^{+  3.3}_{-  2.7}$ & $ \left. \! \right) \cdot 10^{-2}$ & $   4.4\cdot 10^{-2}$ &
$  719$ & $ 1230$ \\
\multicolumn{2}{|c|}{} & $0.230$ & $0.530$ & $$ & $$ & $0.400$$$ & $\left( \right.\! \! $ & $  -0.5$ & $^{+   1.6}_{-  1.5}$ & $^{+  1.4}_{-  0.6}$ & $ \left. \! \right) \cdot 10^{-1}$ & $   0.2\cdot 10^{-1}$ &
$  213$ & $  373$ \\
\hline $  2500$ & $  3500$ & $0.037$ & $0.060$ & $$ & $ 3000$ & $0.050$$$ & $\left( \right.\! \! $ & $   8.0$ & $^{+   6.2}_{-  5.7}$ & $^{+  2.0}_{-  2.1}$ & $ \left. \! \right) \cdot 10^{-2}$ & $   6.6\cdot 10^{-2}$ &
$   95$ & $  140$ \\
\multicolumn{2}{|c|}{} & $0.060$ & $0.230$ & $$ & $$ & $0.130$$$ & $\left( \right.\! \! $ & $   6.2$ & $^{+   6.6}_{-  6.2}$ & $^{+  2.6}_{-  1.3}$ & $ \left. \! \right) \cdot 10^{-2}$ & $   7.6\cdot 10^{-2}$ &
$  200$ & $  321$ \\
\multicolumn{2}{|c|}{} & $0.230$ & $0.750$ & $$ & $$ & $0.400$$$ & $\left( \right.\! \! $ & $   3.0$ & $^{+   1.4}_{-  1.3}$ & $^{+  0.3}_{-  1.0}$ & $ \left. \! \right) \cdot 10^{-1}$ & $   0.3\cdot 10^{-1}$ &
$   77$ & $   93$ \\
\hline $  3500$ & $  5600$ & $0.040$ & $0.230$ & $$ & $ 5000$ & $0.130$$$ & $\left( \right.\! \! $ & $  16.9$ & $^{+   4.0}_{-  3.7}$ & $^{+  1.8}_{-  0.8}$ & $ \left. \! \right) \cdot 10^{-2}$ & $  10.8\cdot 10^{-2}$ &
$  253$ & $  296$ \\
\multicolumn{2}{|c|}{} & $0.230$ & $0.530$ & $$ & $$ & $0.400$$$ & $\left( \right.\! \! $ & $   7.5$ & $^{+   8.7}_{-  7.7}$ & $^{+  3.0}_{-  2.6}$ & $ \left. \! \right) \cdot 10^{-2}$ & $   4.7\cdot 10^{-2}$ &
$   65$ & $   91$ \\
\hline $  5600$ & $  9000$ & $0.070$ & $0.230$ & $$ & $ 8000$ & $0.180$$$ & $\left( \right.\! \! $ & $  15.5$ & $^{+   4.9}_{-  4.4}$ & $^{+  1.9}_{-  1.4}$ & $ \left. \! \right) \cdot 10^{-2}$ & $  13.4\cdot 10^{-2}$ &
$   94$ & $   91$ \\
\multicolumn{2}{|c|}{} & $0.230$ & $0.750$ & $$ & $$ & $0.400$$$ & $\left( \right.\! \! $ & $   4.6$ & $^{+   6.8}_{-  5.8}$ & $^{+  1.3}_{-  1.7}$ & $ \left. \! \right) \cdot 10^{-2}$ & $   6.0\cdot 10^{-2}$ &
$   42$ & $   60$ \\
\hline $  9000$ & $ 15000$ & $0.110$ & $0.230$ & $$ & $12000$ & $0.180$$$ & $\left( \right.\! \! $ & $  13.0$ & $^{+   6.9}_{-  5.7}$ & $^{+  2.1}_{-  0.9}$ & $ \left. \! \right) \cdot 10^{-2}$ & $  16.0\cdot 10^{-2}$ &
$   29$ & $   25$ \\
\multicolumn{2}{|c|}{} & $0.230$ & $0.530$ & $$ & $$ & $0.400$$$ & $\left( \right.\! \! $ & $   3.8$ & $^{+   5.9}_{-  4.7}$ & $^{+  1.2}_{-  1.5}$ & $ \left. \! \right) \cdot 10^{-2}$ & $   7.1\cdot 10^{-2}$ &
$   18$ & $   22$ \\
\hline $ 15000$ & $ 50000$ & $0.180$ & $0.750$ & $$ & $30000$ & $0.400$$$ & $\left( \right.\! \! $ & $   9.3$ & $^{+   3.3}_{-  2.7}$ & $^{+  0.7}_{-  0.3}$ & $ \left. \! \right) \cdot 10^{-2}$ & $   9.0\cdot 10^{-2}$ &
$   23$ & $   13$ \\
\hline
\end{tabular}
}

  \end{center}    
  \setlength{\localtextwidth}{18.5cm} 
  \caption[this space for rent]{
    The structure function $xF_3$  for the reaction $e^{\pm} p \rightarrow e^{\pm} X$. 
    The following quantities are given for each bin: the \qq\ and $x$ ranges,
    the values at which $xF_3$ is quoted, \qqc\ and $x_c$, the
    value of the measured structure function and the $xF_3$ value predicted
    by the SM using CTEQ5D PDFs.
    The first error of the measured value gives the statistical error and the
    second is the systematic uncertainty. The last two columns contain the number
    of events in each of the $e^-p$ and $e^+p$ data samples.
  }
  \label{tab-xF3}
\end{sidewaystable}
\clearpage

%
%
\begin{figure}
  \begin{center}
    \includegraphics[width=.9\textwidth]{figures/fig.2}
  \end{center}
  \caption{Double-differential bins at $\sqrt{s} = 318 \gev$ in the
    $x$-$Q^2$ plane. The heavy solid line $y=1$ marks the kinematic
    limit. The two solid diagonal lines are lines at $y = 0.1$ and
    $y=0.01$, whereas the curved line is a line at $E'_e = 10 \gev$.
    The dashed lines are lines of constant $\theta_e$ and mark the
    transition regions between R/BCAL ($\theta_e = 2.25 \rad$) and
    B/FCAL ($\theta_e = 0.64 \rad$).  The dash-dotted line
    $y(1-x)^2=0.004$ indicates the validity limit of the MC
    simulation. The number of data events is displayed in each bin.}
  \label{fig-KinePlan}
\end{figure}
%
%
%
%

%
%
%
%

%
%
\end{document}